\documentclass[fleqn,usenatbib]{mnras}

\usepackage{newtxtext,newtxmath}

\usepackage[T1]{fontenc}

\DeclareRobustCommand{\VAN}[3]{#2}
\let\VANthebibliography\thebibliography
\def\thebibliography{\DeclareRobustCommand{\VAN}[3]{##3}\VANthebibliography}

\usepackage{graphicx}	
\usepackage{amsmath}	
\usepackage{csquotes}
\usepackage{graphicx}
\usepackage{subfigure}
\usepackage{enumerate}
\usepackage[dvipsnames]{xcolor}
\usepackage{mwe}
\usepackage{soul}
\usepackage{comment}

\title[Type Ia SNe feedback in GCs]{On the role of Type Ia supernovae in the second generation star formation in globular clusters}

\author[E. Lacchin et al.]{
E. Lacchin$^{1,2}$\thanks{E-mail: elena.lacchin@inaf.it},
F. Calura $^{1}$, 
E. Vesperini $^{3}$\\
$^{1}$ INAF - OAS, Osservatorio di Astrofisica e Scienza dello Spazio di Bologna, via Gobetti 93/3, I-40129 Bologna, Italy\\
$^{2}$ Department of Physics and Astronomy, University of Bologna, via Gobetti 93/3, 40129 Bologna, Italy\\
$^{3}$ Department of Astronomy, Indiana University, Swain West, 727 E. 3rd Street, Bloomington, IN 47405, USA}
\date{Accepted XXX. Received YYY; in original form ZZZ}

\pubyear{2021}

\begin{document}
\label{firstpage}
\pagerange{\pageref{firstpage}--\pageref{lastpage}}
\maketitle

\defcitealias{calura2019}{C19}

\begin{abstract}

By means of 3D hydrodynamic simulations, we study how Type Ia supernovae (SNe) explosions affect the star formation history and the chemical properties of second generation (SG) stars in globular clusters (GC).  
SG stars are assumed to form once first generation asymptotic giant branch (AGB) stars start releasing their ejecta; during this phase, external gas is accreted by the system and SNe Ia begin exploding, carving hot and tenuous bubbles. Given the large uncertainty on SNe Ia explosion times, we test two different values for the \enquote{delay time}. We run two different models for the external gas density: in the low-density scenario with short delay time, the explosions start at the beginning of the SG star formation, halting it in its earliest phases. The external gas hardly penetrates the system, therefore most SG stars present extreme helium abundances (${\rm Y>0.33}$). The low-density model with delayed SN explosions has a more extended SG star formation epoch and includes SG stars with modest helium enrichment. On the contrary, the high-density model is weakly affected by SN explosions, with a final SG mass similar to the one obtained without SNe Ia. Most of the stars form from a mix of AGB ejecta and pristine gas and have a modest helium enrichment.
We show that gas from SNe Ia may produce an iron spread of $\sim 0.14$ dex, consistent with the spread found in  
about $20\%$  of Galactic GCs, suggesting that SNe Ia might have played a key role in the formation of this sub-sample of GCs. 

\end{abstract}




\section{Introduction}
\label{sec:1}
Over the last decades many photometric and spectroscopic studies have shown that Galactic globular clusters (GCs) host multiple stellar populations, therefore they can no longer be considered a single stellar population (SSP). First evidences of the presence of different populations within the same GC came from studies on the chemical composition of the host stars, a fraction of which showed chemical abundances rarely found in field stars; these anomalies concern mainly light elements, such as C, N, Na, O, Mg and Al (see e.g. \citealt{carretta2009a}, \citealt{carretta2009b}, \citealt{carretta2009c}, \citealt{gratton2012}, \citealt{masseron2019}, \citealt{gratton2019} and references therein). Later, photometric studies revealed that GCs are characterized by a splitting of sequences in color-magnitude diagram, which have provided further evidences of the presence multiple populations within the same GC \citep{lee1999,pancino2000,bedin2004,piotto2005, piotto2007,piotto2009,marino2008,piotto2015,milone2017,marino2019}. Stars with chemical composition similar to field stars are considered to represent the first generation (FG) population whereas \enquote{anomalous} stars i.e. stars showing enhanced in He, N and N and depleted in O and C, are associated to subsequent stellar generations. However, how and when these peculiar stars were formed is still an open question. 
 Several scenarios have been proposed so far to account for the origin of multiple populations in GCs, however none of them is able to explain all the features and the trends observed in GCs without encountering some difficulties (see \citealt{renzini2015, bastian2018,gratton2019}). 
Among the various theories proposed to date, the most thoroughly studied one is the asymptotic giant branch (AGB) scenario \citep{dantona2004,dercole2008,bekki2011}, in which the first generation stars are supposed to be formed at once and with the same chemical composition (i.e. an SSP). Then the feedback from massive stars, both in the pre-supernova (SN) and SN phases, has cleared the system from all the gas, therefore no enriched material, coming from the ejecta of massive stars, has been left in the system \citep{calura2015}. At this stage, a new stellar generation can form from the ejecta of FG AGB stars plus diluted gas with same composition as FG stars (see \citealt{dercole2010,dercole2012,dantona2016}).  

Other scenarios assume that second generation (SG) stars forms from gas ejected by fast-rotating massive stars \citep{decressin2007},  massive interactive binaries \citep{demink2009}, massive stars \citep{elmegreen2017}, very massive ($ m{\rm \sim 10^4 M_{\odot}}$) stars \citep{denissenkov&hartwick2014,gieles2018} or black holes accretion discs \citep{breen2018}. Many are the observational constraints that these scenarios should reproduce, like the chemical patterns of the SG stars, the relative number of FG stars and SG stars and their spatial distribution. 

The work we are presenting here is accomplished in the AGB framework and is aimed at extending the works of \citet{dercole2008} and \citet[hereafter \citetalias{calura2019}]{calura2019}. 
Focusing on the AGB scenario, \citet{dercole2008} shows that the gas ejected by AGBs collects in a cooling flow towards
the cluster core and later condenses forming SG stars, which are therefore more centrally concentrated than the FG ones. In addition, they show that the AGB yields alone are not able to reproduce the observed SG abundance patterns, especially when light elements are concerned. In particular, AGB models predict a direct correlation between sodium and oxygen, while observationally an anticorrelation is obtained \citep{carretta2009a}. The proposed solution is to assume a dilution of the AGB ejecta with pristine gas (i.e. same chemical composition of the gas out of which the FG stars were formed) during the formation of the SG stars \citep{dercole2008,dercole2012, dercole2016,dantona2016}.  Dilution is required by all the other models as well in order to reproduce the observed chemical abundance patterns.
  Moreover, since the physical processes which cause the end of the star formation (SF) are still not known, \citet{dercole2008} ran simulations including also Type Ia SNe to test whether they could halt the SF in a GC. They performed 1D hydrodynamic simulations taking into account the feedback from Type Ia SNe comparing the results with a case without such feedback. They found that, assuming a constant SN rate, the gas is rapidly wiped out from the system almost immediately after the first explosions, which results in a sudden halt of the SF. 

Recently, \citetalias{calura2019} performed 3D hydrodynamic simulations of a star-forming massive GC moving through a uniform distribution of gas to study the formation of SG stars, including also mass return from AGB stars and studying the He content of SG stars. They varied the density of the pristine gas, studying its effects on the spatial distribution of SG stars. They concluded that the most He-rich SG stars were concentrated in more central regions than the less enriched ones. However, in this study the feedback of Type Ia SNe was not included.
Their effect on the gas removal might be different from the one obtained in 1D simulations but also from what has been found by \citet[]{calura2015} and more similar to what has been recently obtained
by \citet{romano2019}. \citet{romano2019} found that the filling factor of the superbubbles created by sparse energy sources in an ultra-faint dwarf galaxy is quite small and the bulk of the cold gas is not removed. We intend to verify whether stellar feedback from Type Ia SNe of the FG could have the same effect in denser systems like GCs.


In general, since the bulk of GCs are characterized by a narrow internal iron dispersion of ${\rm \sigma_{[Fe/H]}<0.1 dex}$ \citep{carretta2009c}, Type Ia SNe belonging to the FG, which are significant Fe producers, should not provide a significant contribution to the gas out of which SG stars are formed but they could provide the mechanism responsible for halting the SG formation. On the other hand, about 20 per cent of the Galactic GC (referred to as Type II clusters in \citealt{milone2017}; see also \citealt{johnson2015}) are instead characterized by a significant dispersion in Fe; the origin of this spread and its link with the SG formation is still unknown.

In this paper we present, for the first time, a series of 3D hydrodynamic simulations to explore the effects of the feedback of SN Ia explosions on the
duration of the SG star formation phase and on the chemical properties of the SG stars.

This paper is organized as follows: in Section \ref{sec:simuset} we describe the set-up of the model we are adopting, together with the main ingredients we have taken into account focusing in particular on the Type Ia SN feedback. In Section \ref{sec:results} we present the results we have obtained which are then discussed in  Section \ref{sec:discuss}. Finally, in Section \ref{sec:conclu}, we draw our conclusions.

\begin{table}
\centering
\caption{Parameters of the simulations.}
\hspace{-0.5cm}
\resizebox{1.05\columnwidth}{!}{
\renewcommand{\arraystretch}{1.40}
\begin{tabular}{llc} 
\hline
\hline
Parameter & Description & Adopted values \\
\hline
$M_{\rm FG}$      & Stellar mass of the FG                           & $10^7 {\rm M_{\odot}}$\\
$r_{\rm plum}$   & Plummer radius of FG stellar distribution        & 23 pc\\
$\rho_{\rm pg} $  & Density of the pristine gas                      & ${\rm 10^{-24}; 10^{-23} g\ cm ^{-3}} $ \\
$v_{\rm pg}  $    & Pristine gas velocity relative to the cluster    & ${\rm 2 \times 10^6 cm\ s^{-1}}$   \\
$Z_{\rm pg}  $    & Metallicity of the pristine gas                  & 0.001  \\
$X_{\rm Fe} $     & Iron mass fraction of the pristine gas           & $3.77 \times 10^{-5}$   $^{(a)}$    \\
$X_{\rm He} $    & Helium mass fraction of the pristine gas         & 0.246  \\
$T_{\rm pg} $     & Temperature of the pristine gas                  & $10^4 {\rm  K}$ \\
$T_{\rm floor}$   & Minimum temperature                              & $10^3{\rm K}$ \\
$t_{\star} $      & Star formation time-scale                        & 0.1 Gyr \\
\hline
\hline
\end{tabular}
}
\label{tab:param}
$^{(a)} $ taken from \citet{dercole2012} 
\end{table}

\begin{table*}
\caption{Models description. {\it Columns}: (1) name of the model, (2) description of the model, (3) pristine gas density, (4) time of pristine gas reaccretion, (5) starting time of Type Ia SN explosions, (6) resolution. Times listed in (4) and (5) are expressed assuming $t_{\rm AGB}=39$ Myr as the time zero.} 
\begin{tabular}{llcccc} 
\hline
\hline
 Model & Description& ${\rm \rho_{pg}[g\ cm^{-3}]}$ & $t_{\rm inf}\mathrm{[Myr]} $&${t_{\rm Ia}\mathrm{[Myr]}} $ & Resolution [pc]  \\
\hline
LD                   & Low density                                              & $  10^{-24}$            & 21  & 0 & 0.6    \\
LD\_HR$^{a}$ & Low density at high resolution    & $  10^{-24}$  & 21 & 0 & 0.3  \\
LD\_DS & Low density with delayed Type Ia SNe           & $  10^{-24}$   & 21  & 25  & 0.6  \\
 LD\_DSI $^{b}$& Low density with delayed Type Ia SNe           & $  10^{-24}$   & 21  & 25  & 0.6  \\
 \hline
HD                  & High density                                             &  $  10^{-23}$  & 1  & 0  & 0.6    \\
HD\_DC       & High density with delayed cooling                 &  $  10^{-23}$  & 1  & 0 & 0.6     \\
\hline
\hline
\multicolumn{1}{l} {$^a$ truncated after 17 Myr} \\
\multicolumn{6}{l} {$^b$ infall stopped once the first SN bubble reaches the boundary from which the infalling gas is entering the box ($\sim 28$ Myr).}\\
\end{tabular}

\label{tab:simu}
\end{table*}

\section{Simulation set-up}
\label{sec:simuset}

The initial configuration of our work is the same presented in \citetalias{calura2019} with the addition of Type Ia SNe. Therefore, the starting point of our simulations $t_{\rm AGB}$ lies 39 Myr after the formation of the FG, when the gas has been already wiped out from the system because of the FG core-collapse SNe (CC-SNe) explosions. For computational reasons, however, we assume that the computational box is filled with a negligible amount of gas. The cluster is composed only by low- and intermediate-mass stars (i.e. $ m<8 {\rm M_{\odot}}$), the more massive of which are undergoing the AGB phase and are therefore starting to return mass and energy to the system. 
However, as it has been discussed in Section \ref{sec:1}, the AGB ejecta alone are not able to reproduce the observed SG abundance patterns therefore they are generally assumed to be diluted with pristine gas. As in \citetalias{calura2019}, we assume that the cluster is moving into a background distribution of gas representing the disk of a star-forming high-redshift galaxy \citep{kravtsov&gnedin2005,bekki2012, kruijssen2015,mckenzie2021}. The consequence of this motion is an asymmetric accretion of gas on the system from the side toward which the cluster is moving \citep{Naiman2011}. We assume two values for $\rho_{\rm pg}$, the density of the infalling pristine gas. The low density case is characterized by $\rho_{\rm pg}=10^{-24} { \rm g\ cm^{-3}}$ while for the high density case we assume a gas density 10 times greater. Ours is one of the first \enquote{wind tunnel} experiment on cluster scale (\citealt{priestley2011}, \citetalias{calura2019}), in which the effects of feedback are investigated in 3D (together with \citealt{chantereau2020}, who included photoionization). In our setup, the cluster is maintained fixed in its position and, at time ${t_{\rm inf}}$, gas is allowed to flow into the computational box from one of the boundaries. The onset of the infall event does not correspond with the beginning of the simulation but happens after a time ${ t_{\rm inf}}$ since, following \citet{dercole2016}, we suppose that the subsequent explosions of FG massive stars have generated a wind which has carved a large cavity surrounding the system. This hypothesis is supported by numerical simulations \citep{maclow1988} and more recently by hydrodynamic ones, both in one \citep{sharma2014} and in three-dimensions \citep{hopkins2012,creasey2013,walch2015,KIM2017} which study the evolution of SN driven superbubbles in galactic disks. The maximum radius achieved by this cavity, corresponding to the radius at which the wind ram pressure equals the pressure of the ambient medium, is:

\begin{equation}
 R_{\rm eq,2}=41.43\left( \frac{L_{41}}{n_0 V_{w,8}(\sigma^2_{0,6}+v^2_{\rm pg,6})}\right)^{1/2} 
\end{equation}
expressed in units of 100 pc. The term $L_{41}$ is the mechanical luminosity of the supernova-driven wind in units of ${\rm 10^{41} erg\ s^{-1}}$ and depends on the number of FG massive stars. For a cluster mass of $ M_{\rm FG}{\rm =10^7M_{\odot}}$ and a standard initial mass function (IMF) the number of CC-SNe is $\sim 10^5$ which means $L_{41} \sim 1$. The quantity $n_0 $ represents the number density of the ambient medium while $V_{w,8}$ is the velocity of the wind in units of ${\rm 10^{8} cm\ s^{-1}}$ which we assume to be $V_{w,8} \sim 2$. The two velocities $\sigma_{0,6} \sim 1$ and $v_{\rm pg,6} \sim 2$, both expressed in units of ${\rm 10^{6} cm\ s^{-1}}$, represent the isothermal sound speed, namely the velocity dispersion of the cluster within the galaxy, and the velocity of the recollapsing ambient medium relative to the system, respectively. 

 At solar metallicity, \citet{silich2004} found that if the radiative cooling is efficient, SN explosions do not form a stationary radiative wind. Through hydrodynamic simulations, \citet{wunsch2017} showed that, in young, massive and compact clusters, stellar winds become thermally unstable  leading to the formation of dense and cool clumps (see also \citealt{wunsch2008}). Inside these clumps the gas is able to self-shield from the extreme ultraviolet stellar radiation and form new stars giving rise to a SG. The SN energy in our setup is, instead, assumed to be mostly thermalized and a strong wind is established as derived in the adiabatic case \citet{Chevalier1985}. We follow the results obtained by \citet{calura2015}, where, through 3D hydrodynamic simulations, they found that, the residual gas, out of which the FG formed, is expelled from the cluster by the combined action of stellar winds and SN explosions. 

Once CC-SNe stop exploding, the surrounding gas starts to fall towards the system. Depending on the size of ${\rm R_{eq}}$ the gas will reach the system at different times. In particular, the time at which the gas starts to be re-accreted by the cluster is given by:

\begin{equation}
t_{\rm inf}=t_{\rm SNe}+\frac{R_{\rm eq}}{\sigma_{0}+v_{\rm pg}}
\label{eq:infall}
\end{equation}

where ${\rm t_{SNe}=30Myr}$ is the lifetime of the least massive CC-SN progenitor (corresponding to a mass of ${\rm \sim 9 M_{\odot}}$), which means that after that time no more massive stars are exploding. Assuming that ${\rm n_0 \sim \rho_{pg}/m_p}$, we obtain ${\rm t_{inf}\sim 60Myr}$ for ${\rm \rho_{pg}=10^{-24} g\ cm^{-3}} $ while ${\rm t_{inf}\sim 40Myr}$ for ${\rm \rho_{pg}=10^{-23} g\ cm^{-3}} $.

In Table \ref{tab:param} we have summarized the main parameters we have adopted in our simulations while Table \ref{tab:simu} contains a more detailed description of each model we have run.

\subsection{Physical ingredients}
\label{sec:phy_ing}
In this work, we use a customized version of the adaptive mesh refinement code RAMSES \citep{teyssier2002}, which solves the Euler equations of hydrodynamics with an unsplit second-order Godunov method, to perform 3D simulations of a young GC. The fluid follows the adiabatic equation of state for an ideal monoatomic gas with adiabatic index $\gamma=5/3$. The model includes the mass and energy return from AGB stars, radiative cooling, star formation and the feedback from Type Ia SNe. 
Newborn star are treated as collisionless particles and their trajectories are computed thanks to a particle-mesh solver taking into account of the mutual gravitational interactions between FG, SG stars and the gas. The gravitational effect of FG stars is considered assuming a static external potential, while for the SG stars and the gas self-gravity is taken into account. Moreover, we follow the chemical composition both of the gas and of the newborn stars through passive tracer focusing on iron, helium and metallicity Z. The system is located at the centre of a cubic computation box with a volume ${\rm L^3=(160\ pc)^3}$. In all our simulations we adopt a uniform grid reaching a resolution of ${\rm \sim 0.3\ pc }$ for the run performed at the highest level of refinement.
 We adopt free outflow boundary conditions for all the six faces of the computational box. At $t_{inf}$, the infalling gas enters the box from the $yz$ plane at negative $x$. For simplicity, in all our simulations we assume that gas continues to enter, even when the SN bubbles reach this boundary. This assumption affects neither the chemical composition of newborn stars nor the final stellar mass of the cluster (as explained in Section \ref{sec:discuss}). 

We summarize here the main physical ingredients of our simulations focusing on the Type Ia SN feedback, the novelty introduced in our work (for more details on basic setup see \citetalias{calura2019}). 

The mass return from AGB stars is taken into account adding a source term to the mass conservation equation given by:

\begin{equation}
\dot{\rho}_{\rm \star,AGB}=\alpha \rho_{\star}
\end{equation}

where $\alpha$ is the specific mass return rate and $\rho_{\star}$ is the density of the FG stars, distributed following a \citet{plummer1911} profile with mass $M_{\rm FG}=10^7 \mathrm{M_{\odot}}$ and Plummer radius $a=23 \mathrm{pc}$. 

In order to trace the helium abundance we adopt the yields of \citet{ventura&dantona2011} (for AGBs with a progenitor mass of $8\ \mathrm{M_{\odot}}$ we have chosen an average value between the model of \citealt{ventura&dantona2011} and \citealt{siess2010}) calculated for a metallicity of $Z=10^{-3}$. Since AGBs are not producing iron, their ejecta have the same iron mass fraction of the pristine gas. 

The energetic feedback from AGBs it is implemented as a source term of the form:
\begin{equation}
S=0.5\rho_{\star} (3 \sigma^2+ v^2+v_{\rm wind}^2)
\end{equation}
where  $\sigma$ represents the velocity dispersion of the FG stars, $v$ is the gas velocity while $v_{\rm wind}$ is the wind velocity of the AGB stars \citep{dercole2008}. 

Both $\dot{\rho}_{\rm \star,AGB}$ and S are added at each timestep to the fluid density and energy, respectively.

Together with the heating associated with the energetic feedback from AGBs, the source term includes the cooling due to radiative gas losses. We use the RAMSES built-in cooling rates calculated, for temperature $T>10^4 $ K, fitting the difference between the cooling rates at solar and zero metallicity through the photoionisation code CLOUDY \citep{ferland1998},
which involves both atomic (i.e. due to H, He) and metal cooling (see \citealt{few2014}). 
For temperatures $T<10^4$ K, metal-line structure cooling rates are taken from \citet{rosen&bregman1995}.
The temperature of the pristine gas is fixed to $10^4$ K, a typical one for the ISM in a star-forming galaxy which is maintained photoionized by a stable UV radiation field \citep{Haffner2009}.
A temperature floor of $T=10^3$ K is adopted. 

In our work, we use the RAMSES built-in star formation recipe, which is largely described by \citet{rasera&teyssier2006}. Star formation is allowed only in cells in which the temperature $T< 2 \times 10^4 $ K, therefore only where the gas is assumed to be neutral, and where $\nabla \cdot v <0 $, corresponding to the cells in which the net flow is converging. 
Moreover, for numerical reasons, not all the gas inside a cell can be used to form stars; in our code a maximum of 90 per cent of the gas in each cell is available for star formation.
Then, once the cells eligible for SF are known, gas is converted into stars according to the \citet{schmidt1959} law:

\begin{equation}
\dot{\rho}_{\rm \star, SG} =\frac{\rho}{t_{\star}}.
\end{equation}
The quantity $t_{\star}$ represents the star-formation timescale, which is proportional to the local free-fall time, and has been set to be $t_{\star}=0.1\ \mathrm{Gyr}$. 
Every collisionless stellar particle has a mass $M_{\rm p}=N\ m_0$ where $m_0=0.1  \mathrm{M_{\odot}}$ is the minimum mass while $N$ is obtained sampling the Poisson distribution with a mean value of $\lambda_{\rm p}= \left( \frac{\rho \Delta x ^3}{m_0}\right) \frac{\Delta t}{t_{\star}} $.
The chemical composition of each newborn star is the same of the gas of its parental cell.
These particles are located at the centre of the cell in which they are formed with a velocity equal to the local fluid one. The mass, momentum and energy associated to the newborn stars are conservatively removed from the parental cell and also from the passive tracers devoted to follow the chemical composition of the gas. 


\subsection{Type Ia SN feedback}

\begin{figure} 
 \centering
  \hspace{-0.4cm}
  \includegraphics[width=0.5\textwidth]{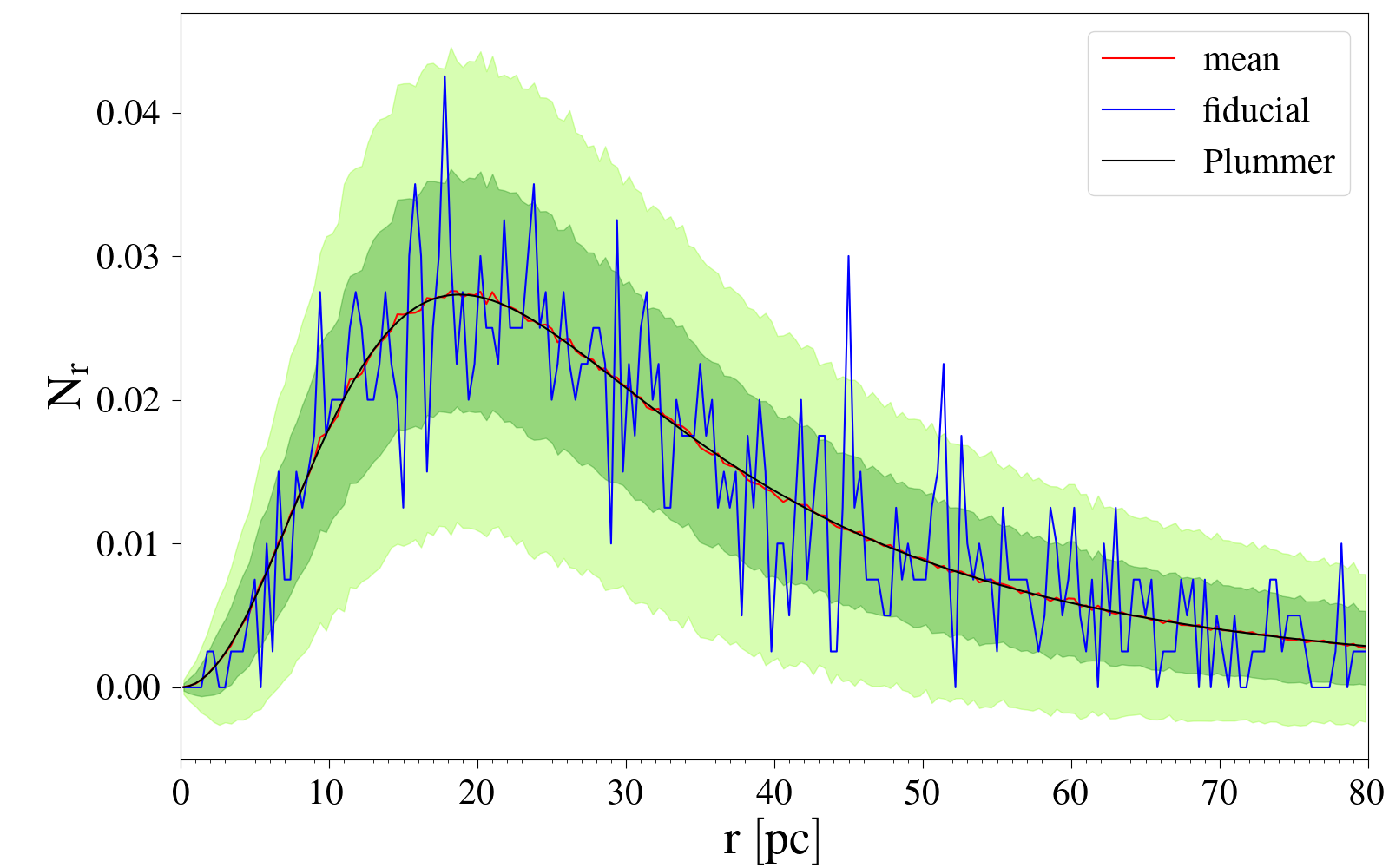}
  \caption{Spatial distribution of Type Ia SNe as a function of the radius. In black we show the analytical result for the Plummer profile assuming $\mathrm{M_{\star}=10^7M_{\odot}}$ and $\mathrm{a=23pc}$. The red line indicates the mean of 1000 realizations with the associated regions within 1 and 2 $\sigma$ in green and light green, respectively. The chosen realization is shown in blue, corresponding to the one that deviates less from the mean among the 1000 realizations. All the plotted lines are normalized imposing $\mathrm{\int_{0}^{r_{max}} N_r\  dr=1}$  
  with maximum radius $\mathrm{r_{max}=80pc}$, which corresponds to half the size of the computational box.}
  \label{fig:dist_rad}
\end{figure}

\begin{figure} 
 \centering
 \hspace{-0.4cm}
  \includegraphics[width=0.5\textwidth]{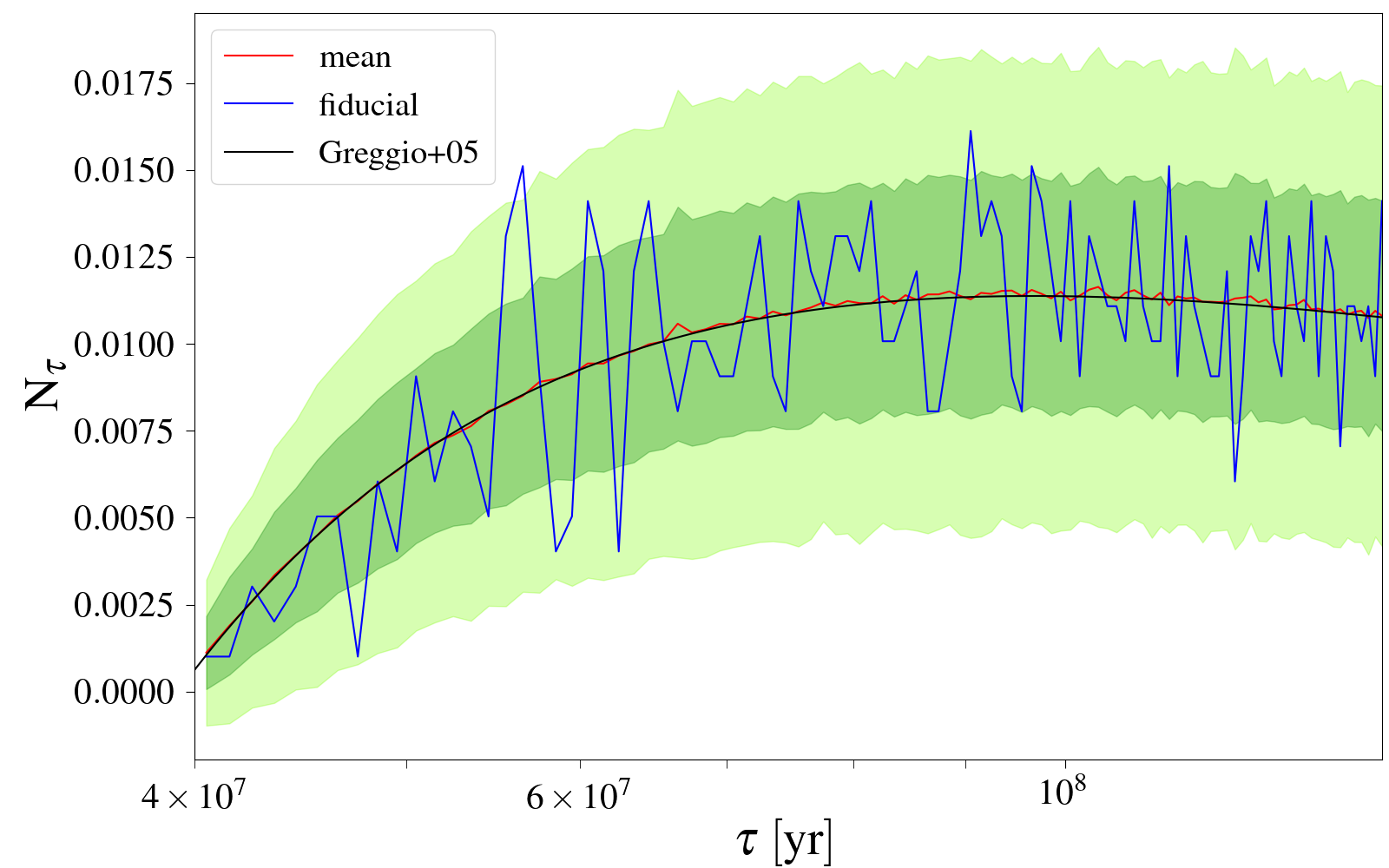}
  \caption{Delay time distribution function. In black we show the theoretical one from Eq.16 of \citet{greggio2005} assuming $\alpha =2.35$ and $\gamma =2$. The red line indicates the mean of 1000 realizations with the associated regions within 1 and 2 $\sigma$ in green and light green, respectively. The chosen configuration among the 1000 realizations is shown in blue, namely the one that deviates less from the mean. The plotted lines are all normalized imposing $\mathrm{\int_{\tau_i}^{\tau_x} N_{\tau} \ d\tau=1}$ 
  where $\mathrm{\tau_i=40Myr}$ is the lifetime of a $\mathrm{8M_{\odot}}$ star and $\mathrm{\tau_x=140Myr}$ in order to cover all the timespan we are focusing on, with the normalization factor in units of $\mathrm{Gyr^{-1}}$.}
  \label{fig:dist_time}
\end{figure}   

The novelty of this work is the introduction of Type Ia SN feedback on the study of a star-forming cluster by means of 3D hydrodynamic simulations.
Type I SNe are originated by the thermonuclear explosion of white dwarfs in binary systems. 
When a Type Ia explodes, we assume that one Chandrasekar mass (1.44 ${\rm M_{\odot}}$) is released into the ISM with an amount of iron of ${\rm 0.5 M_{\odot}}$ \citep{scalzo2014} and a metallicity equal to 1. Each SN will also release $10^{51}$ erg of thermal energy in the ISM. To every Type Ia SN progenitor we have associated an explosion time assuming the delay time distribution (DTD) of \citet[their Equation 16]{greggio2005} for the single degenerate scenario. Therefore the SNe are assumed to start exploding 40 Myr after the FG formation.

Given that the number of Type Ia SNe per unit mass for the Kroupa IMF \citep{kroupa2001} is $10^{-3}$, assuming a FG mass of ${\rm 10^7 M_{\odot}}$ we end up with $10^4$ Type Ia SN explosions in 10 Gyr. In the first 0.1 Gyr, the timespan we are interested in, $\sim 1000$ SNe explosions would take place.
Spatially, SNe have been distributed following the Plummer profile \citep{plummer1911} computed for a cluster mass of ${\rm 10^7 M_{\odot}}$ and a Plummer radius of $a=23$ pc. The maximum radius of the distribution has been set to 80 pc, the radius of the sphere inscribed in the computational box; this means that all the SNe are located inside our region of interest. 

In order to have a good sampling both of the DTD and of the Plummer profile, we have created 1000 random realizations for each function and then selected the realization that deviates less from the mean. This configuration has been then used to derive the positions and the explosion times of the 1000 SNe. In Figure \ref{fig:dist_rad} and \ref{fig:dist_time} we show the mean over all the realizations in red with the uncertainties at 1 and 2 $\sigma$ and the chosen realization in blue. In black we have reported the two functions that have been sampled.

\subsubsection{The \enquote{overcooling} problem}
\label{sec:overcool}
The dynamical evolution of a SN remnant (SNR) can be split in various phases, each of them characterized by a different expansion rate. Immediately after the explosion, the material ejected by the SN expands freely into the ISM. Once the SN ejecta interacts with the circumstellar medium a forward and a reverse shocks are formed; in particular, the reverse shocks moves inward heating up the expanding ejecta which results in high temperatures and pressures. This phase lasts until the mass of the swept up gas becomes comparable to the ejected mass. From then on, the system evolves almost adiabatically during the so called Sedov Taylor (ST) phase as long as radiative losses become important. However, when, as in our case, SN feedback is modelled through thermal energy injection, if the SNR is not well resolved the injected energy can be artificially radiated away very quickly, reducing the duration of the ST phase \citep{katz1992}. As a consequence, the impact of SN explosions on the ISM heavily decreases leading to the \enquote{overcooling problem}. 
\citet{kim&ostriker2015} studied the evolution of a SNR varying the numerical grid resolution in order to find the maximum cell size one has to assume to model the SN feedback without falling into the overcooling regime.
They found that the numerical resolution $\Delta$ has to be at least 3 times smaller than $r_{\rm sf}$, the radius at the shell formation which is given by:

\begin{equation}
 r_{\rm sf}=22.1\  n_0^{-0.4}\ {\rm pc}
 \label{eq:k&o}
\end{equation}
for a uniform medium. 

We have therefore applied the criterion to our simulations to determine whether our SNe are resolved or not. 

\begin{figure} 
 \centering
  \includegraphics[width=0.5\textwidth]{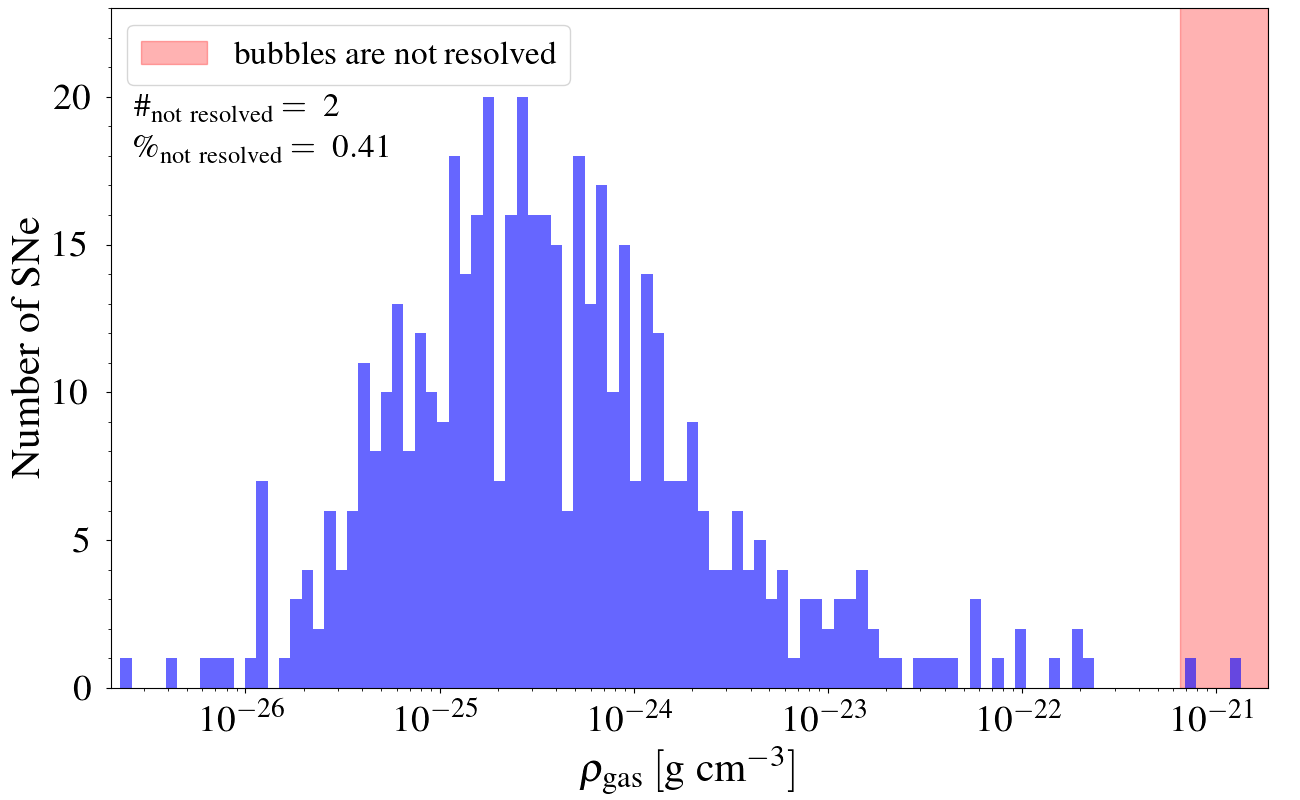}
  \caption{The distribution of the gas density in the cells where the SNe are exploding for the low density simulation. The red shaded area represents the region of densities for which the \citet{kim&ostriker2015} criterion is not satisfied.}
  \label{fig:k&o15_low}
\end{figure}

\begin{figure} 
 \centering
  \includegraphics[width=0.5\textwidth]{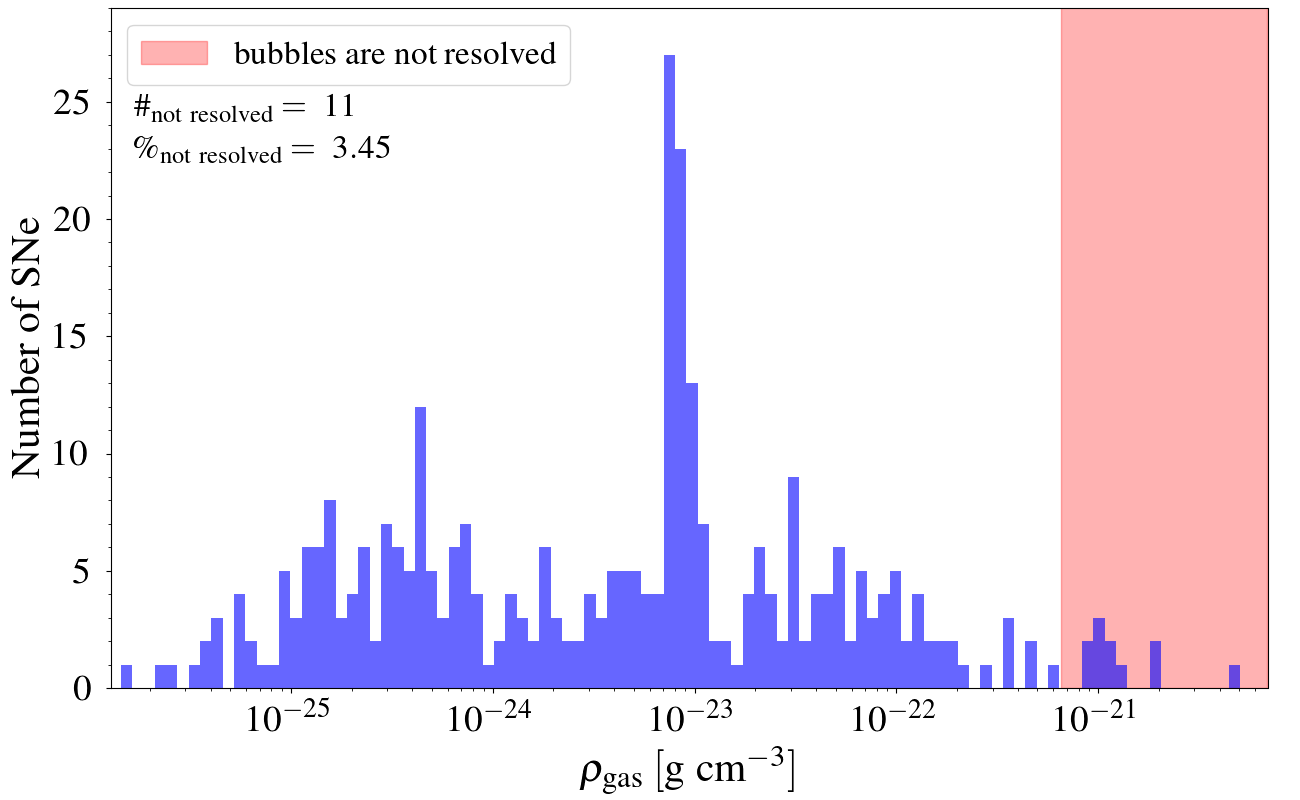}
  \caption{The distribution of the gas density in the cells where the SNe are exploding for the high density simulation. The red shaded area represents the region of densities for which the \citet{kim&ostriker2015} criterion is not satisfied.}
  \label{fig:k&o15_high}
\end{figure}

In Figure \ref{fig:k&o15_low} and \ref{fig:k&o15_high}, we report the results obtained for the LD and HD models, respectively. It has to be said that, given that we have truncated our simulations at different times, in the LD model, more SNe have exploded than in the HD one. 
What emerges is that the number of SNe that do not satisfy the criterion is greater in the HD model than in the LD one. This is a direct consequence of Equation \ref{eq:k&o}: the higher the density of the gas the lower is the radius at shell formation and therefore the required numerical resolution increases. However, for computational limitations, we have assumed the same $\Delta$ for both the two models.

Focusing on the LD case, the fraction of exploded SNe that do not meet the \citet{kim&ostriker2015} criterion is very low, i. e. $0.41\%$. In the HD model this fraction is still very low, although it is nearly a factor $\sim 10$ larger ($3.5\%$) than in the LD model.

 To better investigate this issue, we have decided to perform, for the HD model, a second simulation in which cooling is artificially switched off in the cells surrounding the SNe at the time of their explosion. This method is widely used in literature to prevent the occurrence of the overcooling problem. However, in literature there is not a unique way to switch off cooling manually, neither a definition of the extent of the region of interest in which cooling has to be switched off nor of the duration of this phase \citep{rosdahl2017}. In this work, we turn off cooling using the RAMSES built-in prescription described in \citet{teyssier2013}. Each Type Ia SN is supposed to inject an equal amount of thermal and non-thermal energy into the ISM; the energy density of non-thermal component is stored in a new variable, $e_{\rm turb}$, which acts as a passive tracer. In each cell, the velocity dispersion associated to the non-thermal component is calculated through:

\begin{equation}
\sigma_{\rm turb} = \sqrt{2 \frac{e_{\rm turb}}{\rho}}    
\end{equation}

where $\rho$ is the density of the gas inside the cell. The cooling is switched off, according to \citet{teyssier2013}, in all the cells where $\sigma_{\rm turb} >10 \mathrm{km\ s^{-1}}$. This method allows to overcome the still poorly understood causes leading to the problem of \enquote{overcooling}, being them related either to resolution problems or lack of physics (such as turbulence and radiative feedback).



\subsubsection{The shape of the delay time distribution}
\label{sec:shapeDTD}
Many studies have been done so far aimed at constraining the shape of the DTD, the Type Ia SN rate resulting from a brief burst of star formation. The uncertainties concern both its slope and the time interval, with particular focus on its starting point. While most of the observations agree with $\propto t^{-1}$ dependence at $t>2\ \mathrm{Gyr}$, theoretical works find different shapes, especially at short times as shown in \citet[their Figure 8]{maoz2014}. Some models predict the first explosions after $\sim  30-40 \mathrm{Myr}$ like the one proposed by \citet{greggio2005} while others after some hundreds of Myr. However, the duration of SG formation is assumed to end before stars with $m<{\rm 3M_{\odot}}$ undergo their AGB phase \citep{renzini2015,dantona2016}, corresponding to an age of $\sim 200-300$ Myr, in agreement with the upper limits found by \citet{nardiello2015} and \citet{lucertini2020}, therefore we have focused our attention on the DTD with short delay times. Nevertheless, given the uncertainties of the starting point of the DTD, we have decided to test a case in which Type Ia SN explosions are shifted of 25 Myr with respect to the \citet{greggio2005} formulation, in order to study its effect on the SF of the SG stars. We have chosen this time because we want to study the case in which SNe start exploding when the infall of pristine gas has already started. We have performed one run with delayed Type Ia SN explosions for the low-density case since the effects of SNe for this model are larger than for the high-density one.

\subsubsection{Numerical issues}

Together with the \enquote{overcooling} problem, the introduction of Type Ia SNe feedback leads to the other numerical difficulties. SNe explosions inject fast fluid into the ISM with velocities as high as ${\rm 10^3 km \ s^{-1}}$ which have strong effects of the timestep of the simulation. To satisfy the Courant-Friedrichs-Lewy condition, in fact, the timestep is heavily reduced reaching the order of 10 yr in the highest resolution model (see also \citealt{romano2019}). \citet{emerick2019} overcome this difficulty lowering artificially the velocity of the fluid to be able to follow the evolution of the system for various hundreds of Myr. In all our simulations we do not apply such artefact, therefore we are able to study only a limited timespan. In addition, the high-density models are slower than the low-density ones because of the higher cooling efficiency  characterizing them \citep{romano2019}. Therefore, we have truncated all our low-density models after 48 Myr while the high-density ones after 39 Myr. 

\section{Results}
\label{sec:results}
In this paper we present the results for four models, all of them including the feedback from Type Ia SNe. 

We have performed two runs, the LD and the HD, varying only the density of the pristine gas (see Table \ref{tab:simu} for the details of all the models).
In addition to these models, we have performed, for the low-density one, a run shifting the Type Ia SNe explosions of 25 Myr given the large uncertainties on the shape of the DTD highlighted in Section \ref{sec:shapeDTD}. We have chosen to perform such test for the low density case since, as we will show, the major effects of Type Ia SN feedback are observed in this model.

In addition, we have run a simulation in which cooling is temporarily switched off in regions surrounding the SN explosions, for the high density scenario. This choice has been suggested given the non negligible number of Type Ia SNe not meeting the \citet{kim&ostriker2015} criterion. 

 Lastly, we performed a simulation similar to the ${\rm LD\_DS}$ one but assuming that infall is stopped once the first SN bubble reaches the negative $x$ boundary, namely the face out of which the gas is entering the box. We tested this case since in all our models we are imposing, for simplicity, that gas continues to enter the box even when SN bubbles are pushing it back, far from the cluster. We will not focus in detail on this model, referring to it only for a comparison in Section \ref{sec:discuss}.

Here we show the results obtained at various times during the evolution of the system. It has to be clarified that all the times throughout the paper are expressed assuming $t_{\rm AGB}=39 {\rm Myr}$ as the time zero.


\begin{figure*}
        \centering
       
        \includegraphics[width=0.493\textwidth]{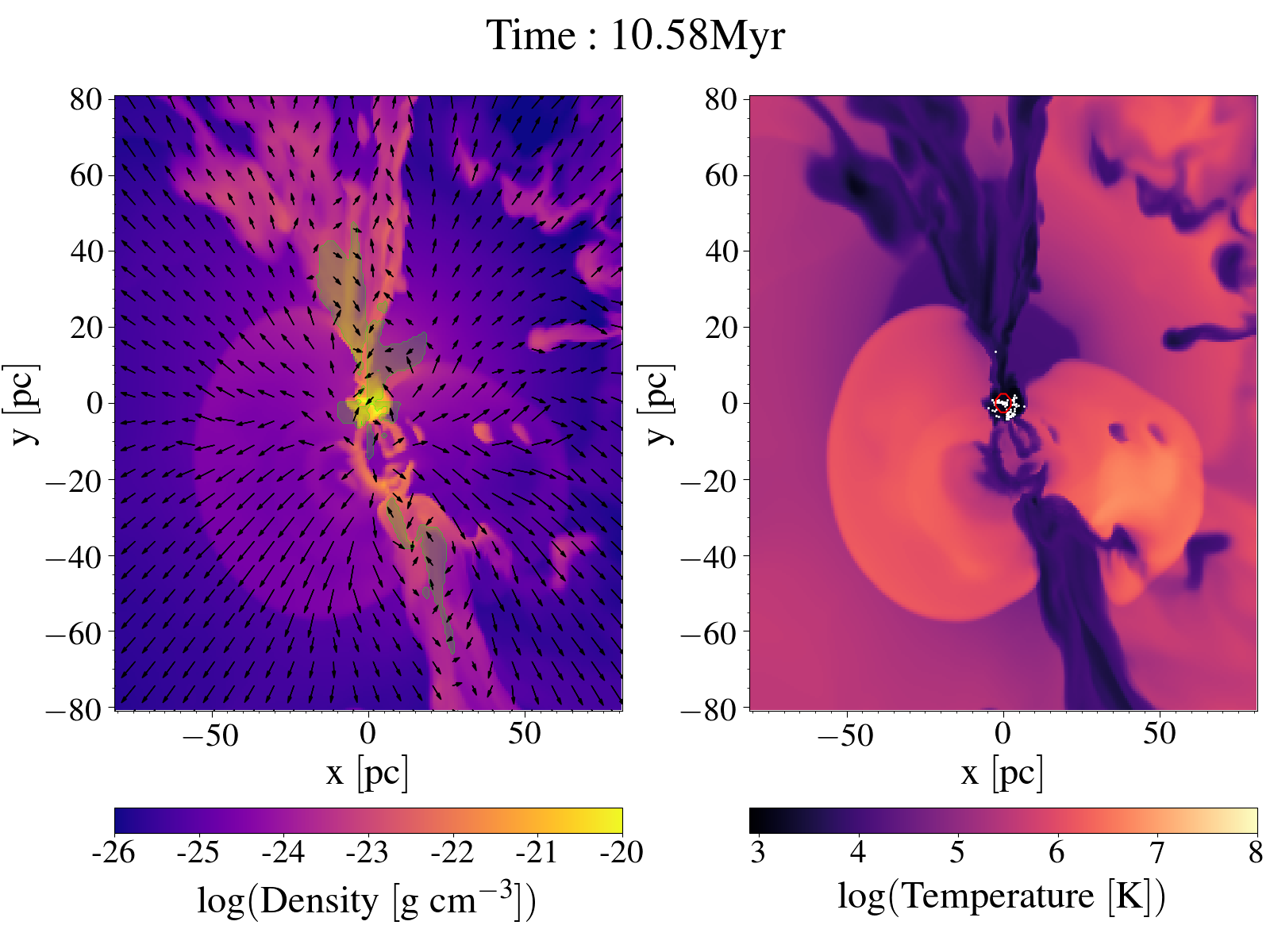}
        \hspace{0.1cm}
        \includegraphics[width=0.493\textwidth]{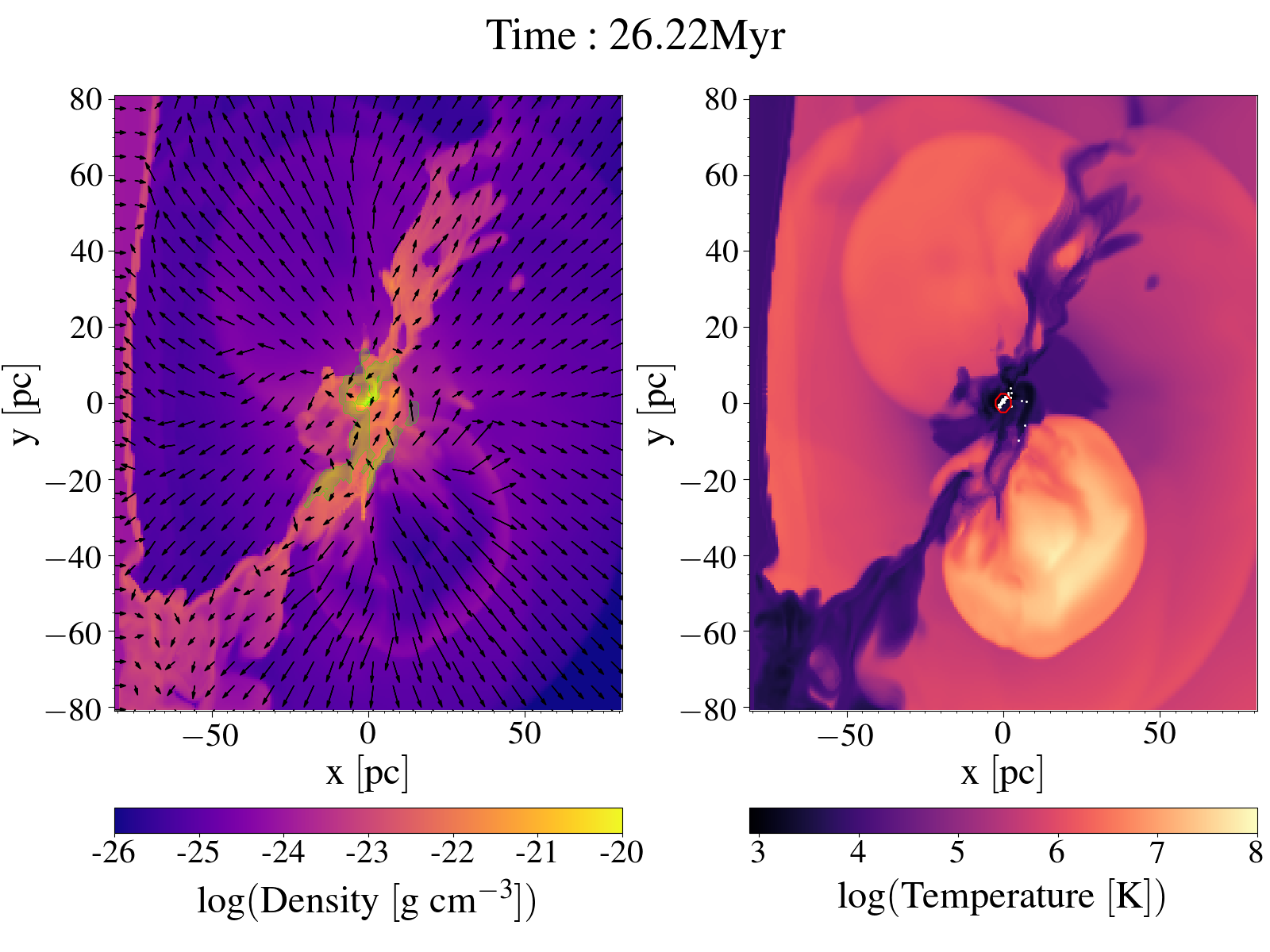}
        \\
        \vspace{0.3cm}
        \includegraphics[width=0.493\textwidth]{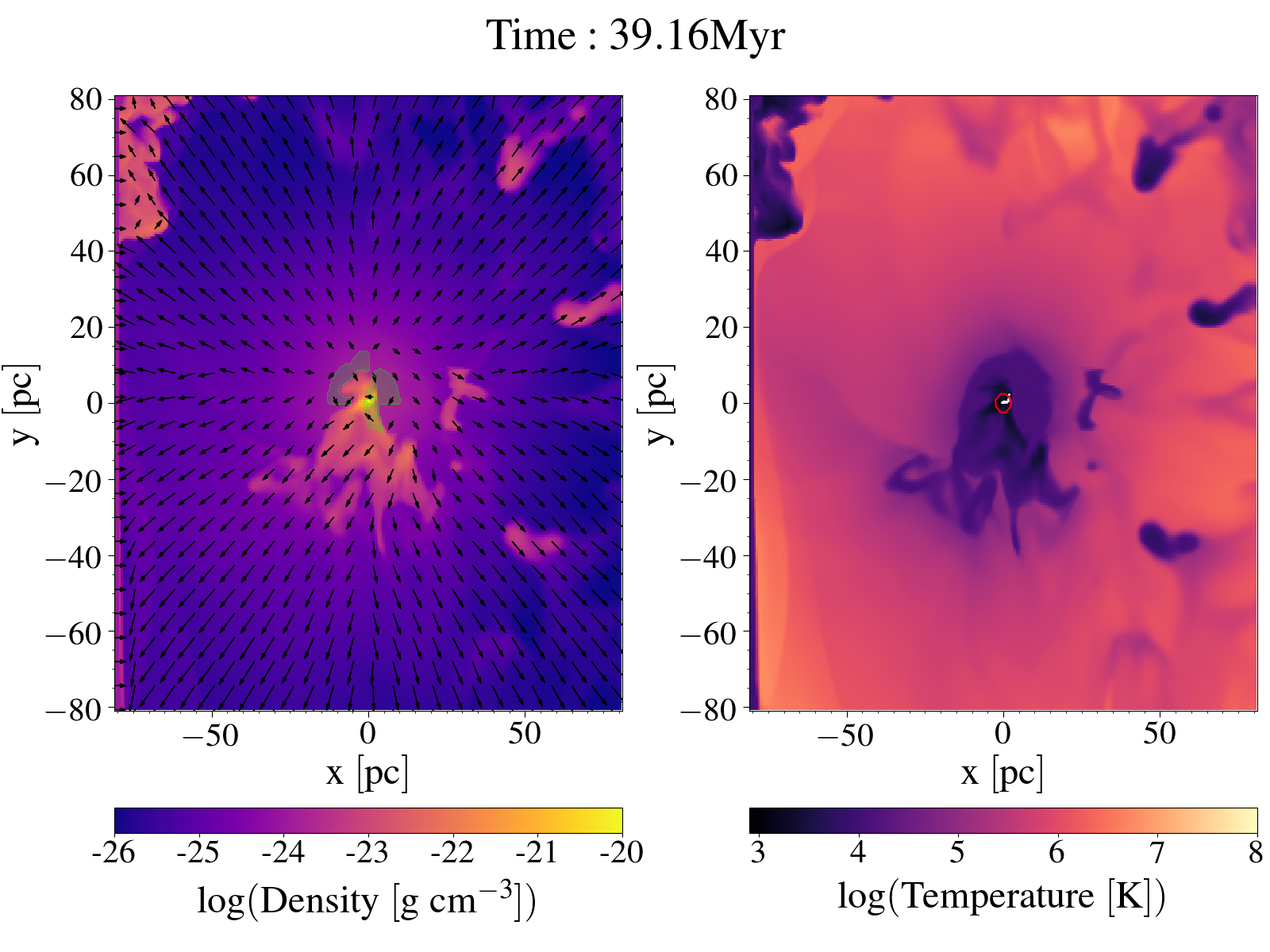}
        \hspace{0.1cm} 
        \includegraphics[width=0.493\textwidth]{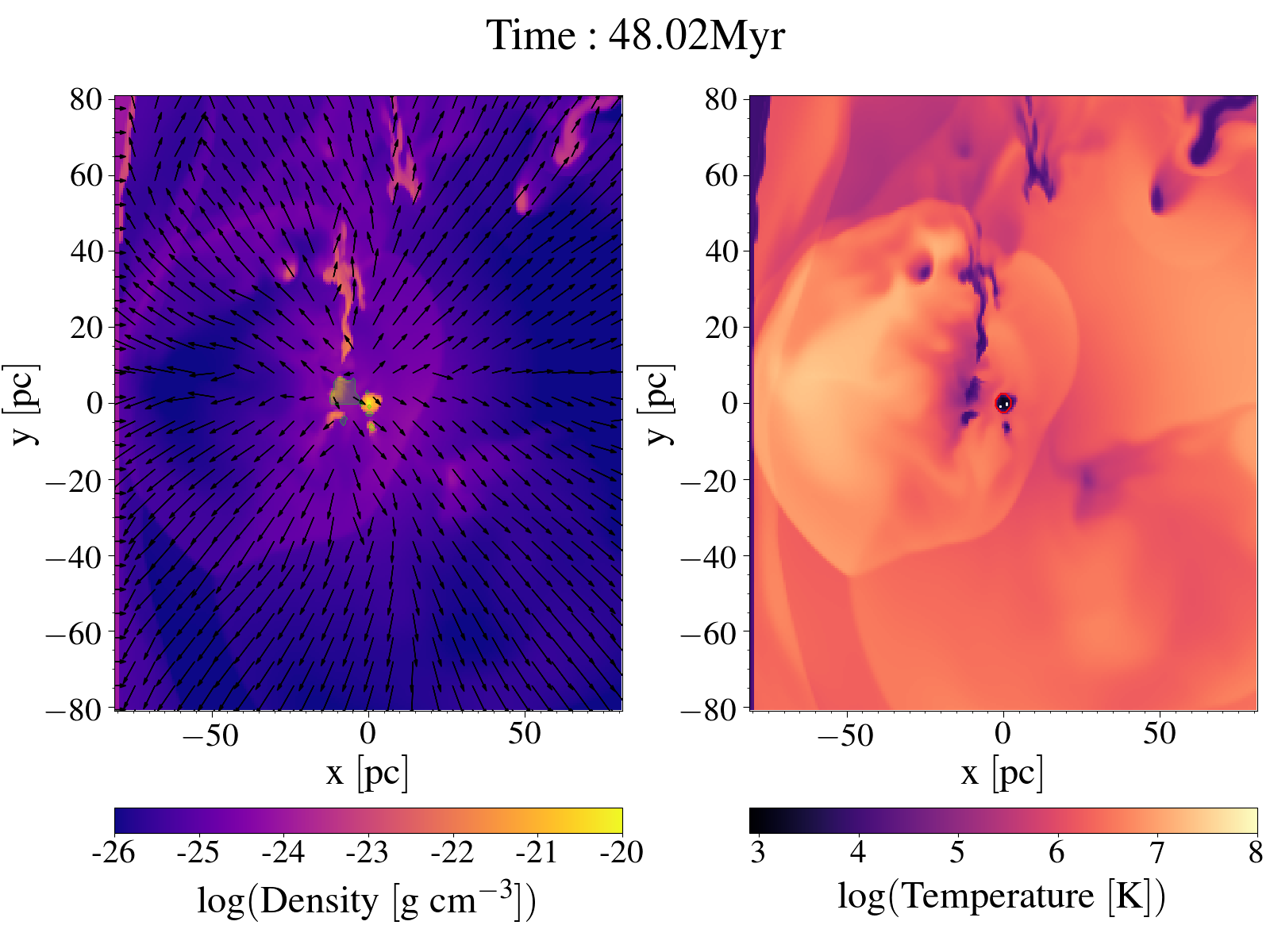}

\caption{Two-dimensional maps of the gas density (left-hand panels) and of the temperature (right hand panels) in the x-y plane at different evolutionary times (reported on top of each pair of panels) for the low density simulation (LD). From the top left to the bottom right: $t=10, 26, 39, 48$ Myr. The black arrows in the density maps represent the velocity field, while the green shaded areas represent the regions of the plane where the gas is moving towards the cluster centre (see the text for details). The white dots in the temperature maps represent the newborn stars (with a lifetime of < 0.05Myr). The red contour in the temperature maps describe the region enclosing 50\% of the SG mass}.
 \label{fig:maps_lowND}
\end{figure*}

\begin{figure*}
        \centering

        \includegraphics[width=0.287\textwidth]{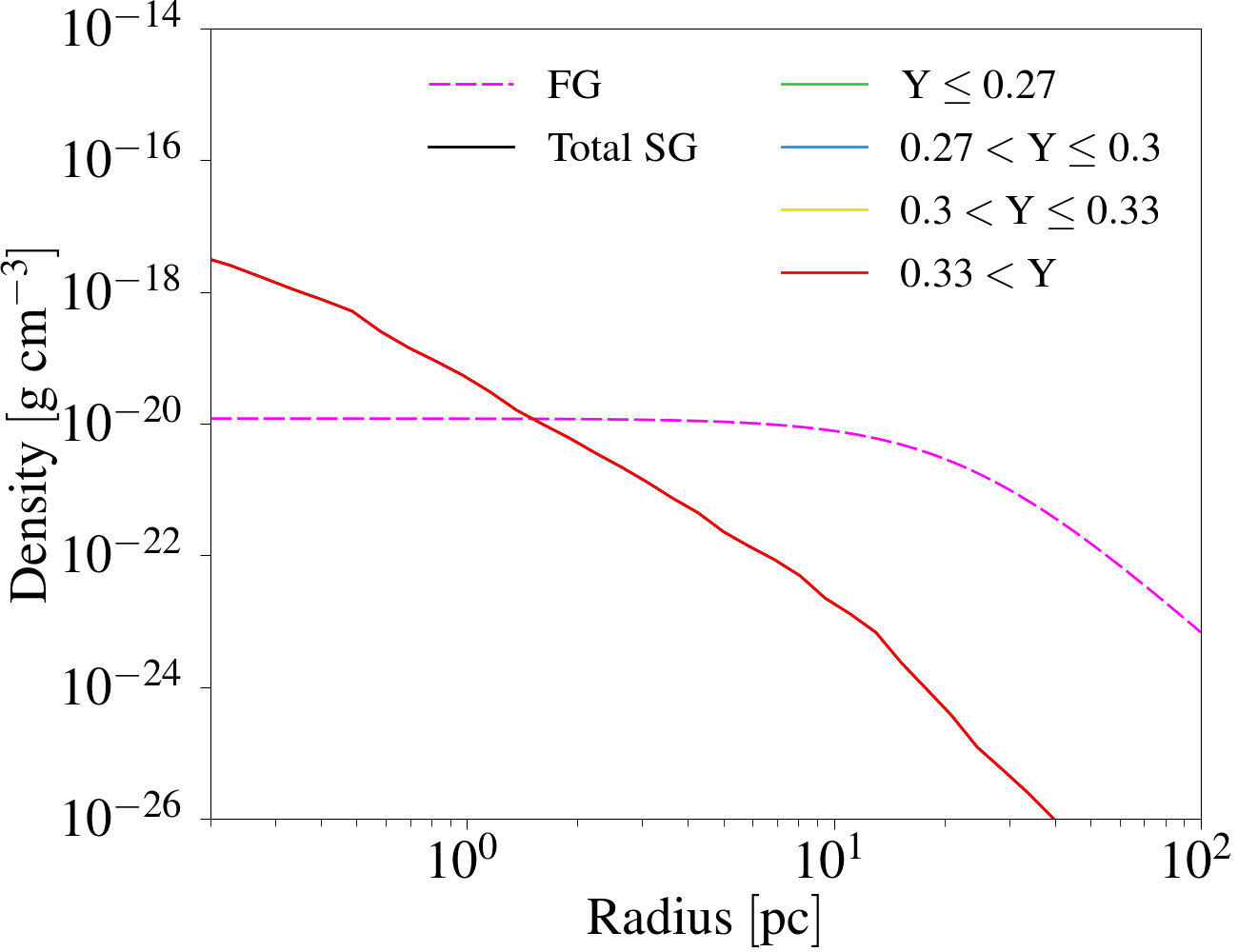}   \includegraphics[width=0.192\textwidth]{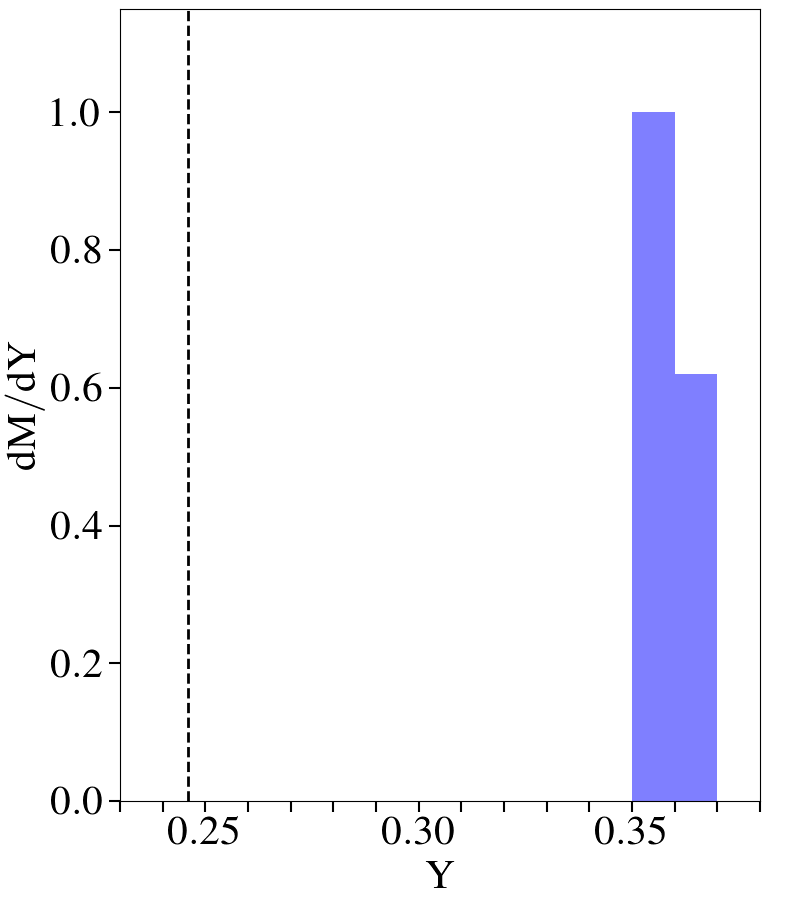}
        \hspace{0.08cm}
        \includegraphics[width=0.287\textwidth]{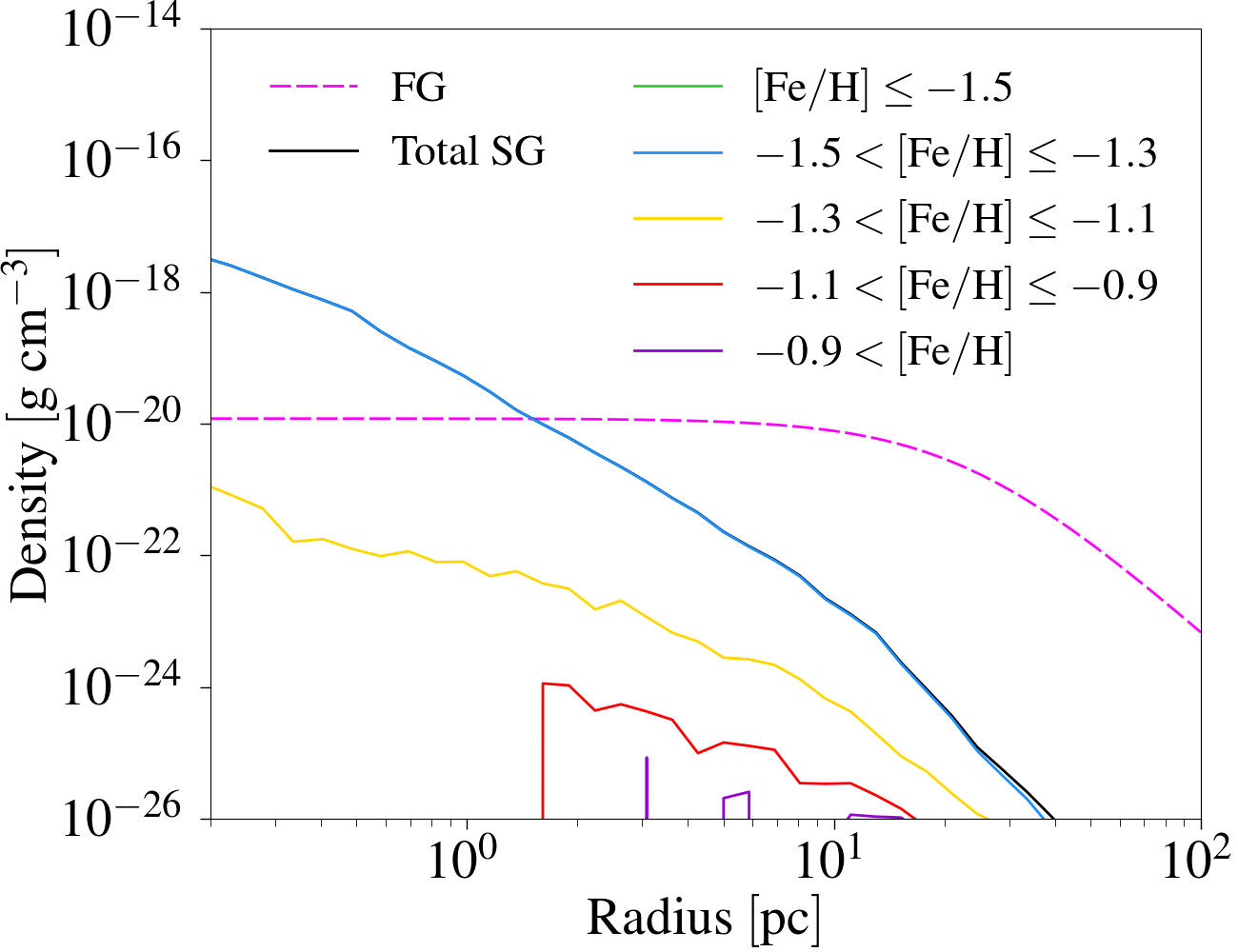}
        \includegraphics[width=0.216\textwidth]{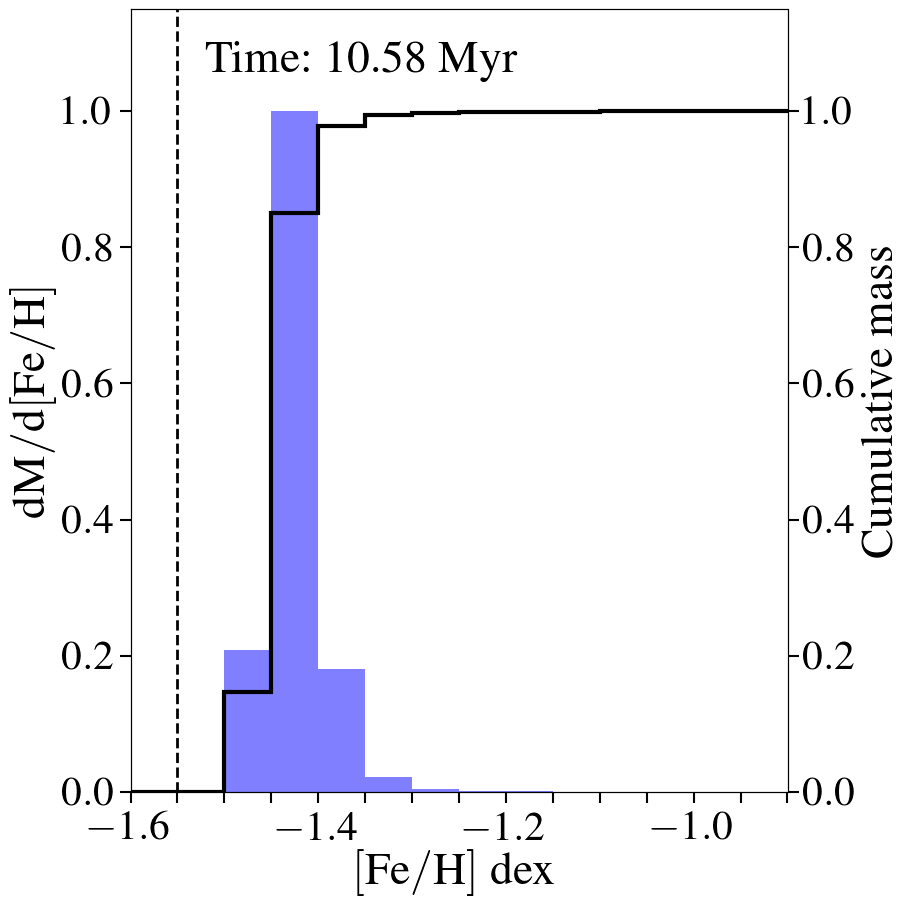}
        \\
        \includegraphics[width=0.287\textwidth]{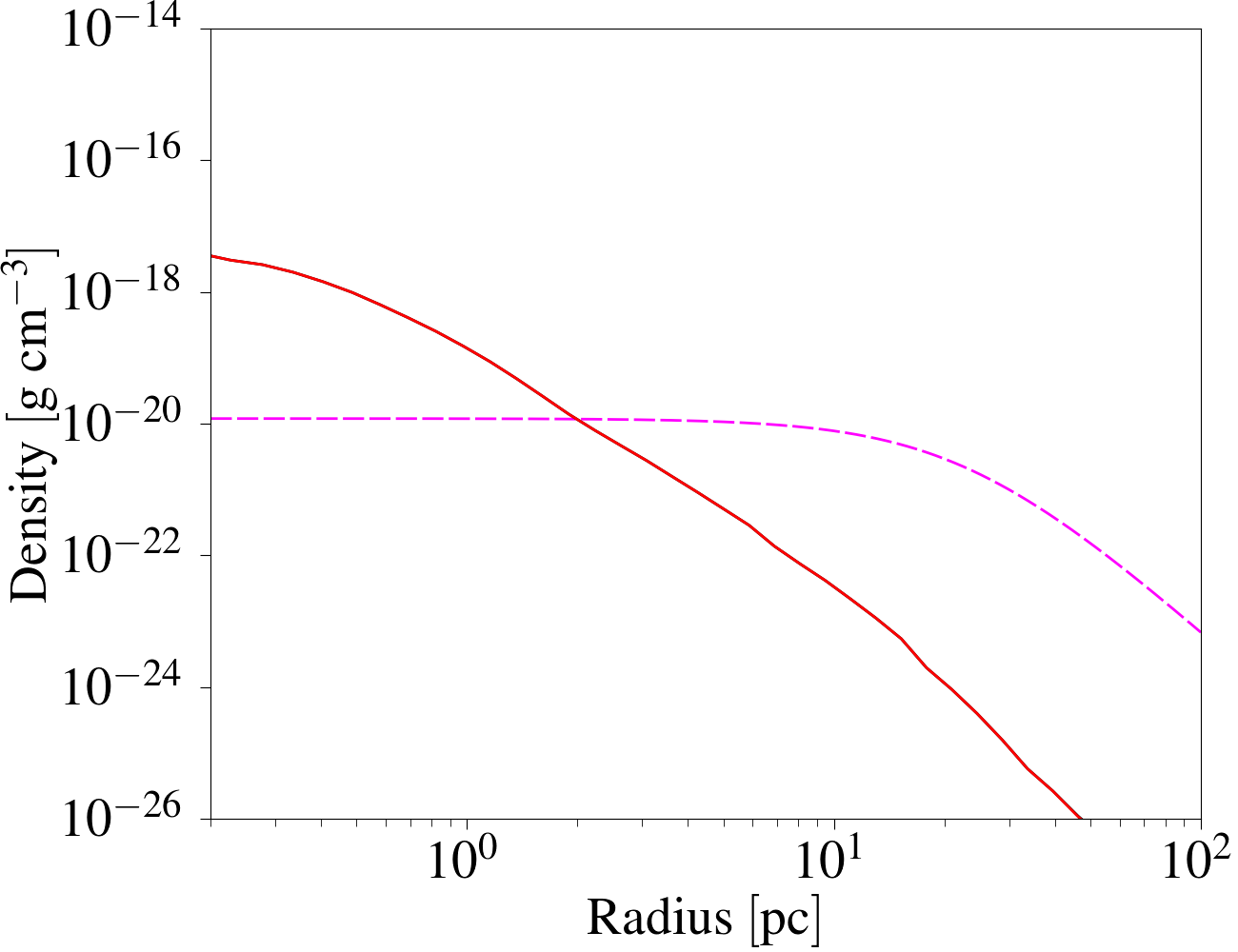}
        \includegraphics[width=0.192\textwidth]{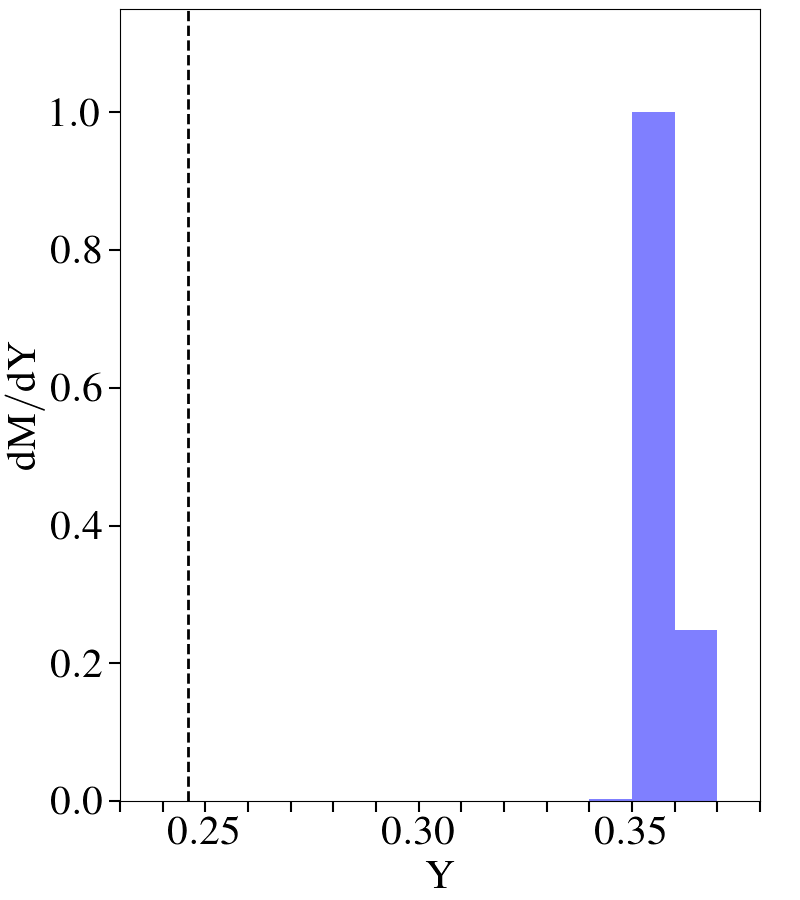}
        \hspace{0.08cm}
        \includegraphics[width=0.287\textwidth]{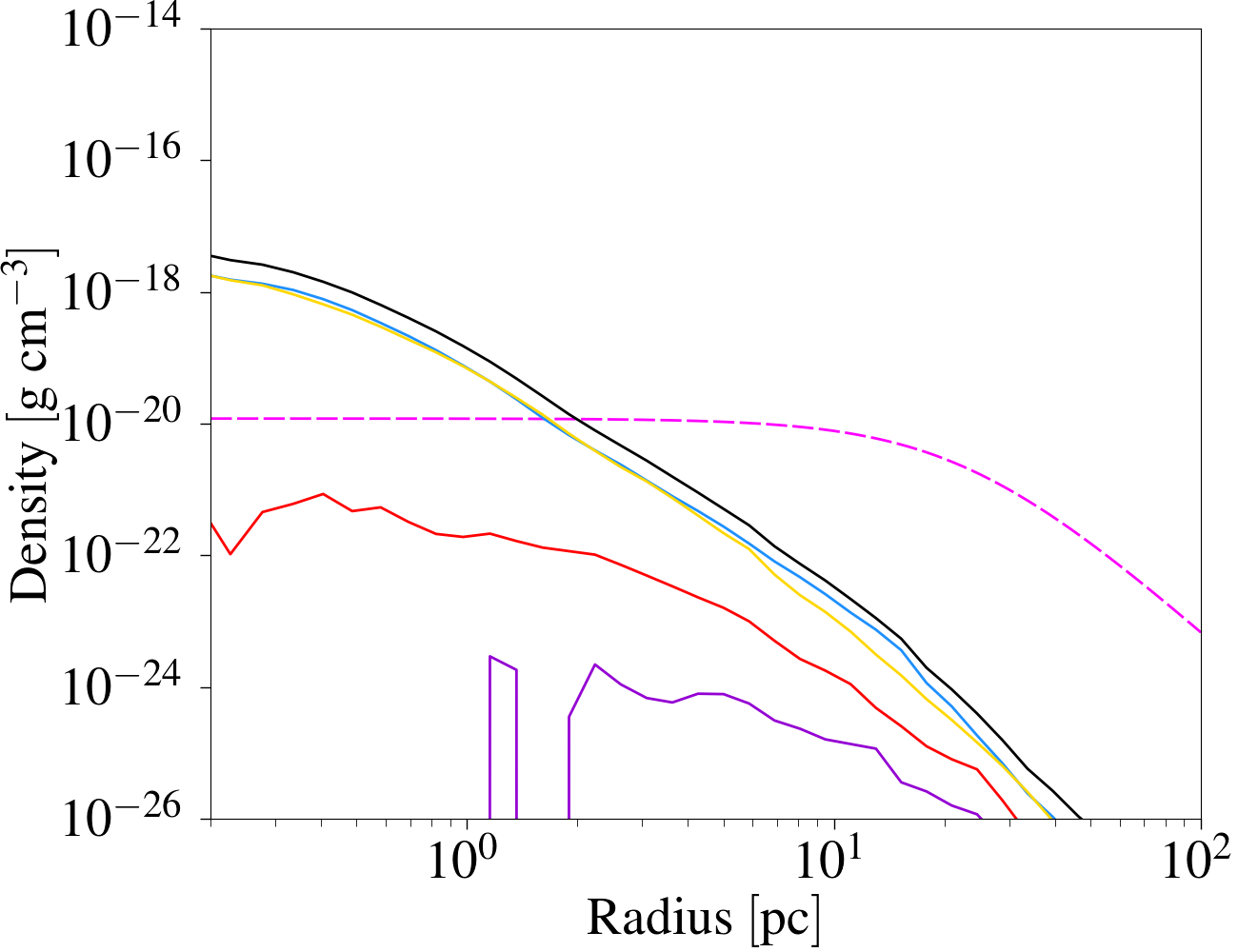}
        \includegraphics[width=0.216\textwidth]{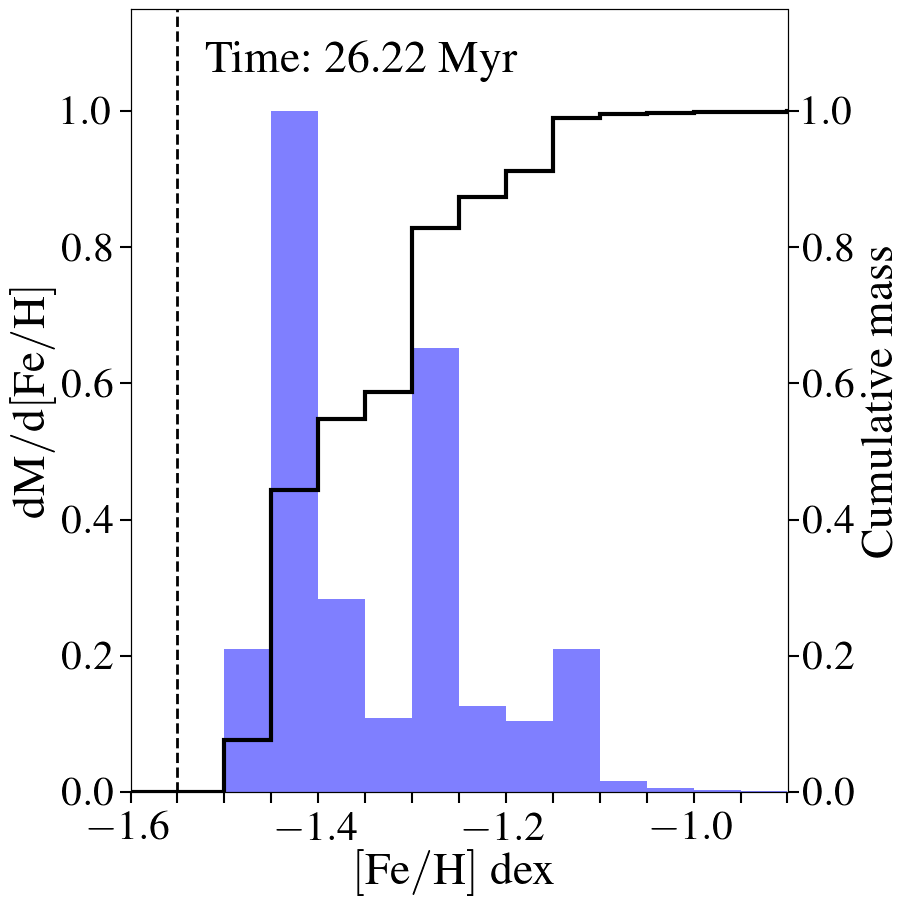}
        \\        
        \includegraphics[width=0.287\textwidth]{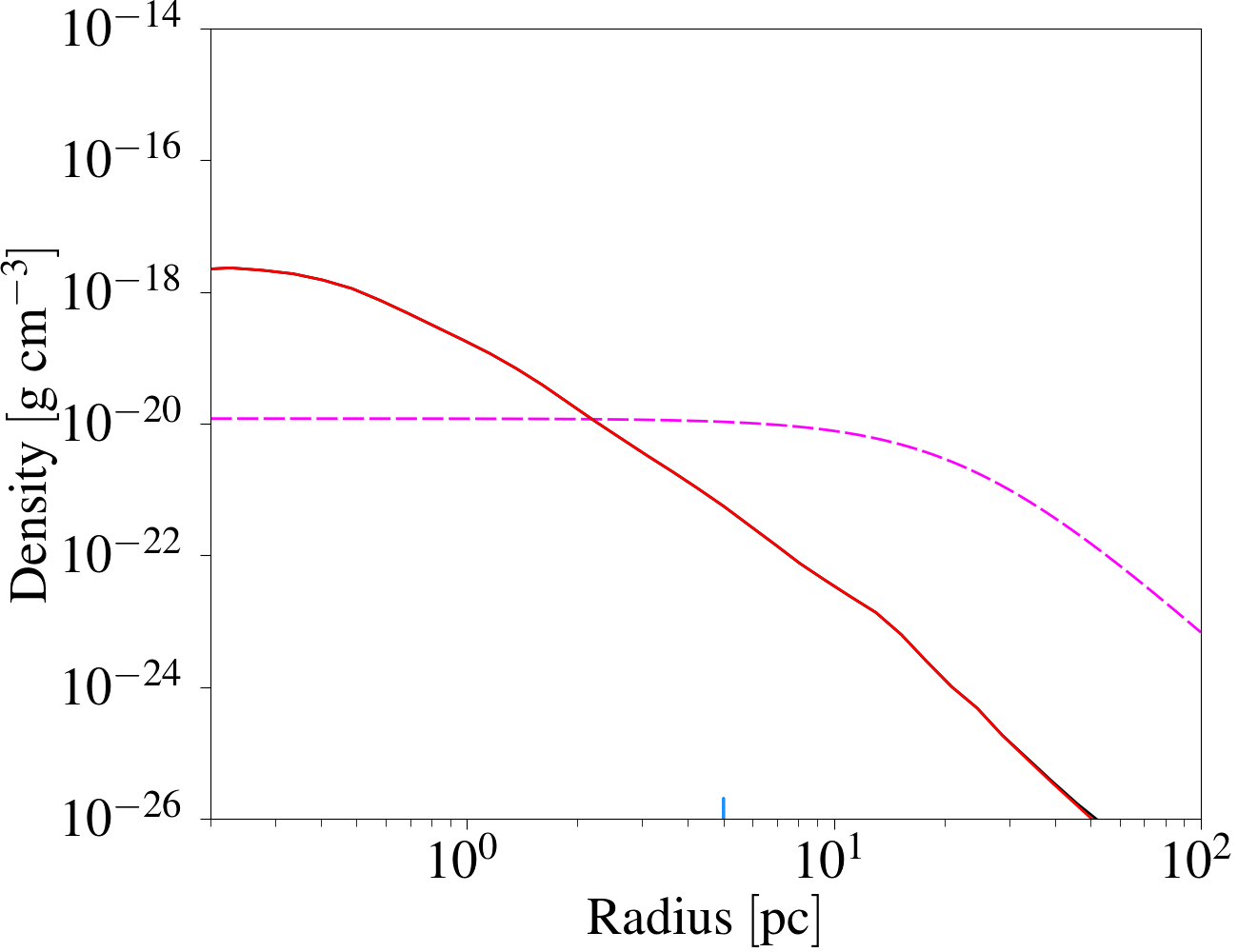}
        \includegraphics[width=0.192\textwidth]{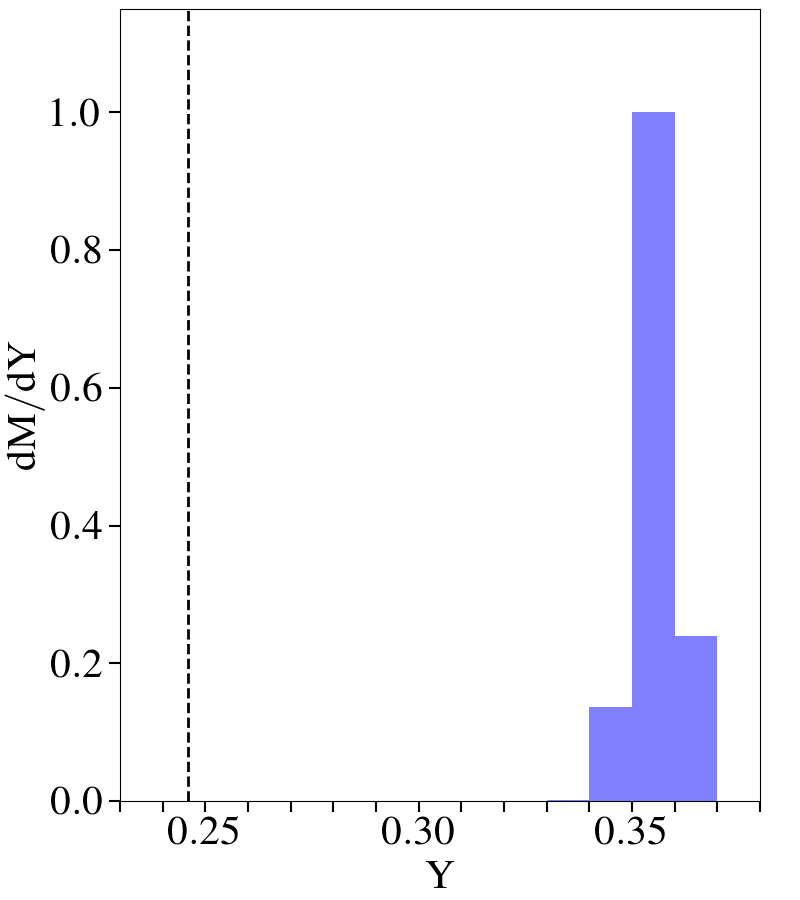}
        \hspace{0.08cm}
        \includegraphics[width=0.287\textwidth]{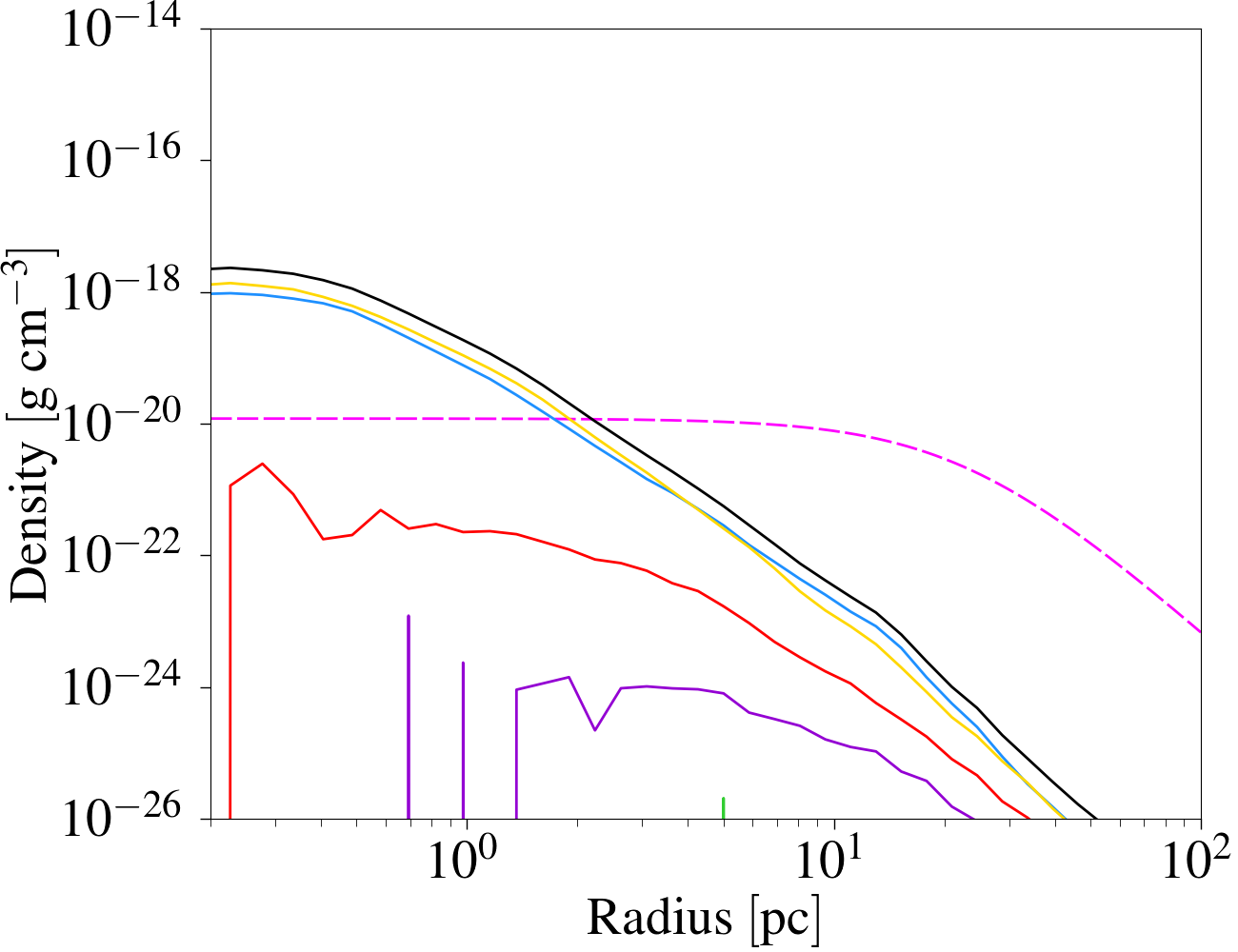}
        \includegraphics[width=0.216\textwidth]{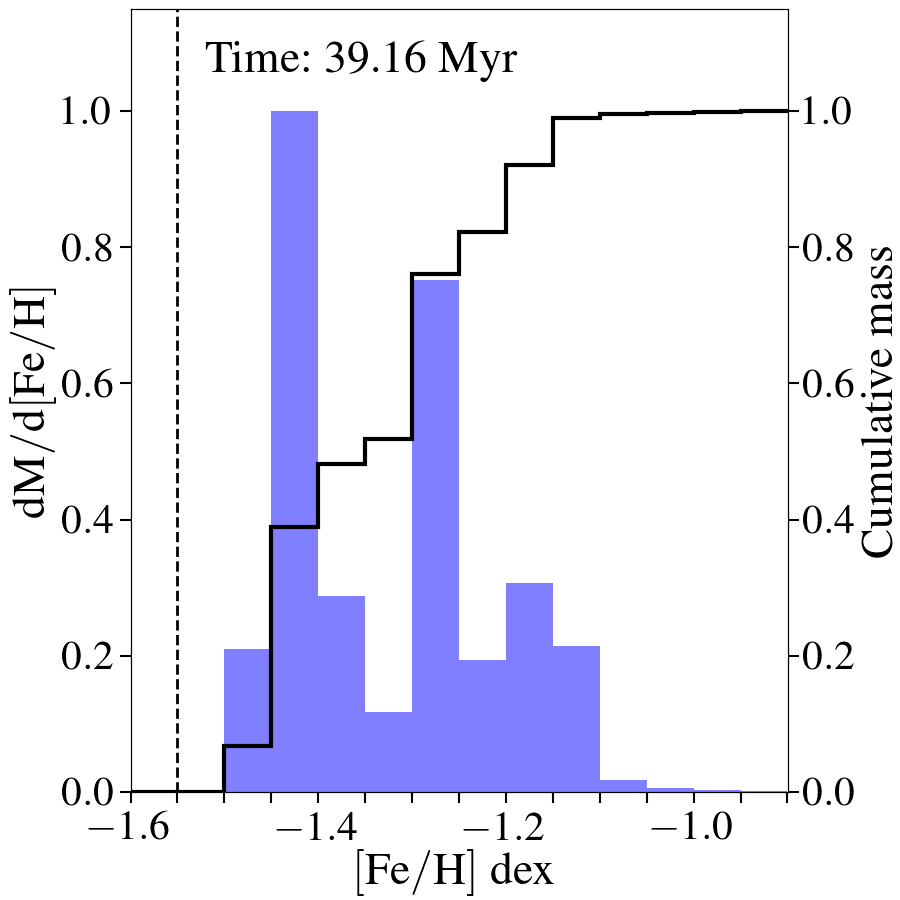}
        \\
        \includegraphics[width=0.287\textwidth]{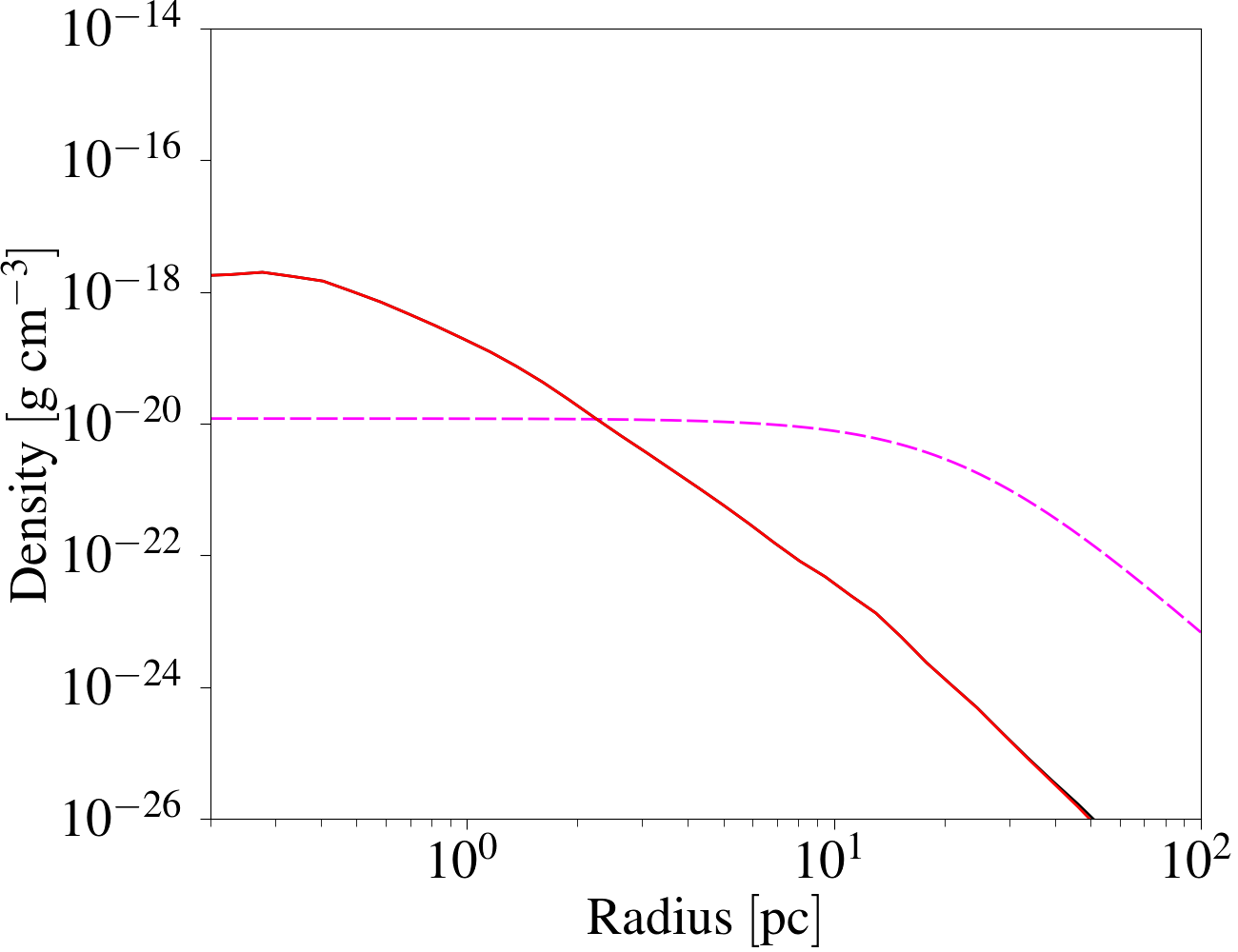}
        \includegraphics[width=0.192\textwidth]{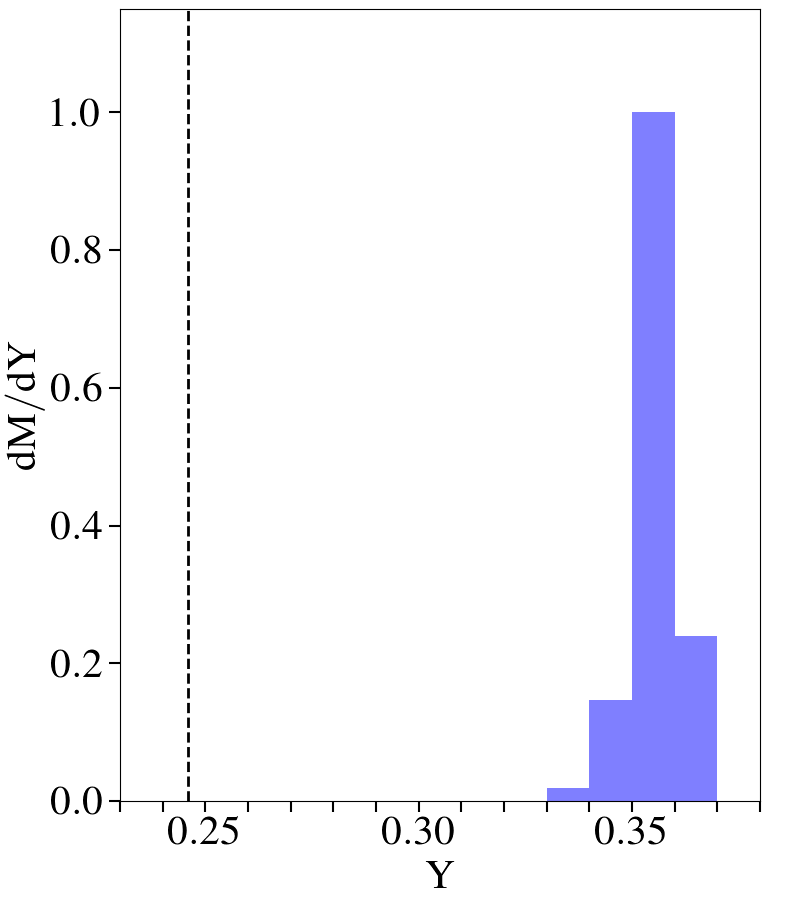}
        \hspace{0.08cm}
        \includegraphics[width=0.287\textwidth]{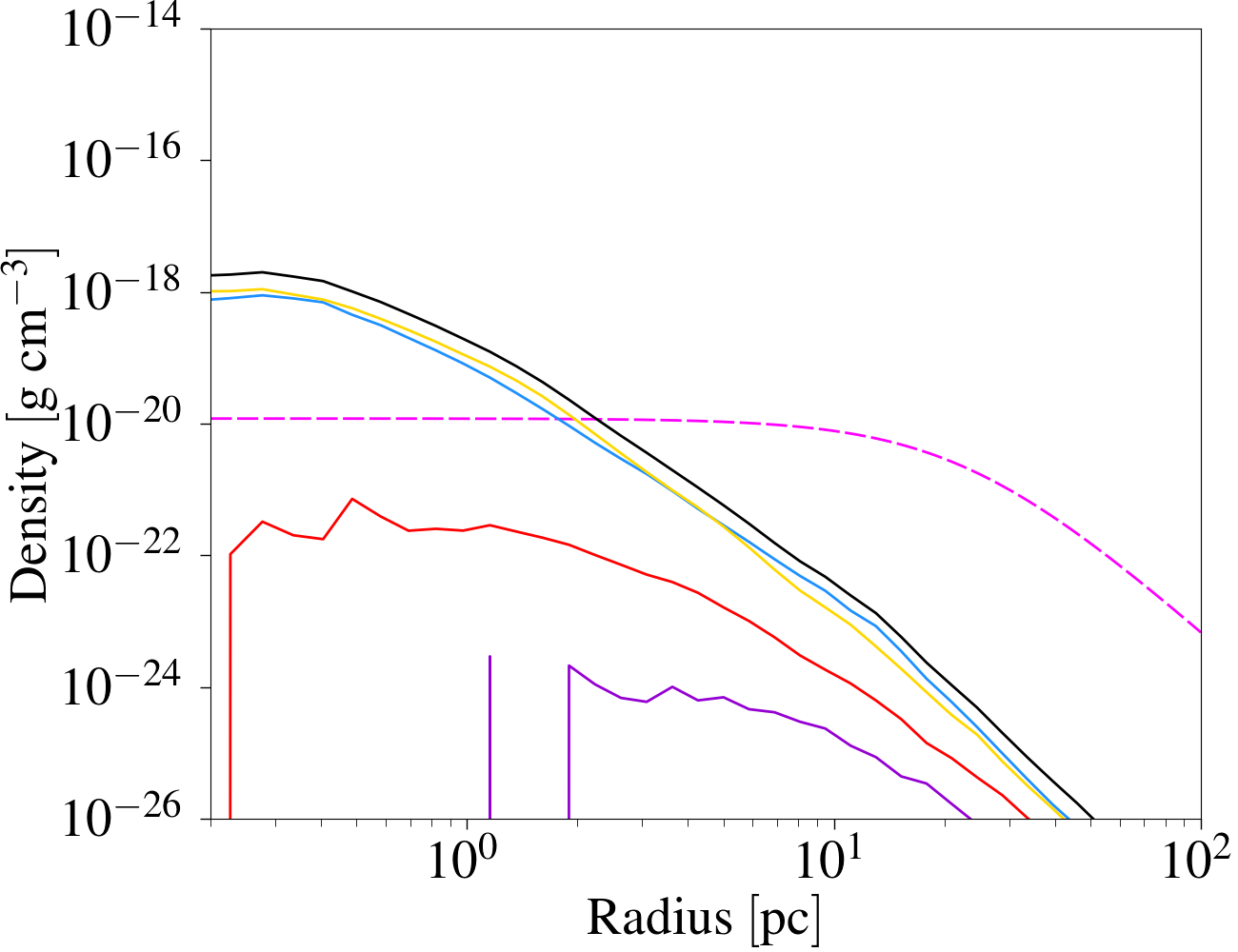}
        \includegraphics[width=0.216\textwidth]{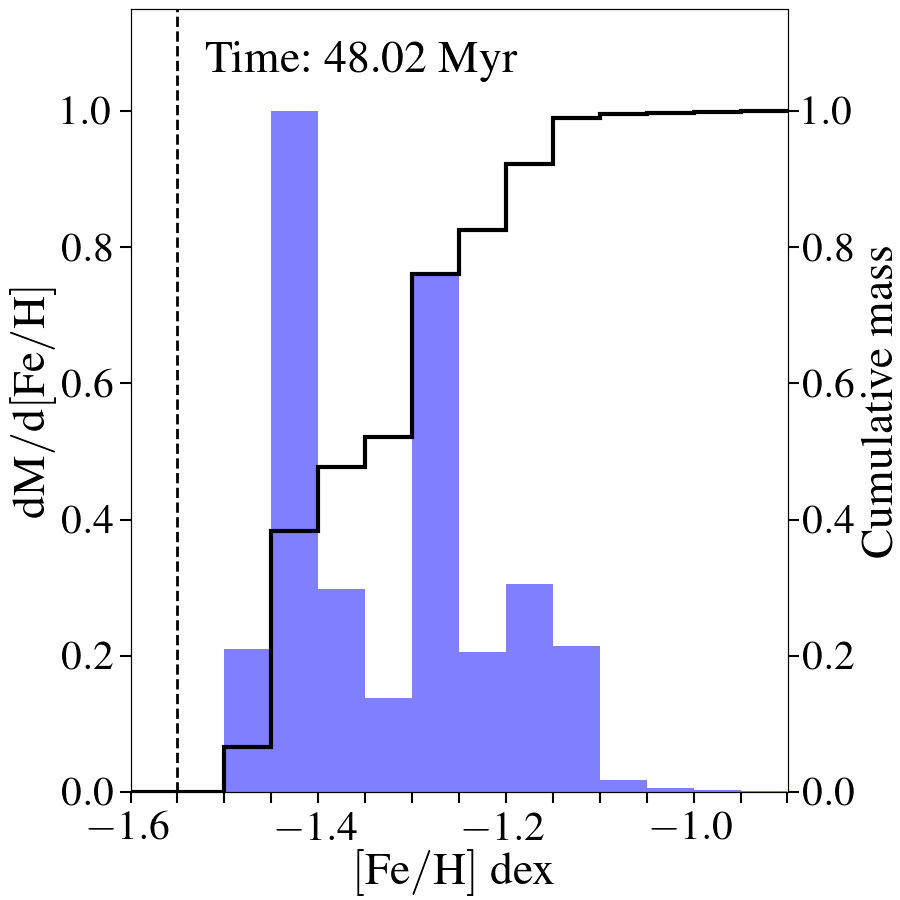}

 \caption{First and third columns: total density profile of SG stars at $t=10, 26, 39, 48$ Myr and density profiles of SG stars for several ranges of the helium mass fraction Y and the [Fe/H] ratio, respectively, for the low-density model (LD). The FG density profile is also plotted (see the legend for the details). Second and forth columns: the mass distribution of Y and [Fe/H] ratio, respectively, in the SG stars at the aforementioned evolutionary times (reported in each panel). The distributions have been obtained summing, in each bin, the masses of the stars belonging to it, and then normalizing every distribution to its maximum value. The black dashed lines represent the pristine gas composition both for Y and [Fe/H] ratio while the solid black line represents the normalized cumulative mass as a function of the [Fe/H] ratio.}   
   \label{fig:den&mdf_lowND}
\end{figure*}

\subsection{Low-density model}
\label{sec:results_LD}
\subsubsection{Dynamical evolution of the gas}

In Figure \ref{fig:maps_lowND}, we show four snapshots of the two-dimensional density and temperature maps at different evolutionary times, for the low density model. The maps have been obtained by selecting all the cells laying on the plane centered in the middle of the computational box and perpendicular to the $z$-axis. 

In the gas density map, the velocity field has been overplotted as black arrows. In addition, we have highlighted in green the regions where the velocity of the gas is pointing towards the cluster centre (the pristine gas infall is not included). It has to be clarified that, to compute such regions, we have used all the cells laying in the selected plane, while the black arrows are drawn only for some, equispaced cells.  Finally, the white dots in the temperature maps represent the newborn stars (with a lifetime of < 0.05Myr), while the red contour describes the region enclosing $50\%$ of the SG mass.

We have decided to show the maps at the same evolutionary times as \citetalias{calura2019} for a comparison. However, for computational reasons, we have truncated our run at 48 Myr, therefore the last map represents the system at this time, at variance with \citetalias{calura2019}. The other three maps are taken at $t=10$ Myr, $t=26$ Myr and $t=39$ Myr.

At $\sim 10$ Myr most of the gas in the system is composed by AGBs ejecta. The remaining gas is coming from Type Ia SNe which create several cavities filled with hot, low-density gas expanding at high velocity of the order of ${\rm 10^7 cm\ s^{-1}}$. Two of these bubbles can be clearly seen in both the maps. The effect of such explosions influences almost all the computational box as it is clearly shown by the velocity field which is pointing towards the boundaries. In the case without Type Ia SNe (taken from \citetalias{calura2019}, renamed ${\rm LD\_C19}$), instead, a cooling flow composed by AGB ejecta is formed . The fast ejecta released by SN explosions push the gas out from the system. However, this outward motion does not happen isotropically because SNe are distributed in space. In this map, two cold and dense filaments are departing from the centre of the cluster, one pointing upward and the other downward. The gas within them has a lower velocity if compared with their surrounding, and cooling flows pointing towards the cluster core are formed as it is highlighted by the green shaded areas. Very few stars are formed in these dense regions, while most of the stars are born in the very central part of the system.  

In the second panel, at 26 Myr, a clear separation between the
pristine gas and the hot material carried by the SN bubbles appears
both in the gas density and temperature maps; as shown in these maps,
the pristine gas and the hot material are pointing in opposite directions.
However, due to the large pressure of the material ejected by the SNe the pristine gas is not able to penetrate deeply into the system as in the case without Type Ia SNe (${\rm LD\_C19}$). Two SN bubbles can be seen with the one located at negative $y$-coordinates reaching a temperature of $T \sim 10^8$ K and high velocities ($\sim {5 \times \rm 10^7 cm\ s^{-1}}$), meaning that the explosion is very recent. Even in this case most of the gas in the computational box is moving outwards at high velocity, imprinted by the SN explosions. Two filaments of cold and dense gas are present, and, also in this case, part of the gas within them is moving towards the centre in a cooling flow. The very central part of the cluster remains dense and stars are still formed but with a lower extent in comparison with the previous evolutionary time. Very few stars are instead formed along the filaments. 

At 39 Myr the pristine gas is still confined at the border of the box by the SN explosions. The velocity field is pointing outwards even in the central part of the cluster, and the gas is generally hot and rarefied as a result of the SN explosions. Here very few filaments of cold and dense gas can be seen at variance with the previous maps. No cooling flow is generated since almost all the gas is highly perturbed by SN explosions whose filling factor is grown with time as a result of the 
increased SN rate (see Figure \ref{fig:dist_time}).

In the last maps, taken at $48$ Myr, almost all the box is filled with hot, rarefied and high-speed gas moving away from the system. The pristine gas remains confined at the boundaries by the SN expanding bubbles. No cooling flow is formed and the regions of cold and dense gas are strongly reduced in comparison with the previous times, and, as a consequence, also the areas where stars can be formed. Only in the core of the system the gas is dense and cold enough to allow star formation.

\subsubsection{Evolution of the stellar component}

In the first column of Figure \ref{fig:den&mdf_lowND} we plot the density profiles of the FG (magenta dashed lines), for the SG both the total (black solid lines) and the profiles computed for different ranges of the helium mass fraction Y. The same has been plotted in the third column but for the [Fe/H] ratio. The density profiles have been computed placing the origin in the centre of mass of the system. In the second and fourth columns the Y and the [Fe/H] mass distributions are plotted, respectively. The vertical black dashed line represents the pristine gas abundances, which are the same of the FG stars while the solid black line in the forth column represents the normalized cumulative mass as a function of the [Fe/H] ratio.

At $\sim 10 $ Myr all the stars are highly enriched in He, meaning that the effects of Type Ia SNe on the helium enrichment is almost negligible. However, the density of SG stars in the centre of the system is around an order of magnitude lower than in ${\rm LD\_C19}$ (see \citetalias{calura2019}, their Figure 3). 
As for iron, the most metal-poor stars, at this stage, have a [Fe/H] $\sim -1.48$ dex, which reflects the chemical composition of the ejecta of the most massive AGBs. The vast majority of the stars are however more enhanced in iron, due to the pollution of the AGB ejecta with the material expelled by Type Ia SNe, as it can be seen from the [Fe/H] mass distribution function. We can therefore define enriched stars the ones with a [Fe/H] >$-1.48 $ dex, while the ones with [Fe/H] <$-1.48 $ dex are produced a mix of AGB ejecta and pristine gas whose [Fe/H]$\sim -1.55$ dex. It has to be said that the AGB yield of He decreases as a function of stellar mass, which means that also the [Fe/H] of the ejecta will decrease during the evolution of the system. However the variation is almost negligible during the time interval we are looking at.  

At $26$ Myr the bulk of the stars are still highly enriched in helium and are therefore formed mostly from AGB ejecta, while only a negligible number of stars are poorly enriched in He. Such stars, formed out of diluted gas, are mainly located far from the centre since, as it can be seen in the maps, the external gas hardly penetrates into the system. 
A large spread is visible in the [Fe/H] ratio where three peaks can be clearly seen. As we will discuss further on, the star formation in this model is not constant during the evolution of the system as it happened in ${\rm LD\_C19}$, namely the case without Type Ia SNe. Some small, but still relevant, bursts of star formation are present which lead to peaks in the [Fe/H] mass distribution where, in general, the higher the [Fe/H] ratio of the peak the younger the stars. The density profile in the third column of Figure \ref{fig:den&mdf_lowND} shows that the bulk of the stars have a [Fe/H] ratio between $-1.5$ and $-1.1$ dex, with an increased number of enriched stars in comparison with the previous snapshot.

At $39$ Myr most of the stars still have an extreme He enrichment, at variance with ${\rm LD\_C19}$ where a peak at intermediate helium enrichment is formed in the Y distribution. This results from the inability of the pristine gas to mix with the AGB ejecta since the SNe are confining it to the border. Stars not showing an extreme He enrichment give a negligible contribution to the mass of the system and therefore also to the density. Focusing instead on the [Fe/H] ratio, the contribution from the two more metal-rich peaks is increased in comparison with the previous snapshot, meaning that newborn stars are formed from gas significantly polluted by Type Ia SNe.

In the last evolutionary time, at $48$ Myr, not much differences can be seen in the graphs if compared with the previous ones. The intermediate helium enriched stars give a negligible contribution to the mass at variance with what has been obtained without Type Ia SNe (${\rm LD\_C19}$), therefore the bulk of the stars still show an extremely He enrichment. Also the [Fe/H] mass distribution is almost unchanged, since the SF is lowered and stars produced at this time give a negligible contribution.

At all the evolutionary times we have selected, SG stars are dominant in the central 2-3 pc, while, at larger radii, most of the stars belong to the FG, in agreement with observations \citep{lardo2011,dalessandro2019}. This result is quite similar to the one obtained by \citetalias{calura2019}, however, the total stellar density of the SG that we obtain in the central region is almost two orders of magnitude lower than in their case. The other major difference regards the helium enrichment in the SG stars; we find that SG stars with an extreme He composition are dominant at all radii, at variance with \citetalias{calura2019} where this population is concentrated in the central part of the system but it is not prevailing in the outskirts.

 At the end of the simulation, the average [Fe/H] of SG stars is $-1.32$ dex with a dispersion of $\sigma^{ \rm SG}_{\rm [Fe/H]}=0.11$ dex. To obtain them, we have converted the ${\rm dM/d[Fe/H]}$ of each bin into number of stars assuming a Kroupa IMF truncated at ${\rm 8 M_{\odot}}$, following the results of \citet{bekki2019}. In order to compare our results with observations, we have estimated the internal iron dispersion of the whole cluster assuming that most of FG stars are lost during the subsequent evolution, ending up with a fraction of SG stars of $\sim 0.7$, as observed in massive GCs \citep{milone2017}. In this way, we derive $\sigma_{\rm [Fe/H]}=0.14$ dex, which is larger than the observed value in the bulk of GCs (\citealt{carretta2009c}, \citealt{bailin2019})  but in agreement with the typical spread found in Type II GCs. 
However, in this model, the infalling gas dilutes negligibly the AGB ejecta and therefore SG stars are mostly displaying an extreme helium abundance, at variance with observations. Moreover, as shown by \citet{dercole2011}, dilution is required by the AGB scenario in order to reproduce the Na-O anticorrelation. Here, instead, Type Ia SNe prevent the accretion of pristine gas and, therefore, the dilution of AGB ejecta. For these reasons, the LD scenario is not viable and has to be discarded.  


\begin{figure*}
        \centering

        \includegraphics[width=0.493\textwidth]{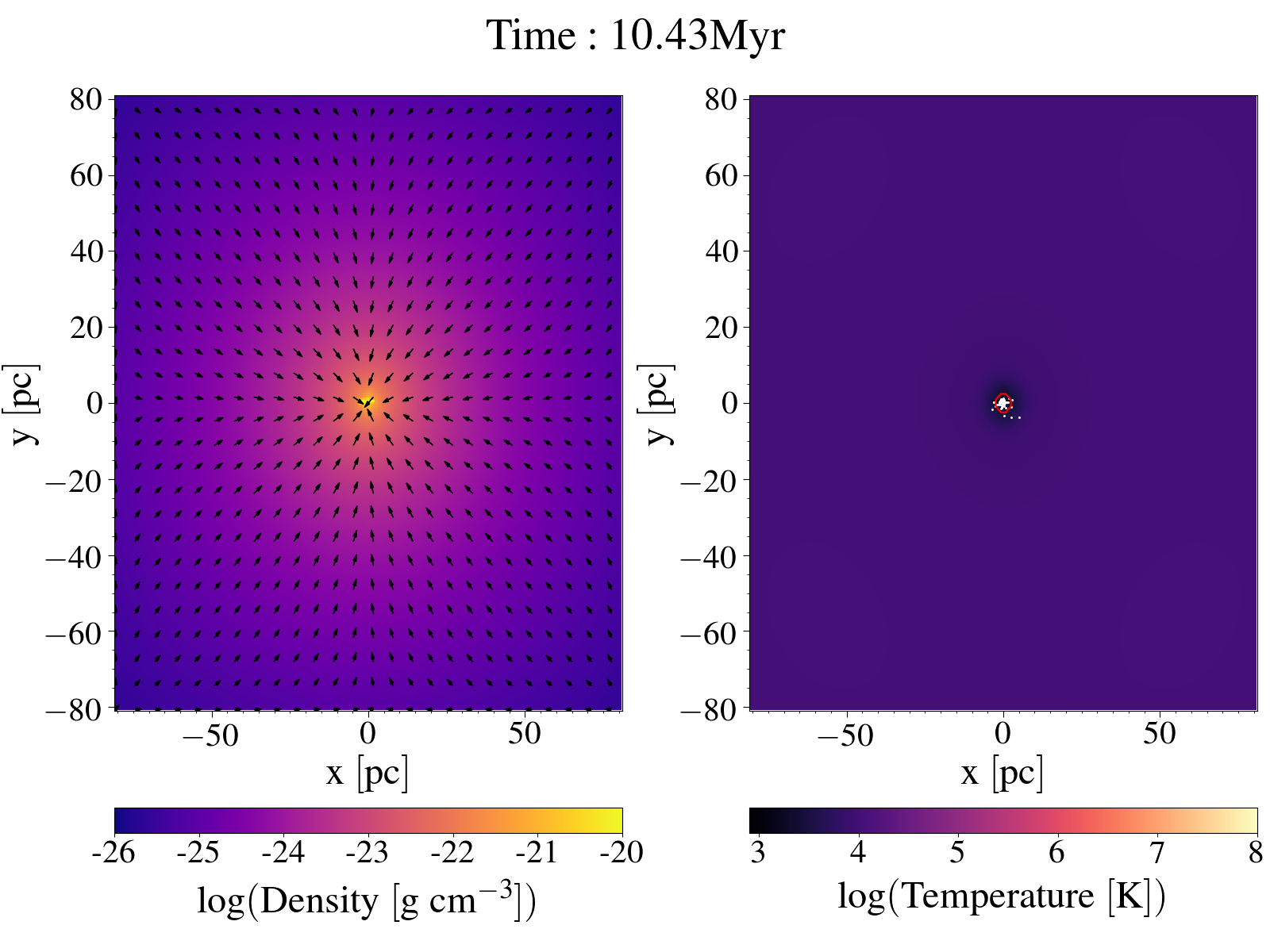}
        \hspace{0.1cm}
        \includegraphics[width=0.493\textwidth]{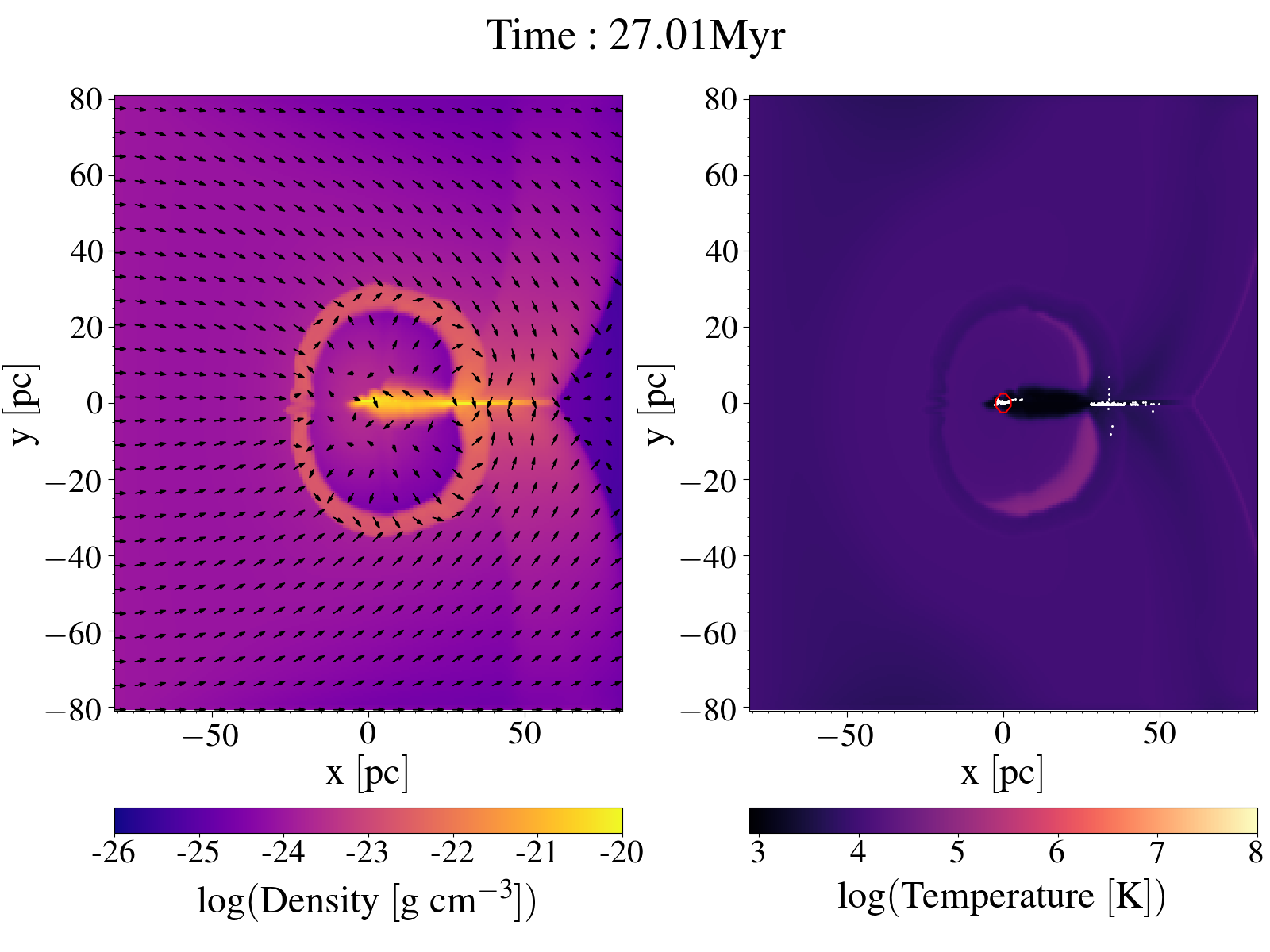}
        \\
        \vspace{0.3cm}
        \includegraphics[width=0.493\textwidth]{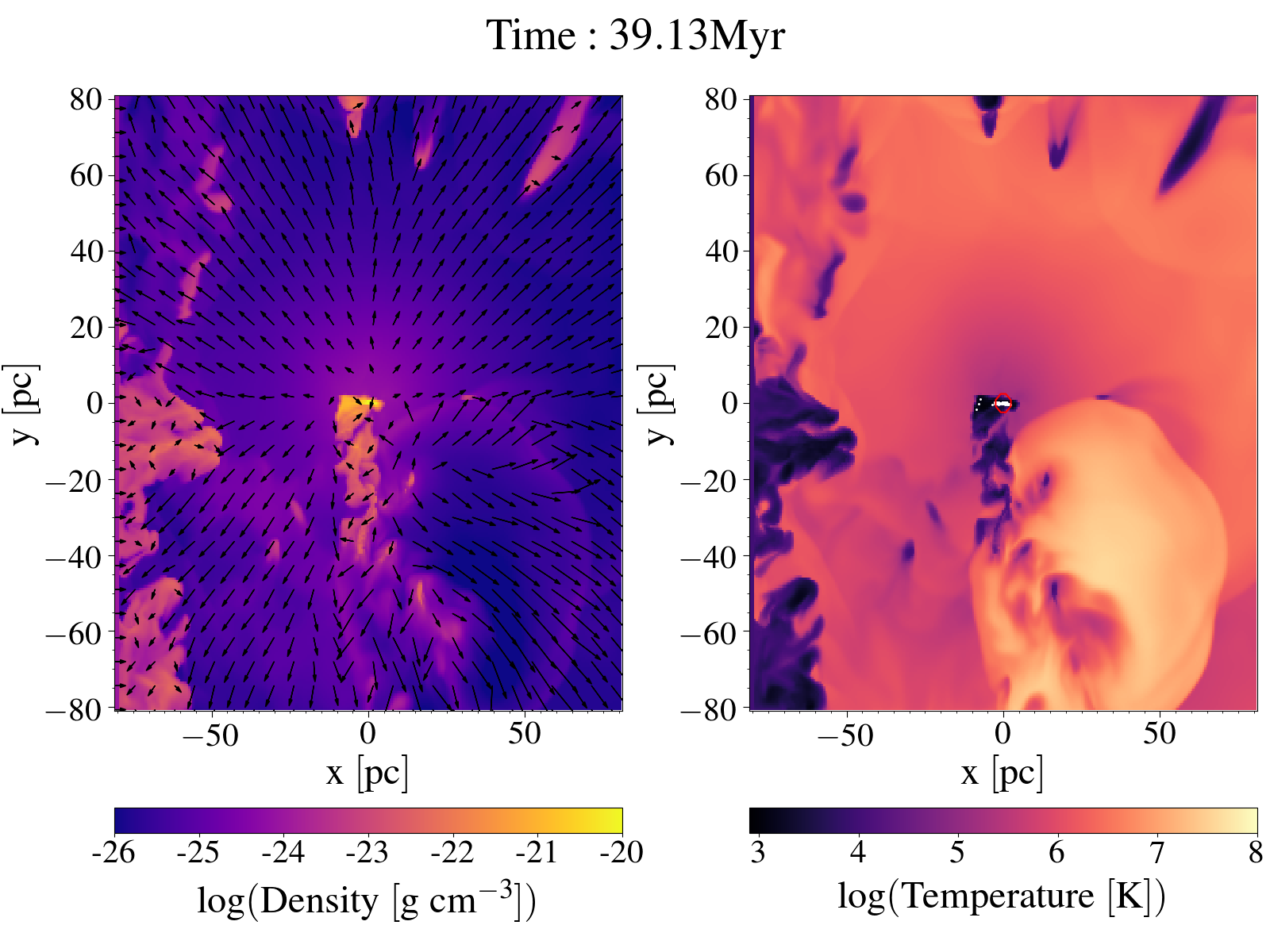}
        \hspace{0.1cm} 
        \includegraphics[width=0.493\textwidth]{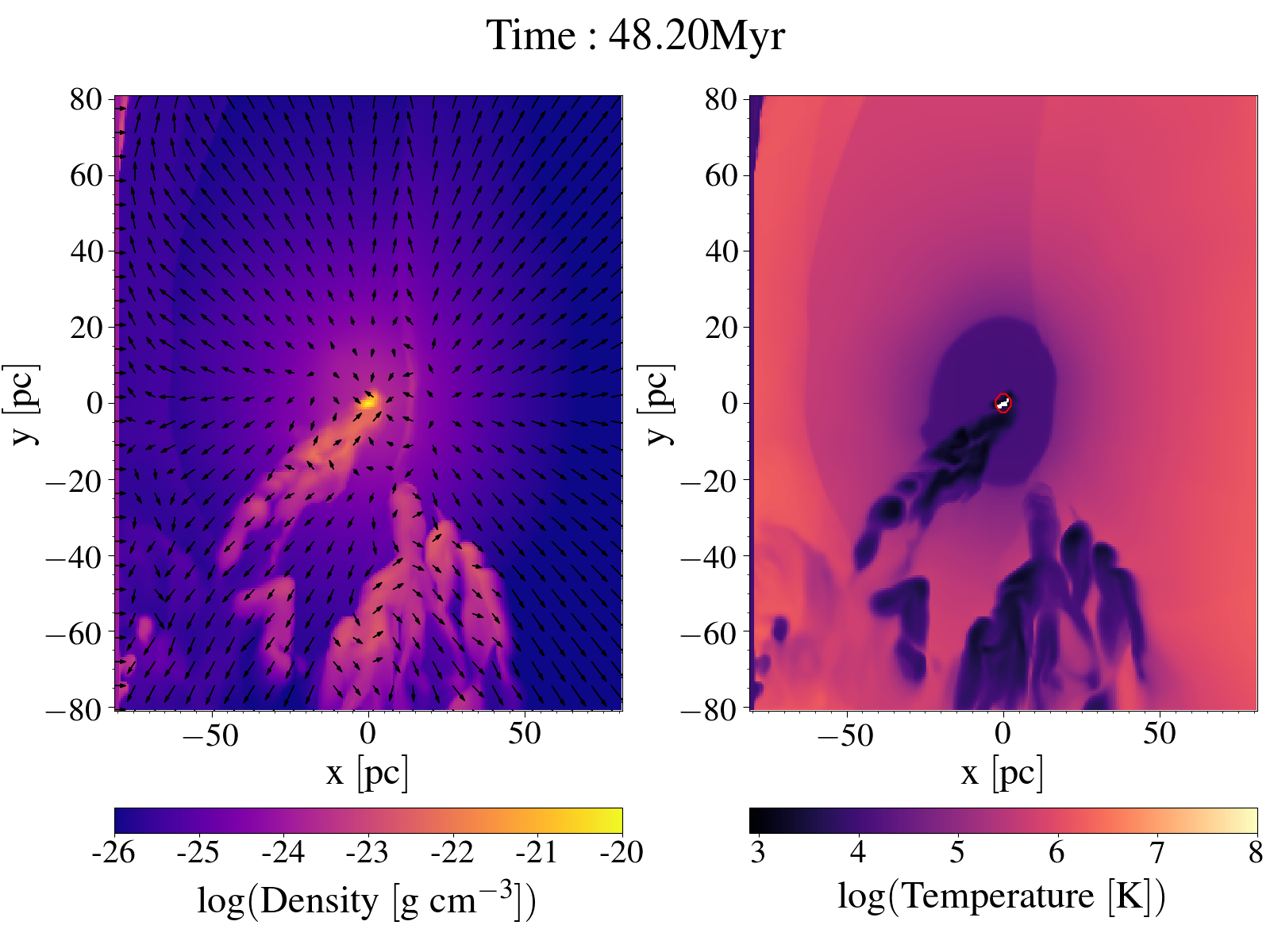}

 \caption{Two-dimensional maps of the gas density on the left-hand panels of each plot and of the temperature on the right hand panels in the x-y plane at different evolutionary times for the low density simulation with delayed Type Ia SNe (${\rm LD\_DS}$). From the top left to the bottom right: $t=10, 27, 39, 48$ Myr. Other symbols and lines as in Figure \ref{fig:maps_lowND}. }   
   \label{fig:maps_lowD}
\end{figure*}

\begin{figure*}
        \centering
        
        \includegraphics[width=0.287\textwidth]{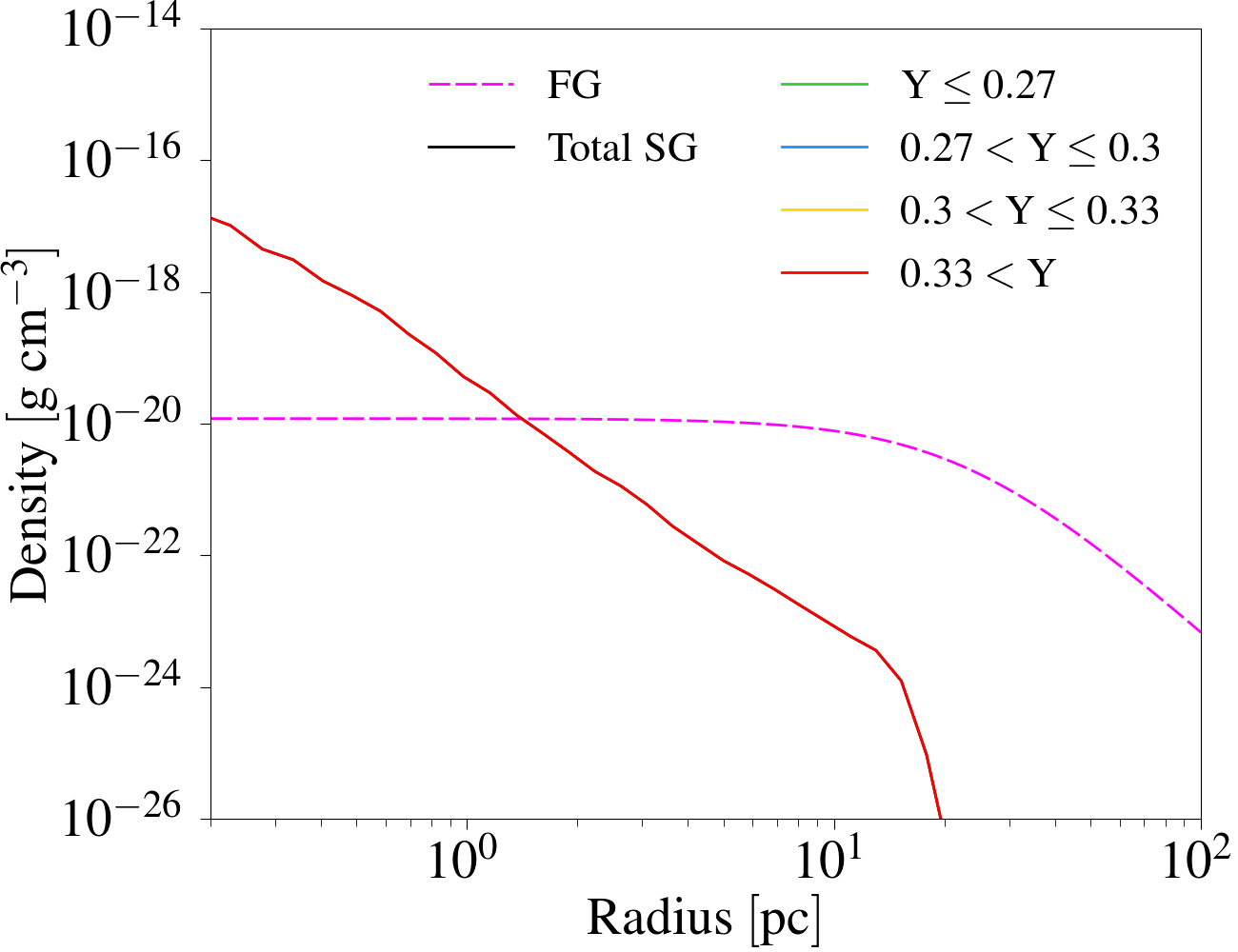}
        \includegraphics[width=0.192\textwidth]{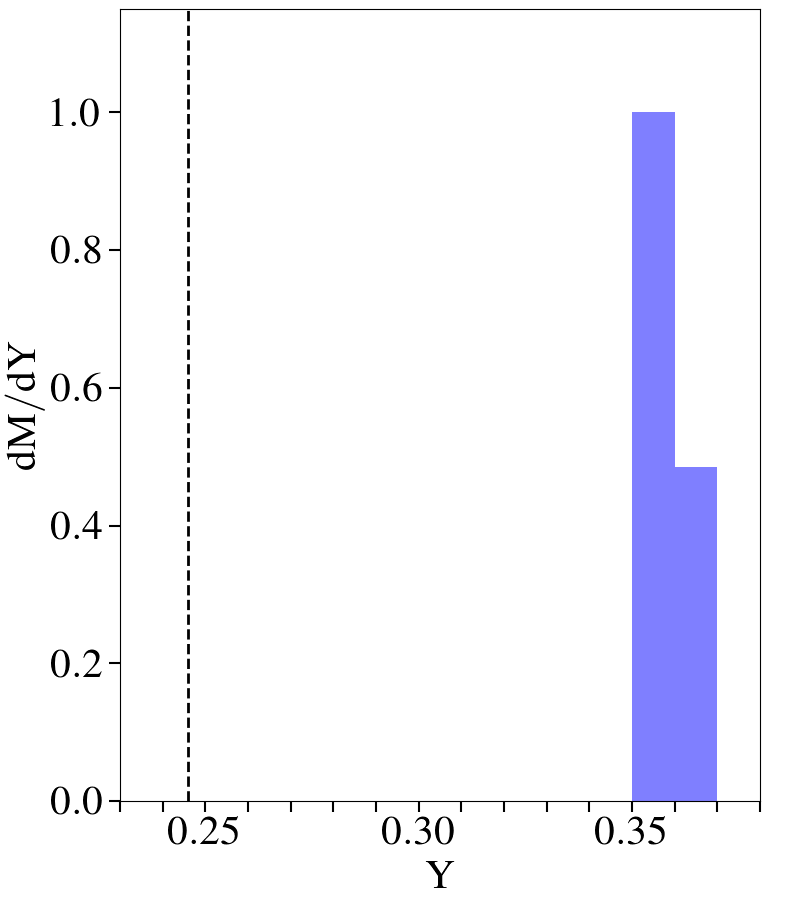}
        \hspace{0.08cm}
        \includegraphics[width=0.287\textwidth]{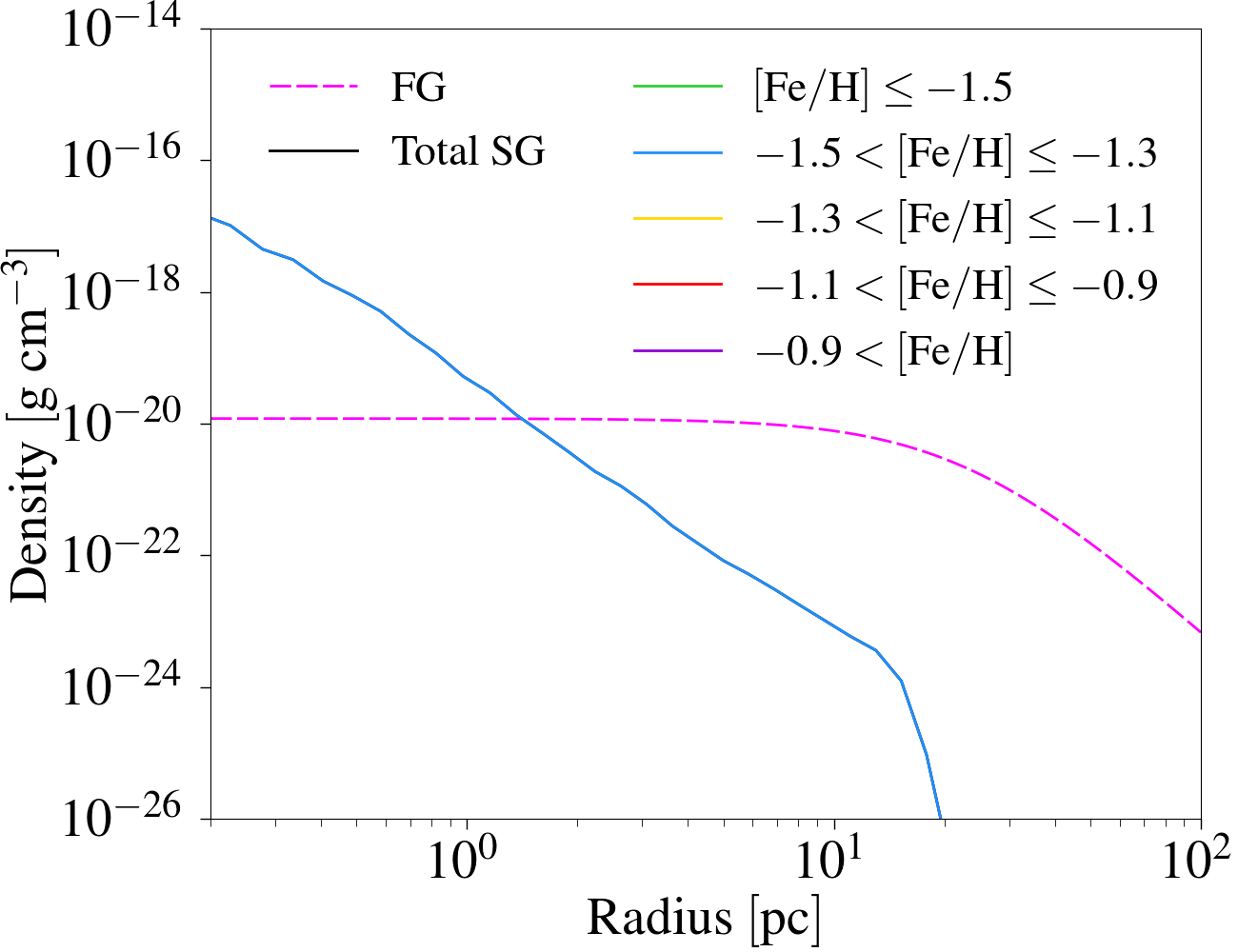}   \includegraphics[width=0.216\textwidth]{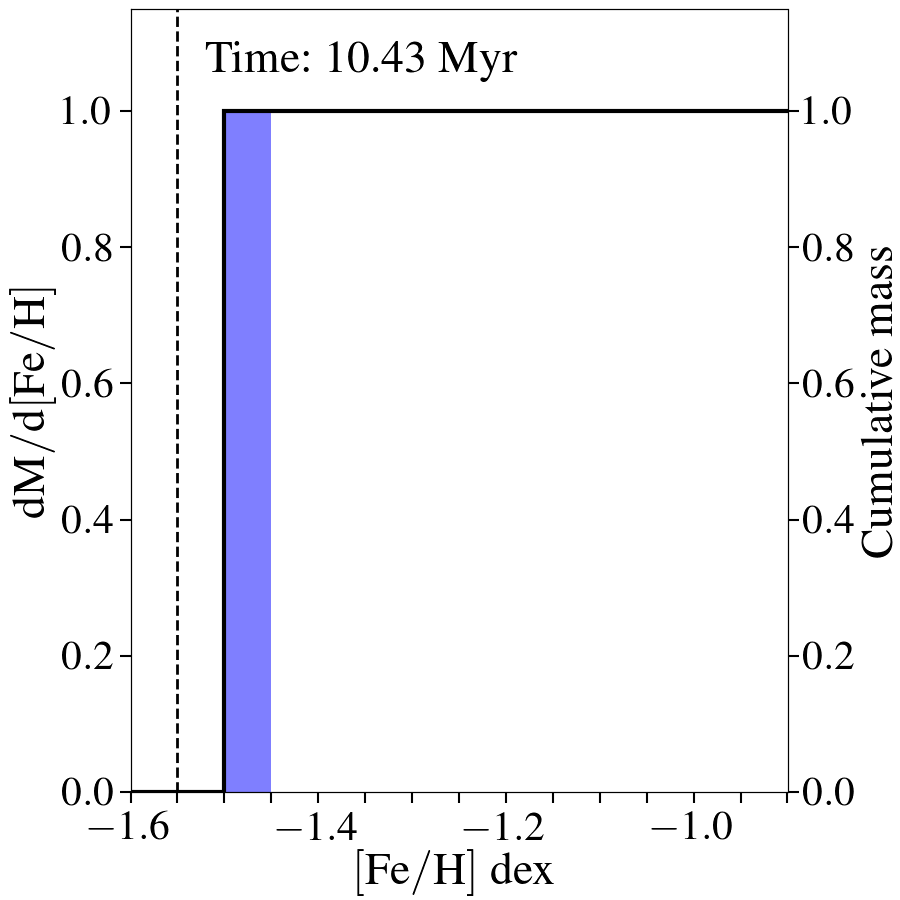}
        \\
        \includegraphics[width=0.287\textwidth]{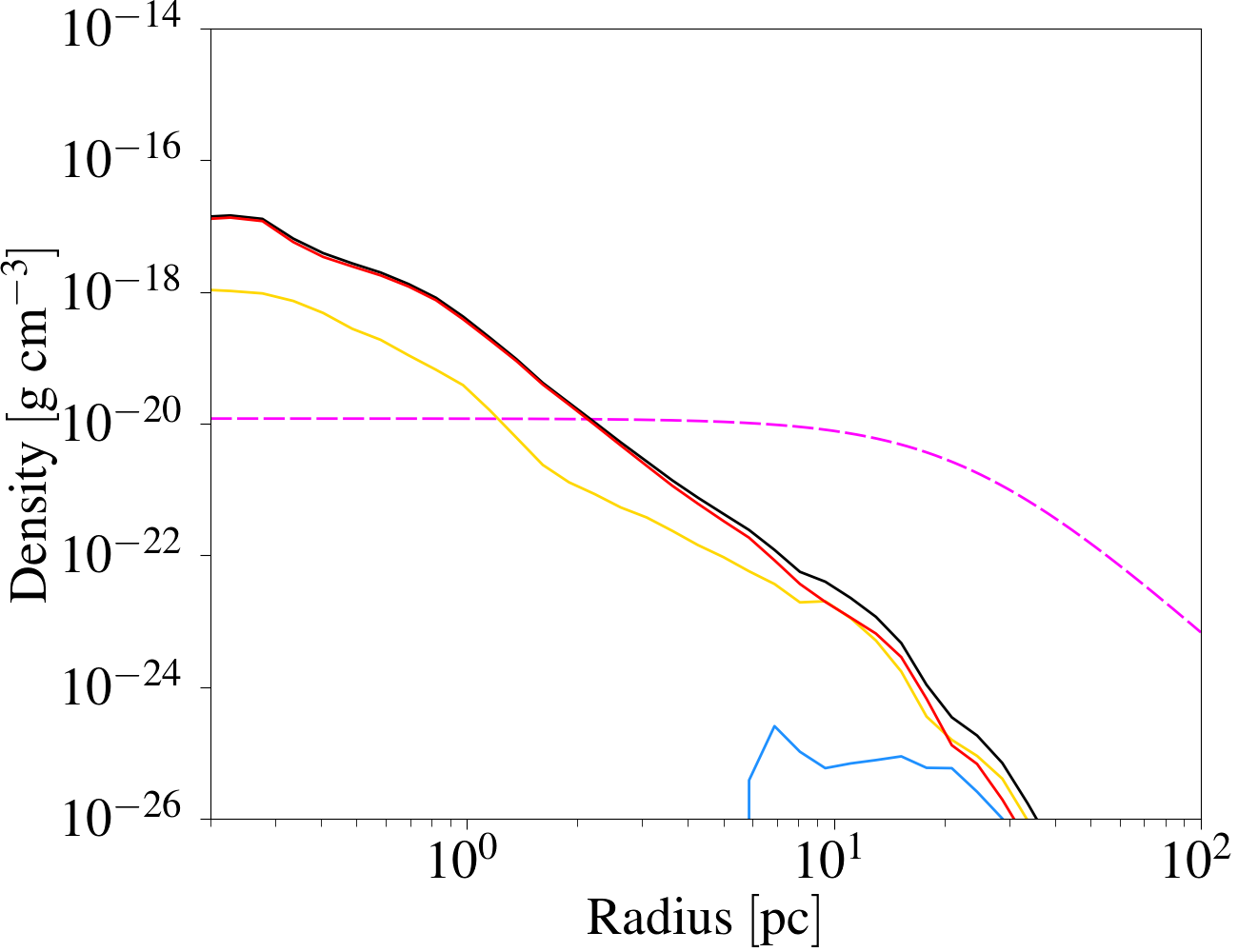}
        \includegraphics[width=0.192\textwidth]{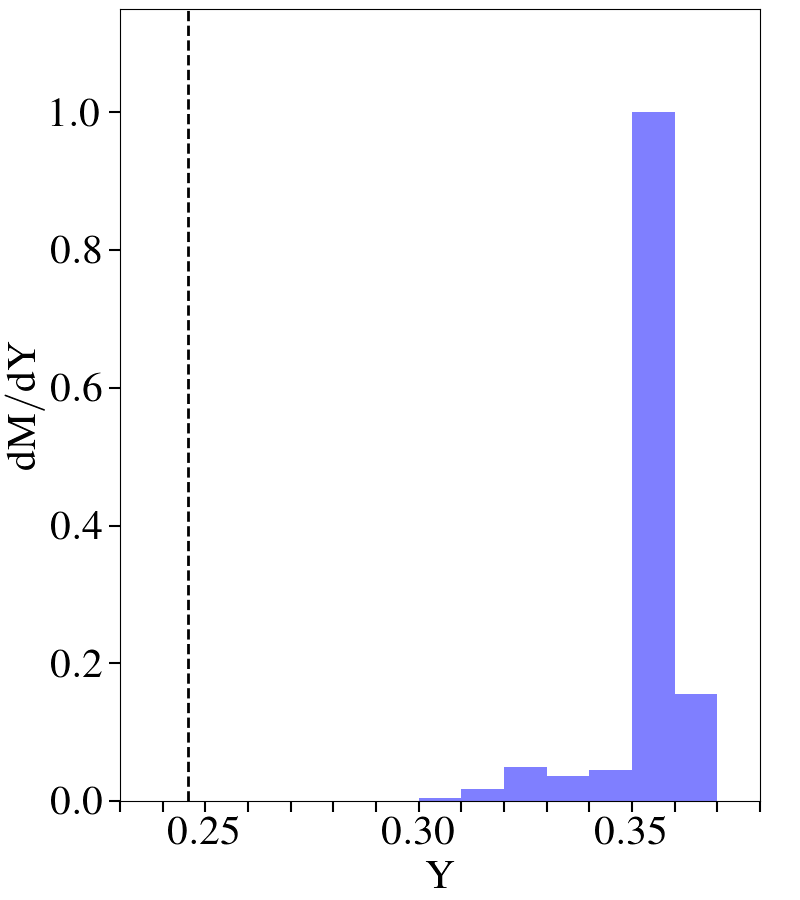}
        \hspace{0.08cm}
        \includegraphics[width=0.287\textwidth]{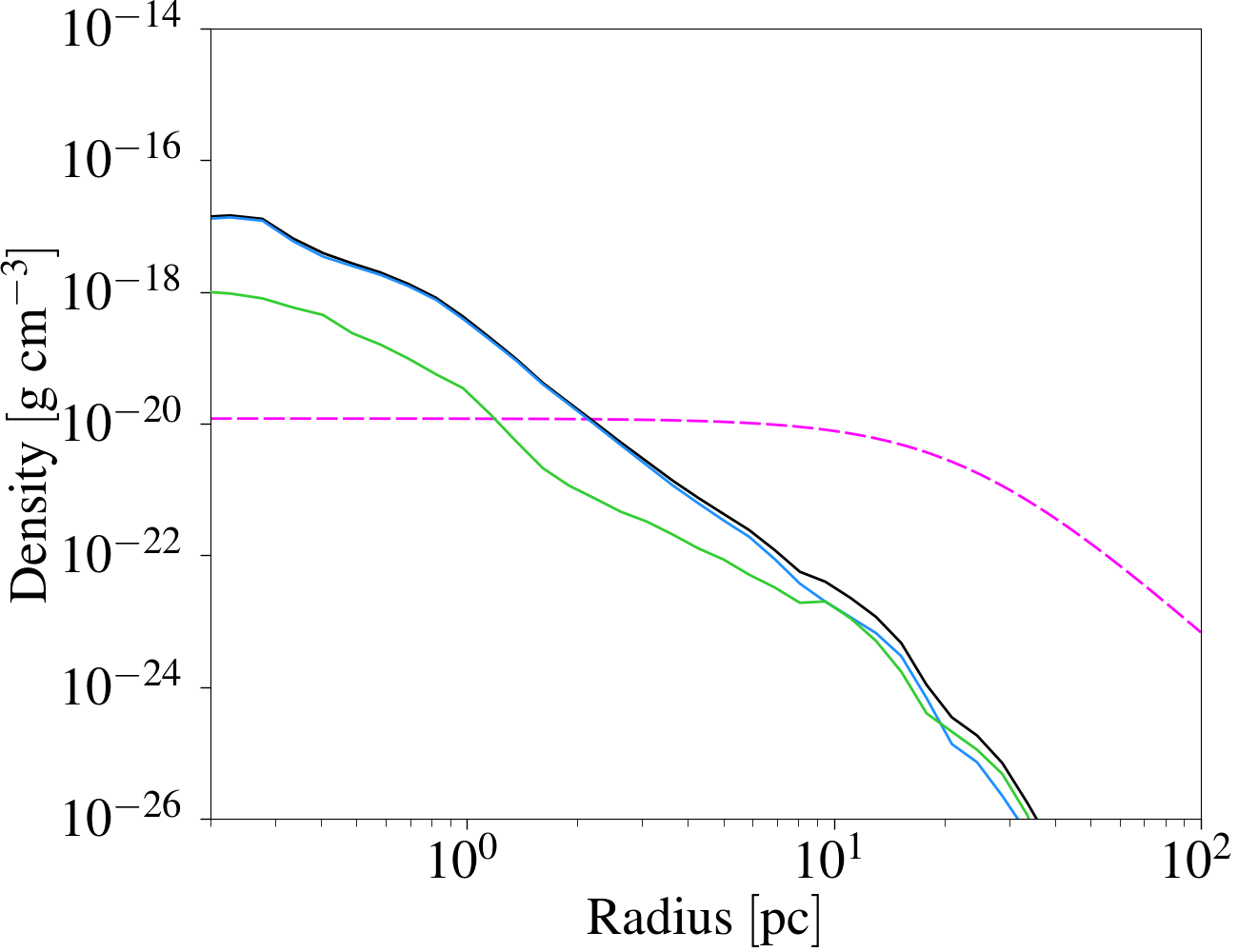}
        \includegraphics[width=0.216\textwidth]{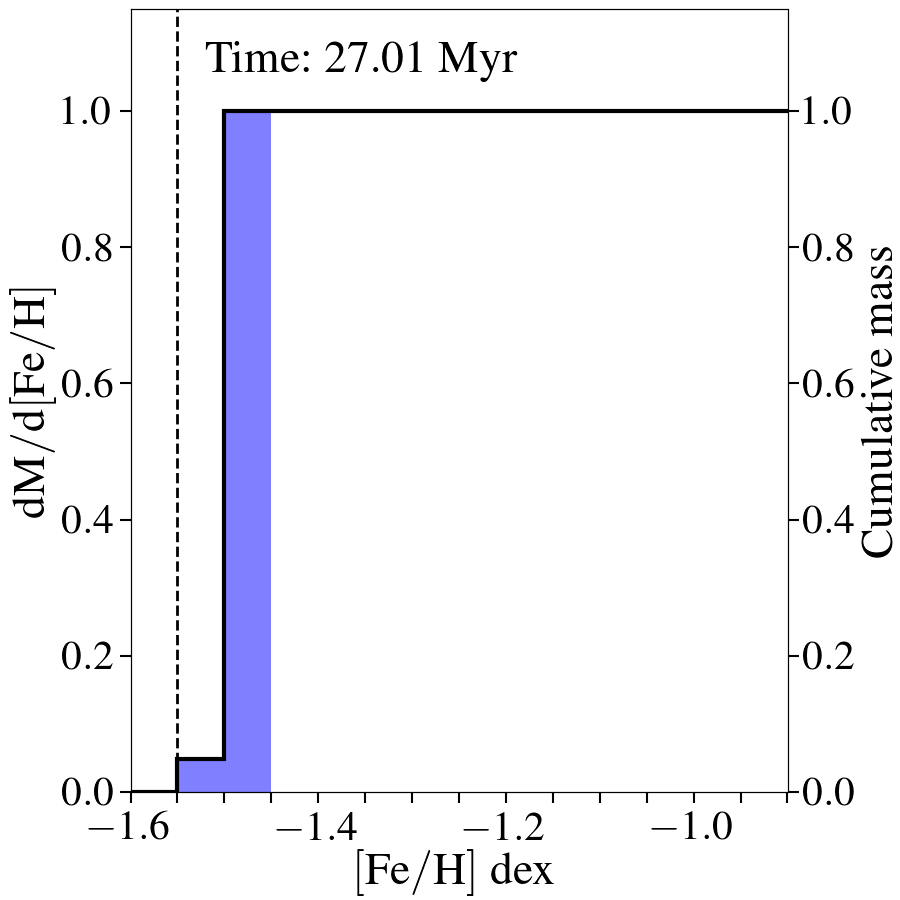}
        \\
        \includegraphics[width=0.287\textwidth]{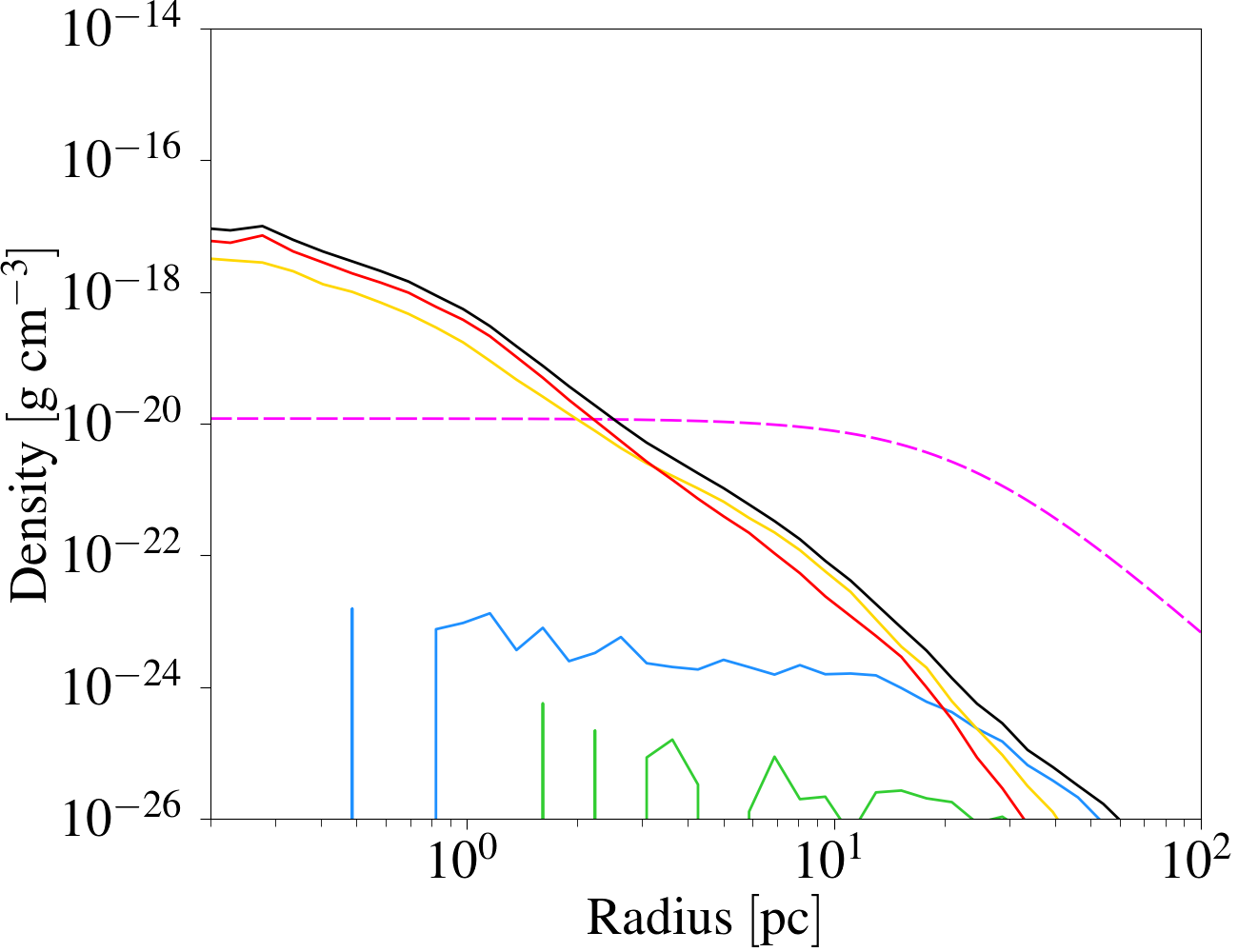}
        \includegraphics[width=0.192\textwidth]{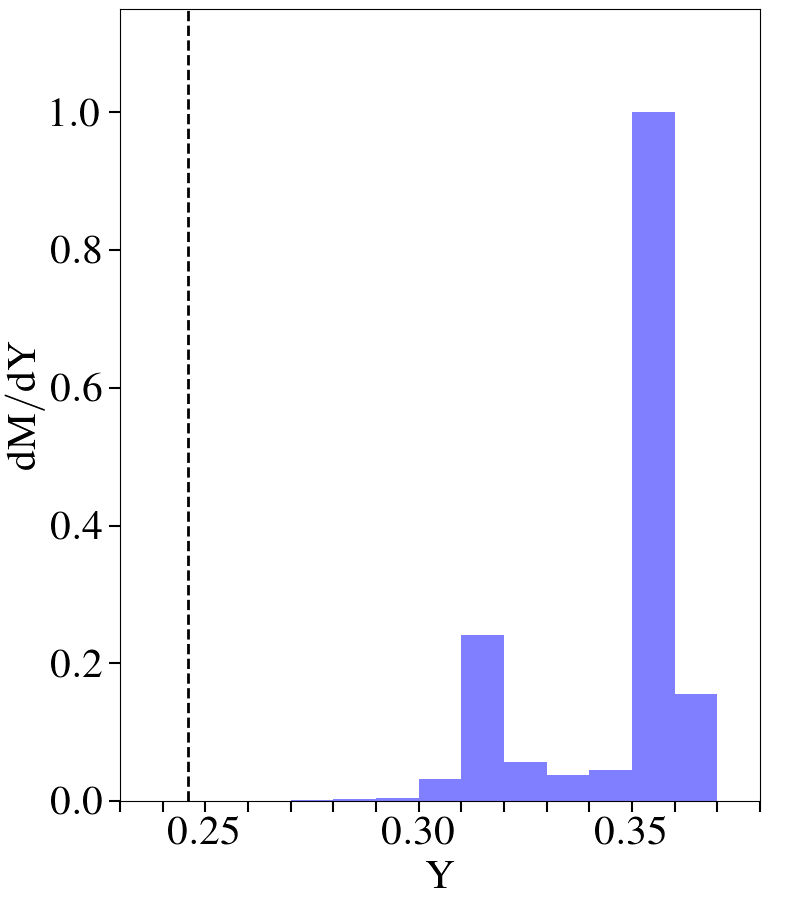}
        \hspace{0.08cm}
        \includegraphics[width=0.287\textwidth]{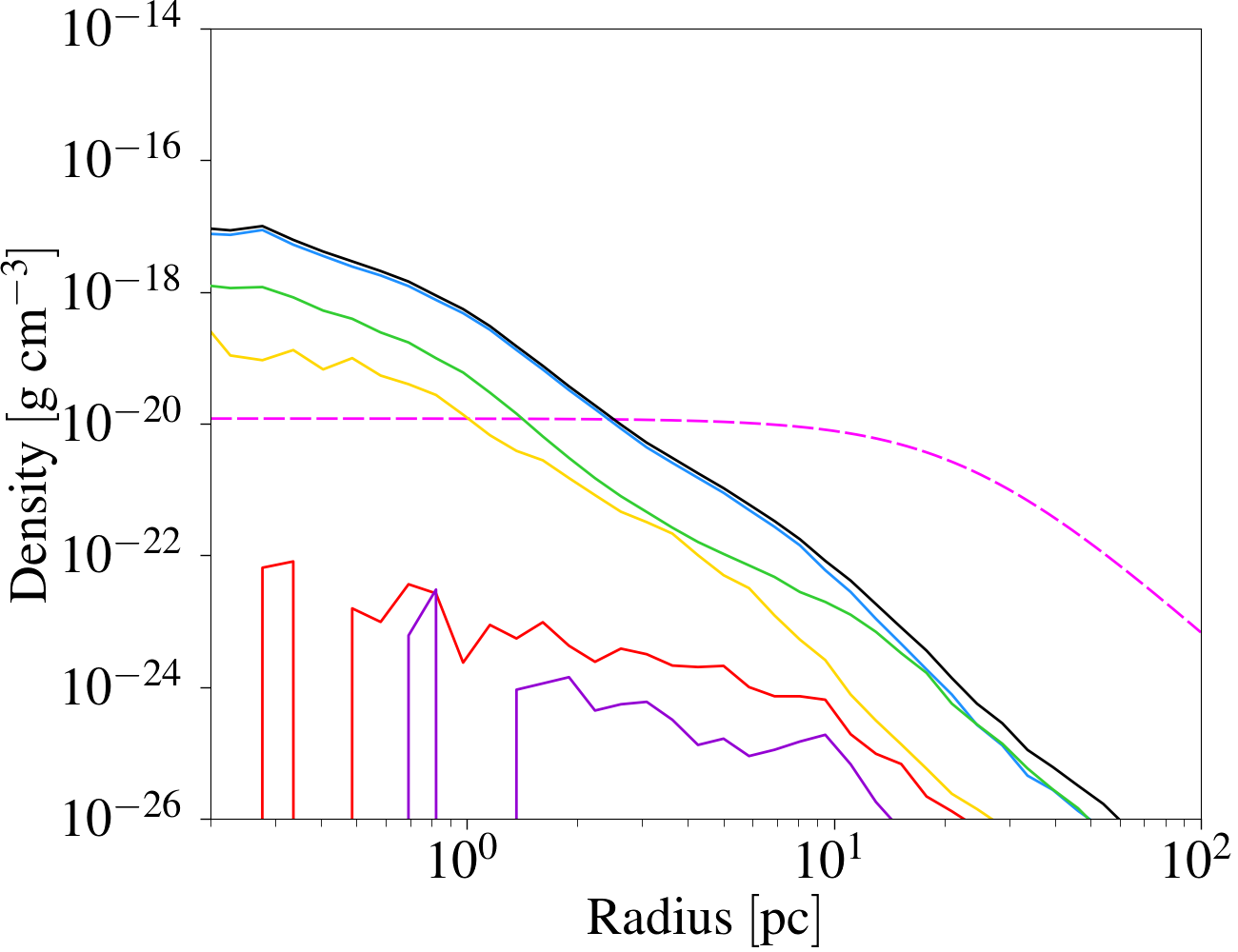}
        \includegraphics[width=0.216\textwidth]{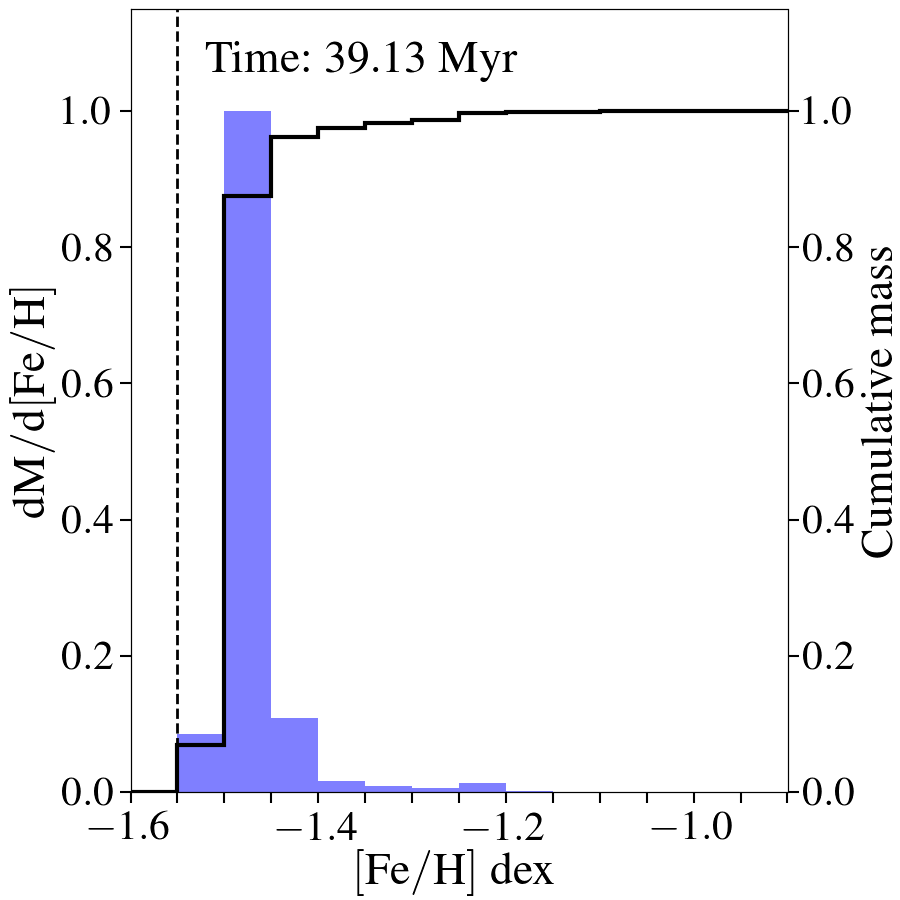}
        \\
        \includegraphics[width=0.287\textwidth]{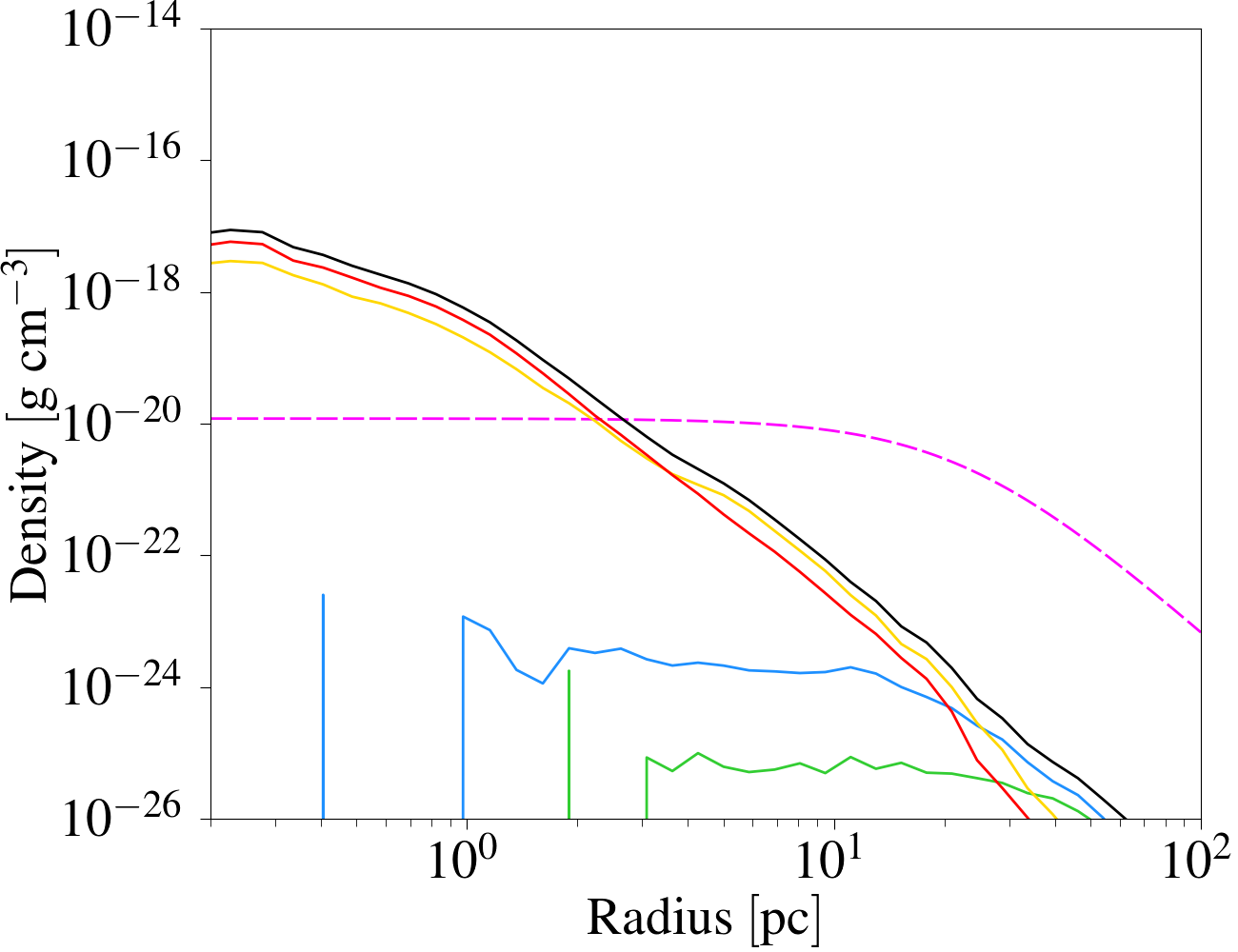}
        \includegraphics[width=0.192\textwidth]{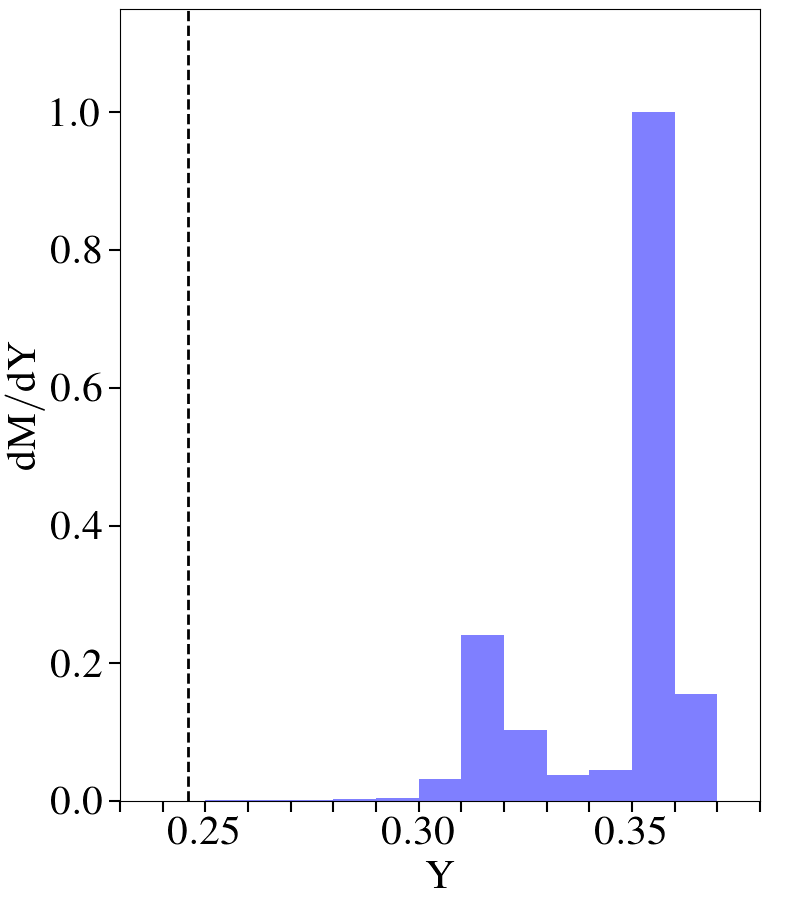}
        \hspace{0.08cm}
        \includegraphics[width=0.287\textwidth]{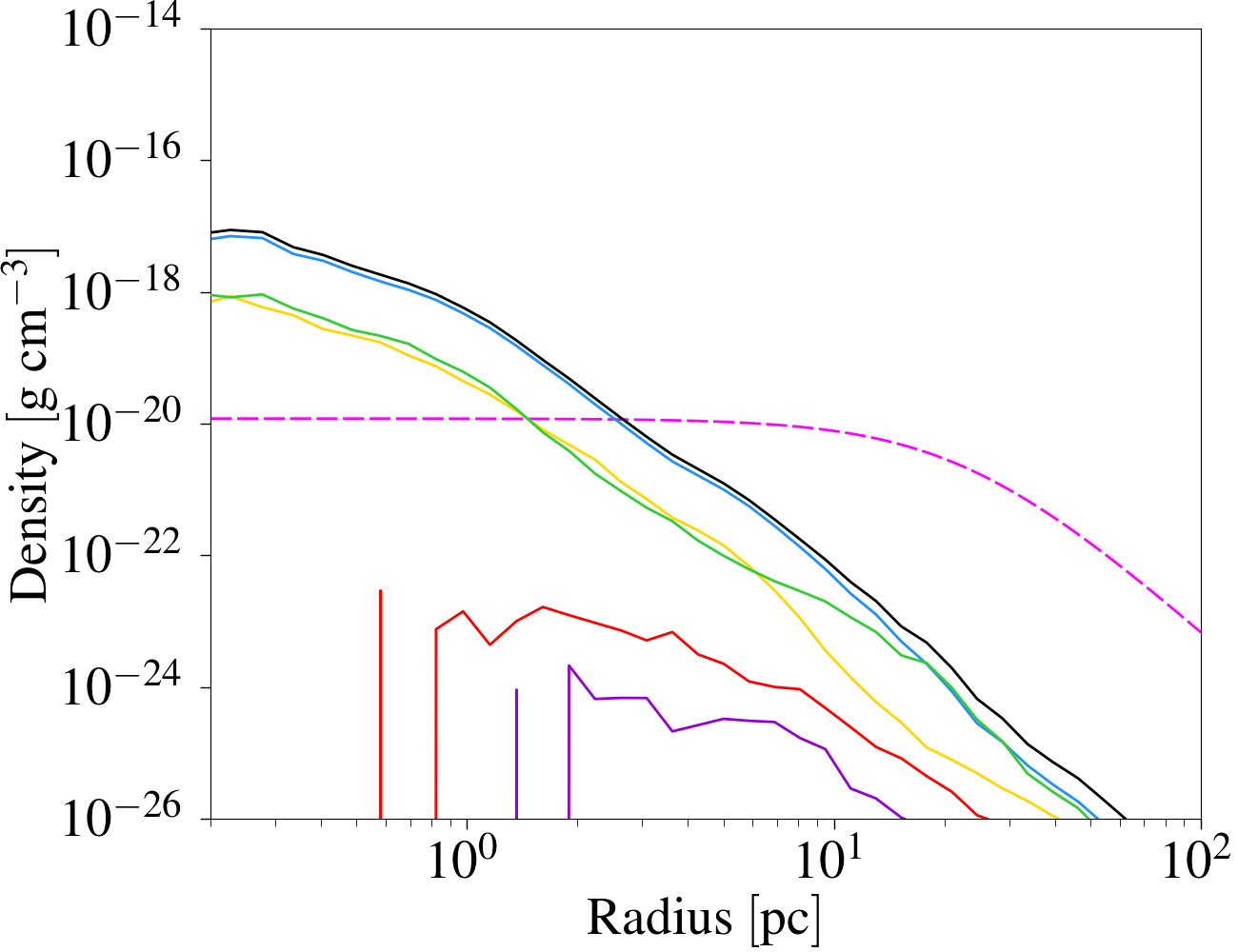}
        \includegraphics[width=0.216\textwidth]{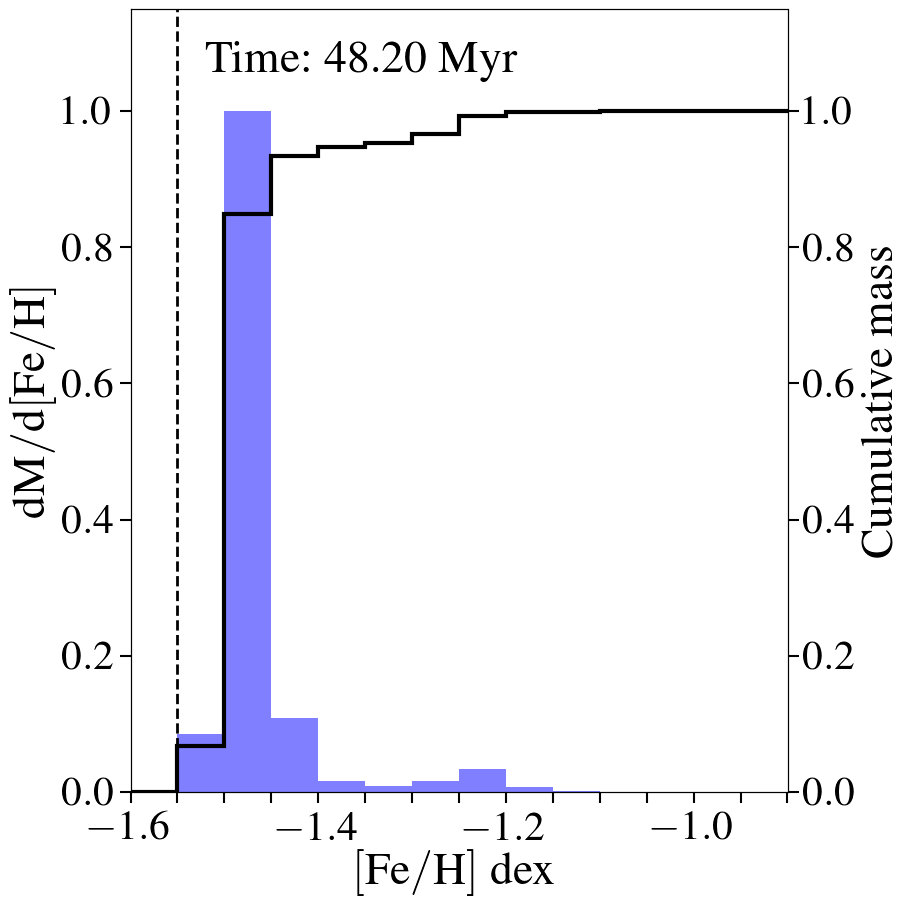}

 \caption{First and third columns: total density profile of SG stars at $t=10, 27, 39, 48$ Myr and density profiles of SG stars for several ranges of the helium mass fraction Y and the [Fe/H] ratio, respectively, for the low-density model with delayed Type Ia SNe (${\rm LD\_DS}$). The FG density profile is also plotted (see the legend for the details). Second and forth columns: the mass distribution of Y and [Fe/H] ratio, respectively, in the SG stars at the aforementioned evolutionary times (reported in each panel). Other symbols and lines as in Figure \ref{fig:den&mdf_lowND}. }   
   \label{fig:den&mdf_lowD}
\end{figure*}

\subsection{Low-density model with delayed Type Ia SNe}
\subsubsection{Dynamical evolution of the gas}

In Figure \ref{fig:maps_lowD} we show the two-dimensional maps for the case in which Type Ia SN are delayed, with respect to the standard \citet{greggio2005} formulation, of $25 $ Myr. In this scenario the infall of pristine gas starts before the first SN explosion, at variance with the ${\rm LD}$ scenario. The maps are taken at the same evolutionary times as for the low-density model, with the exception of the second snapshot for which we have decided to show the maps at $27$ Myr.

At $10$ Myr the system is isolated and the gas, composed only by AGB ejecta, creates a cooling flow towards the cluster core as it can be seen looking at the velocity field. The gas concentrated near the cluster centre has a high density and is cold ( $\sim T_{\rm floor}$) with an almost spherical shape. Stars belonging to the SG have already formed in the centre of the cluster as it is shown both by the contours and by the white dots. This stellar component is more compact and more spherically distributed than the SG in the LD model.

At $27$ Myr the infall of pristine gas is already started, the front shock has crossed the centre of the system and the first Type Ia SN has just exploded in the cluster central regions. In most of the computational volume, the gas velocity field reflects the motion of the pristine gas. Far from the system, the arrows are parallel to the $x$-axis, meaning that the gas is not affected by the presence of the cluster. As we approach the cluster core, the velocities are increasingly distorted in the $y$-direction as a result of the gravitational force exerted by the cluster. Even the front shock is distorted by the presence of the cluster, displaying two symmetric \enquote{wings} downstream of the system. These wings delimit the gas already shocked from the unperturbed one; the latter, besides being less dense, is still pointing towards the cluster centre. The only other part of the box in which the velocity field is not driven by the infall is the central region, where a SN explosion has created a bubble. Near the shell, the density is high, while, inside the bubble, it is almost equal to the pristine gas one with the exception of the very central region, corresponding to the cluster centre, where the density remains high and stars continue to be formed. It has to be noted that the temperature distribution inside the bubble is asymmetric, being hotter next to the shell downstream of the shock, while cold upstream of it. Such an inhomogeneity, which appears also in the density maps, even though it is less evident, is due to the infall of pristine gas. Upstream the density is higher given the direct effect of the infall which leads to an higher efficiency of cooling and consequently to a lower temperature. 

Despite the SN explosion, a clear tail of dense and cold gas moving toward the cluster centre is formed. Through this accretion column \citep{bondi1944,shima1985,moeckel2009}, the gas is able to flow and reach the central part of the system, mixing with the AGB ejecta to form new stars. Newborn stars are present also along the tail, but only in the portion still not perturbed by the SN shock front. In the part of the tail already crossed by the SN expanding shell almost no stars are formed due to the perturbation produced by the shock wave.

 
 At $39 $ Myr, pristine gas is no more falling deeply inside the system, as it happens in the previous snapshots. SNe explosions are confining it to the border of the computational box, as in the LD model. Gas is still falling towards the centre in limited areas of the system, next to the cluster centre. However, the bulk of the gas is pointing outwards, driven by the continuous SN explosions. Even the tail and the related accretion column formed by the pristine gas infall and visible at $27$ Myr have already disappeared. Stars are formed almost only in the very central region where cold and dense gas is still present.
 
 At $48$ Myr, the pristine gas is still confined far from the system by the gas moving outwards, after being perturbed by shock waves generated by SNe. Only in the central region cold gas is falling towards the cluster centre where stars are still formed. 
 
 \subsubsection{Evolution of the stellar component}

In Figure \ref{fig:den&mdf_lowD} we plot the density profiles of SG both total and for various helium composition and [Fe/H] ratio, together with the Y and [Fe/H] mass distributions for the low-density model with Type Ia delayed of $\sim 25$ Myr, at different evolutionary times. 

At $10$ Myr, all the stars are extremely helium-enriched, since the only gas present in the system at this stage is coming from the AGB ejecta. Also the iron composition is peaked at the AGB value because no Type Ia SN has exploded yet. The density in the central region is higher than in the LD model, since star formation is not hindered by SN explosions. In addition, a sharp cut off is present at $\sim 10$ pc because the gas eligible for star formation is concentrated at the centre and is distributed almost isotropically.

Later, at $27$ Myr, the pristine gas is diluting the AGB ejecta, especially in the external regions, where stars with lower helium enrichment and [Fe/H] ratio are formed, being, in the outermost regions, even the dominant component. The pristine gas has, in fact, the same iron mass fraction of the AGB ejecta, but a higher hydrogen one\footnote{  Before and during the AGB phase stars experience various dredge-ups, deepenings of the convective envelope which lead to mixing of the outer envelope with products of the H and He burnings. The consequence, from the chemical point of view, is that the hydrogen mass fraction in the external part of the envelope, and therefore also on the ejecta, is reduced.}. Almost no star with an higher iron composition is already formed, since only one SN is exploded and the iron it has ejected has not been used to create new stars yet. 

At $39$ Myr, instead, an increasing number of stars is formed out of mixed gas with intermediate helium composition, which gives rise to a second smaller peak in the Y distribution. These stars are less centrally concentrated than those with an
extreme helium enrichment. The presence of a small fraction of stars with a higher [Fe/H] ratio indicates that the iron produced by Type Ia SNe has been recycled. Most of these stars are located in the central regions, where, however, the dominant component still remains the one composed by stars formed mainly from AGB ejecta, followed by stars born out of AGB ejecta plus pristine gas ([Fe/H]$<-1.5$ dex).

At $48$ Myr, not many differences are visible in comparison with the previous timestep. The reason lies in the decreased star formation rate (SFR) and, as a consequence, in the almost negligible increase of the total SG stellar mass between the two evolutionary times.


Many are the differences between the model we are describing here and the LD one. Firstly, while in the LD scenario almost all stars are characterized by extreme helium enrichment, here stars with intermediate helium composition contribute significantly, in particular at large radii where they become the dominant component. This result resembles what has been obtained by \citetalias{calura2019} although in our model the contribution from mildly enriched stars is lower. Unlike the ${\rm LD}$ model, in the ${\rm LD\_DS}$ one the pristine gas is able to reach the central part of the system and stars with a mixed composition are formed. As for iron, when SNe explosions are delayed, stars with [Fe/H] ratios similar to the pristine gas one are formed even in the central part of the cluster, while in the LD model this component is absent. On the contrary, very few stars in the scenario with delayed SNe are strongly enriched with respect to the AGB yield value, whereas three peaks at higher [Fe/H] ratios are present in the LD model. In the ${\rm LD\_DS}$ scenario, in fact, no burst of SF is present, at variance with the LD one.

The mean [Fe/H] ratio of SG stars at the end of the ${\rm LD\_DS}$ simulation is $-1.45$ dex with an dispersion of ${\rm \sigma^{\rm SG}_{[Fe/H]}= 0.06 }$ dex. As for the LD model, we have calculated the iron dispersion of the whole cluster assuming that the SG account for the $70\%$ of the stars, obtaining $\sigma_{\rm [Fe/H]}=0.07$ dex, which is within the range of values found in Type II clusters. However, similarly to the LD model, most of the SG stars show extreme helium abundance. Even though a second peak at intermediate $Y$ is formed (see Figure \ref{fig:den&mdf_lowD}), the spread in helium is still too large in comparison with observations, meaning that dilution is still too low.  
The extent of dilution depends on the delay of Type Ia SN explosions with respect to the onset of the infall, becoming larger with greater delays and approaching the results of \citetalias{calura2019}. To better constrain the timescales at play, more chemical elements need to be traced and more delay times to be tested in future works.

Overall, our results and those of \citetalias{calura2019} suggest that in the context of low-density
models, longer delay times, possibly leading to the first SN Ia explosions around $80-100$ Myr after the FG formation, are required to produce multiple-population properties consistent with those observed in Galactic GCs.
Concerning the central SG density, the LD model predicts a lower value than the case with delayed SNe due to the larger number of explosions characterizing it. 


\begin{figure*}
        \centering

        \includegraphics[width=0.493\textwidth]{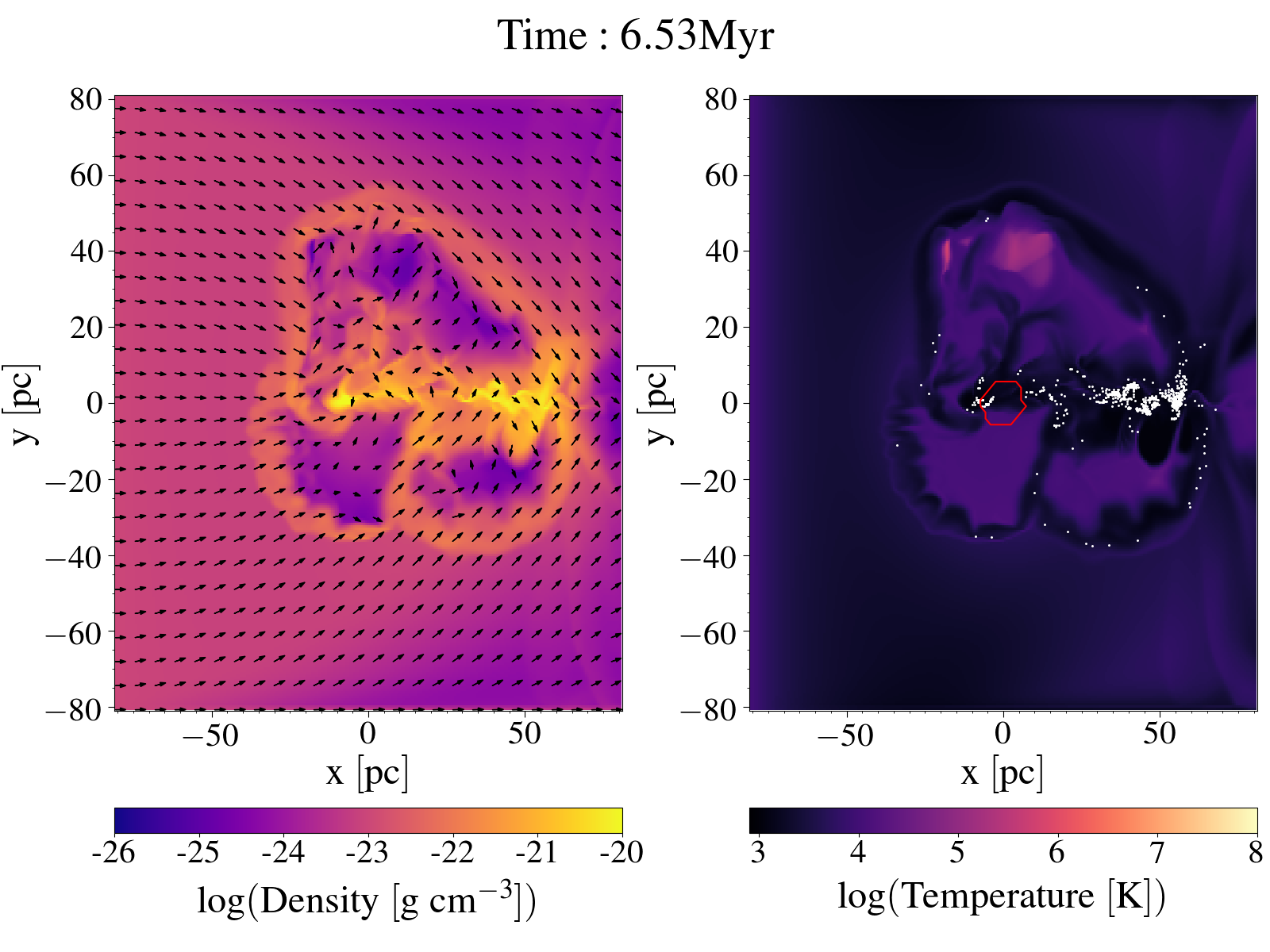}
        \hspace{0.1cm}
        \includegraphics[width=0.493\textwidth]{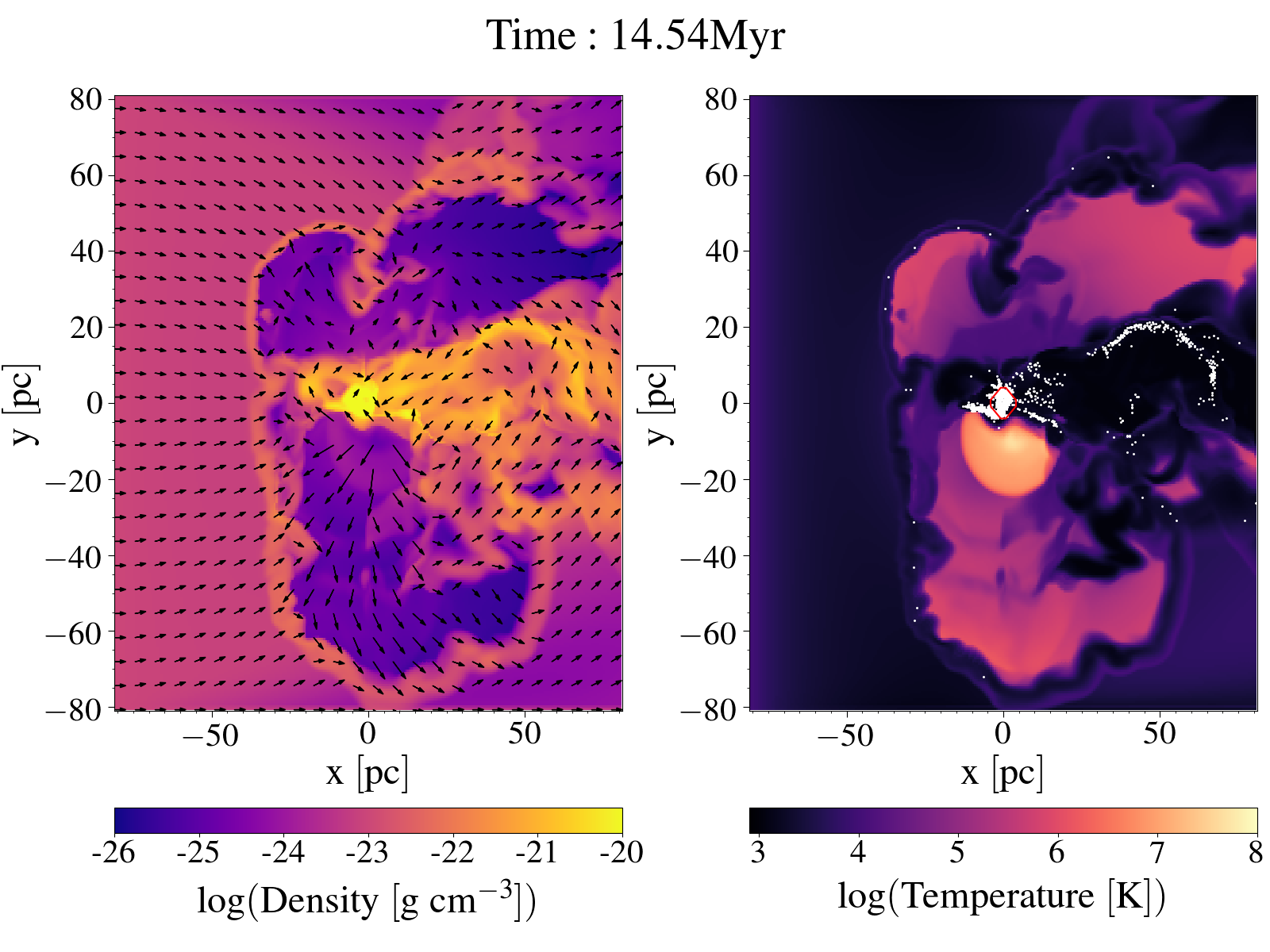}
        \\
        \vspace{0.3cm}
        \includegraphics[width=0.493\textwidth]{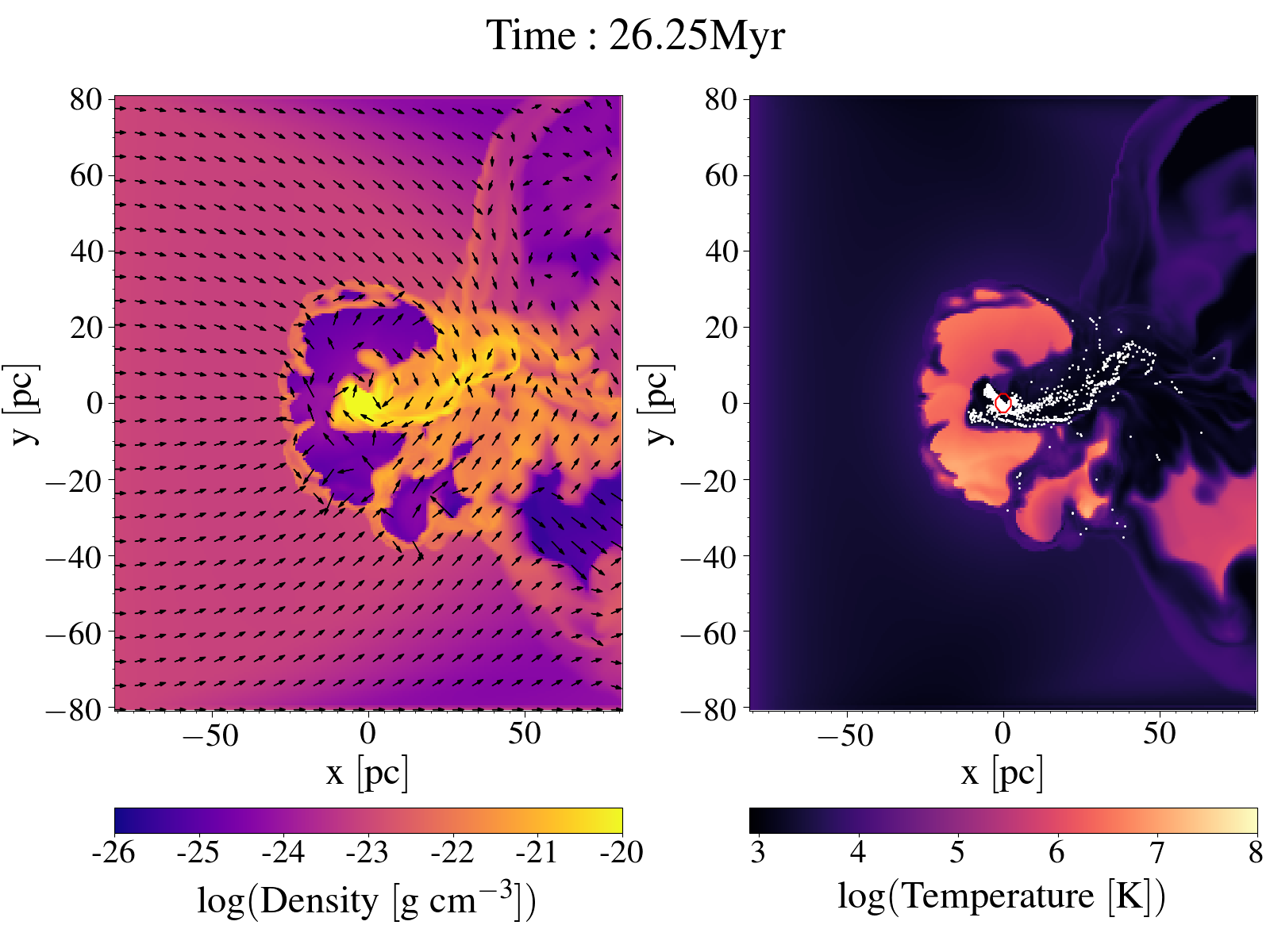}
        \hspace{0.1cm} 
        \includegraphics[width=0.493\textwidth]{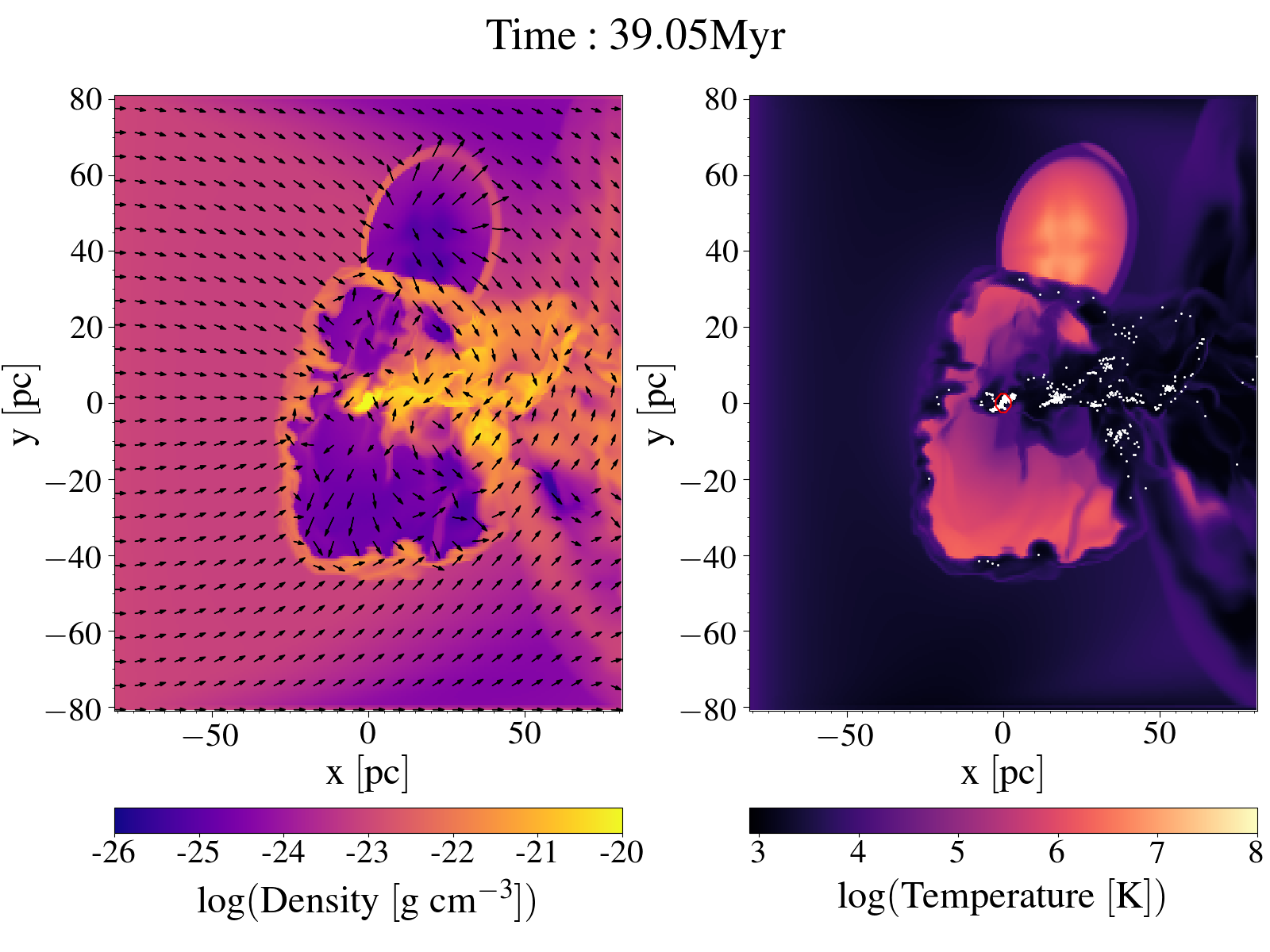}

\caption{Two-dimensional maps of the gas density on the left-hand panels of each plot and of the temperature on the right hand panels in the x-y plane at different evolutionary times for the high density (HD) simulation. From the top left to the bottom right: $t=6, 14, 26, 48$ Myr. Other symbols and lines as in Figure \ref{fig:maps_lowND}.}
  \label{fig:maps_high_ND}
\end{figure*}

\begin{figure*}
        \centering
        
        \includegraphics[width=0.287\textwidth]{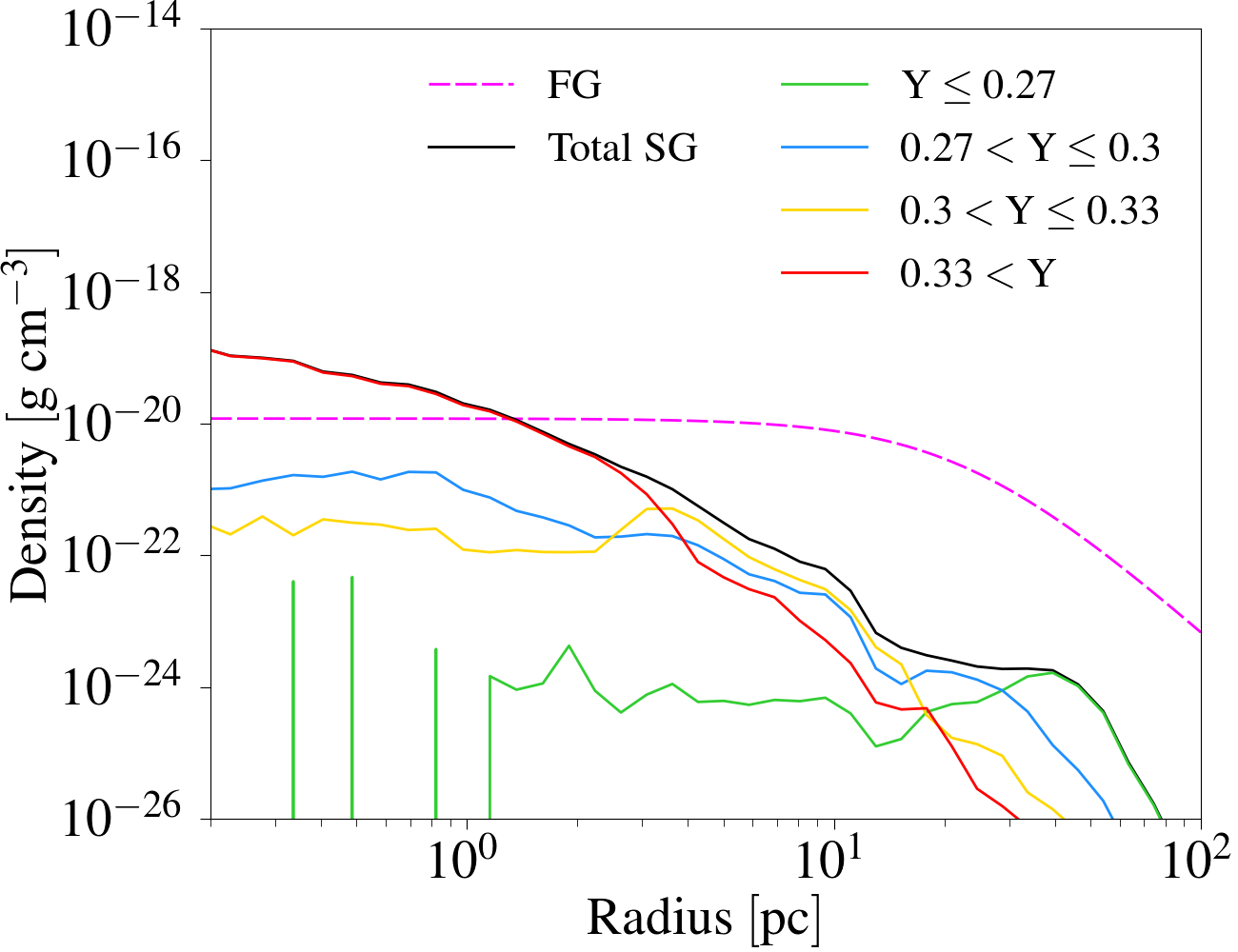}
        \includegraphics[width=0.192\textwidth]{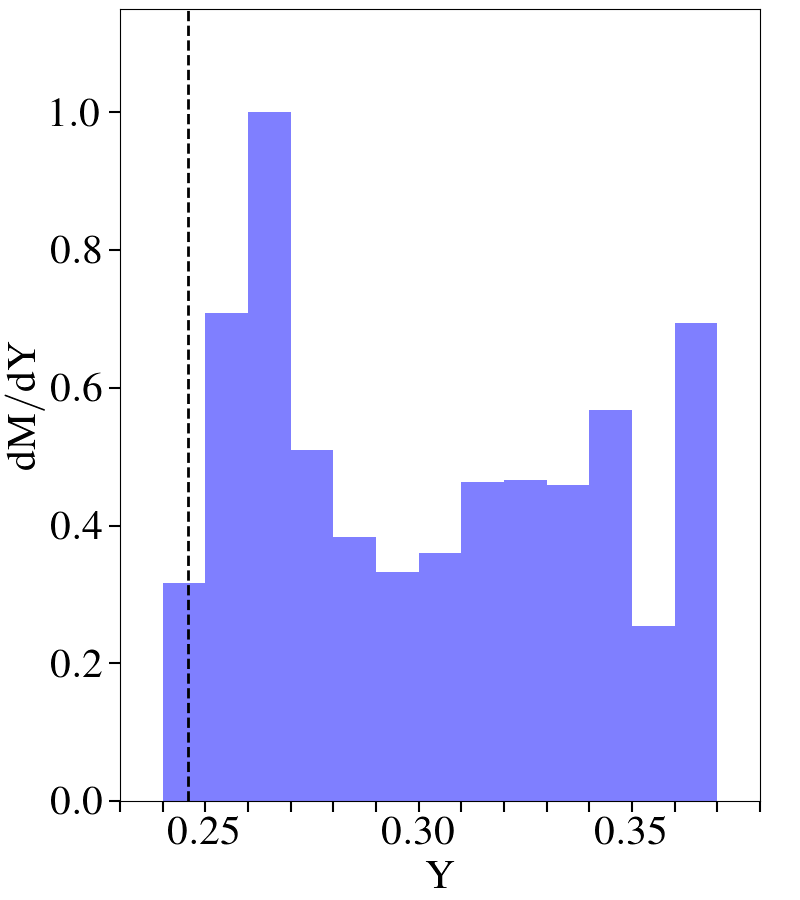}
        \hspace{0.08cm}
        \includegraphics[width=0.287\textwidth]{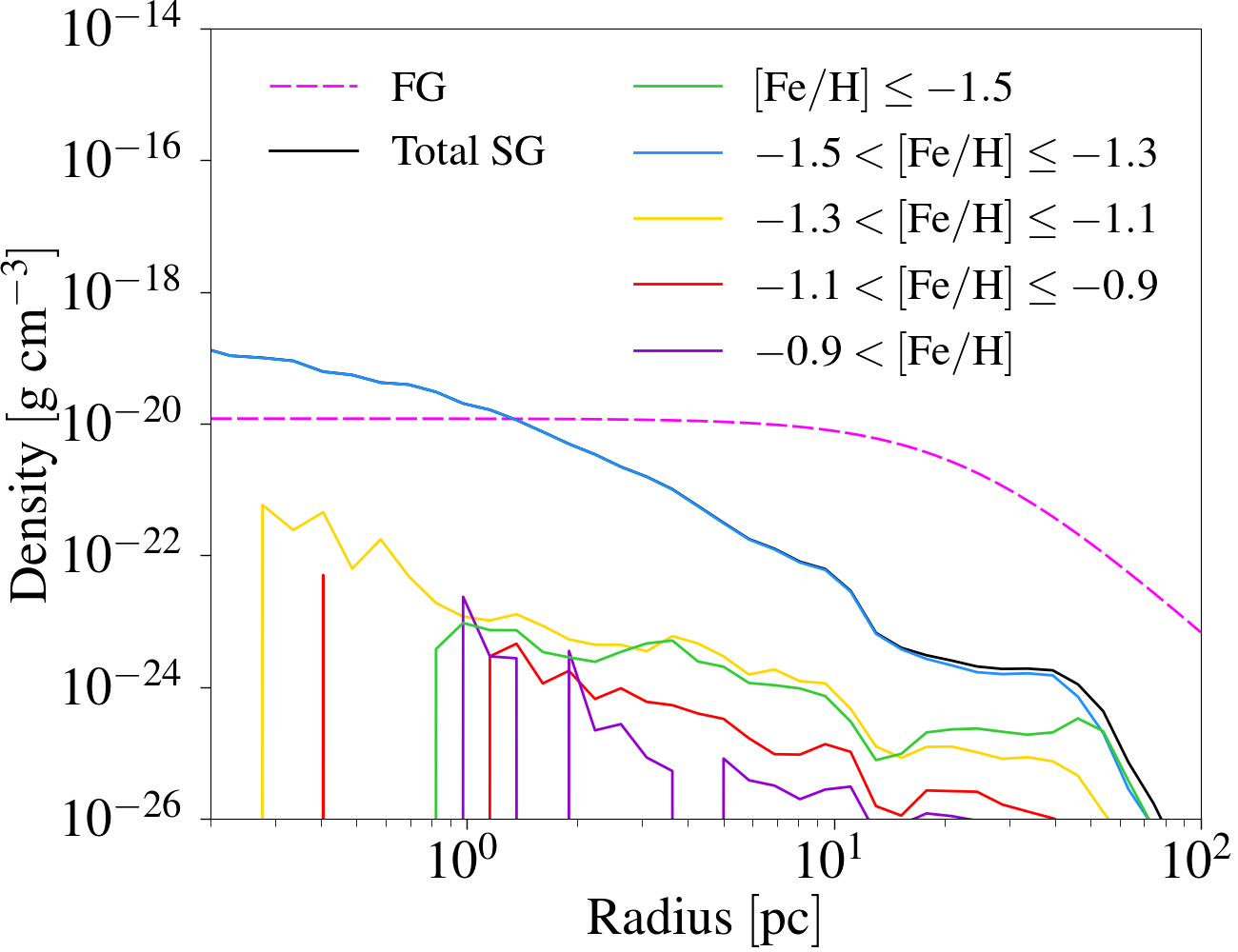}   \includegraphics[width=0.216\textwidth]{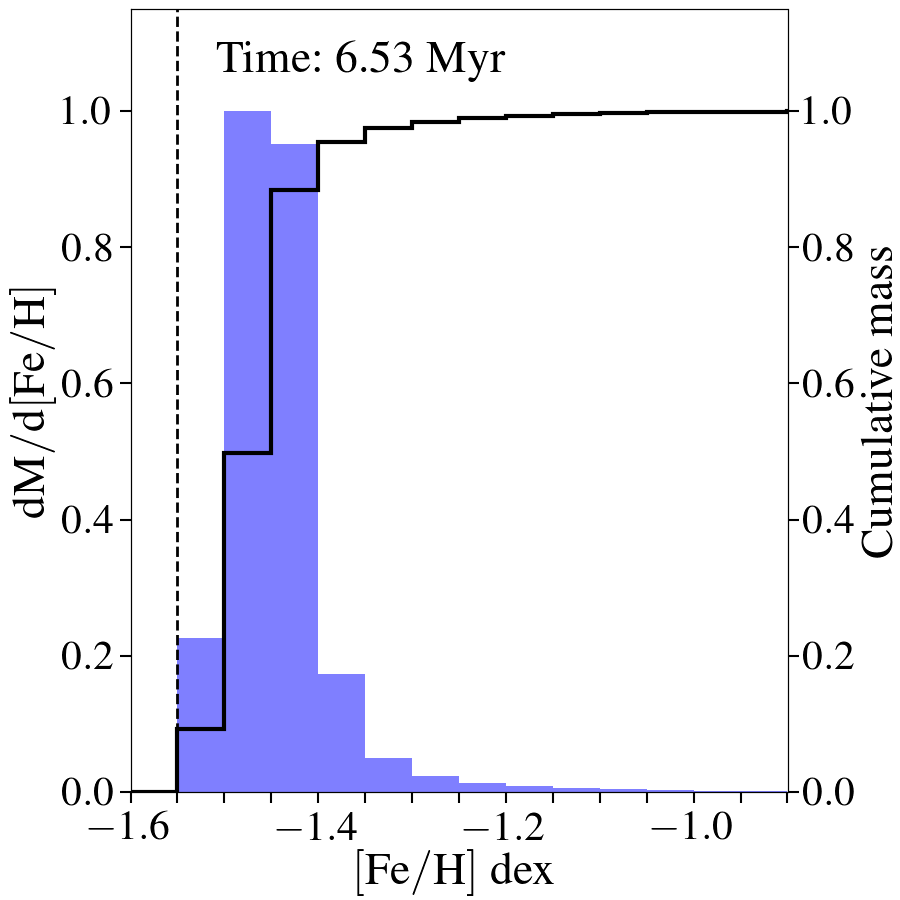}
        \\
        \includegraphics[width=0.287\textwidth]{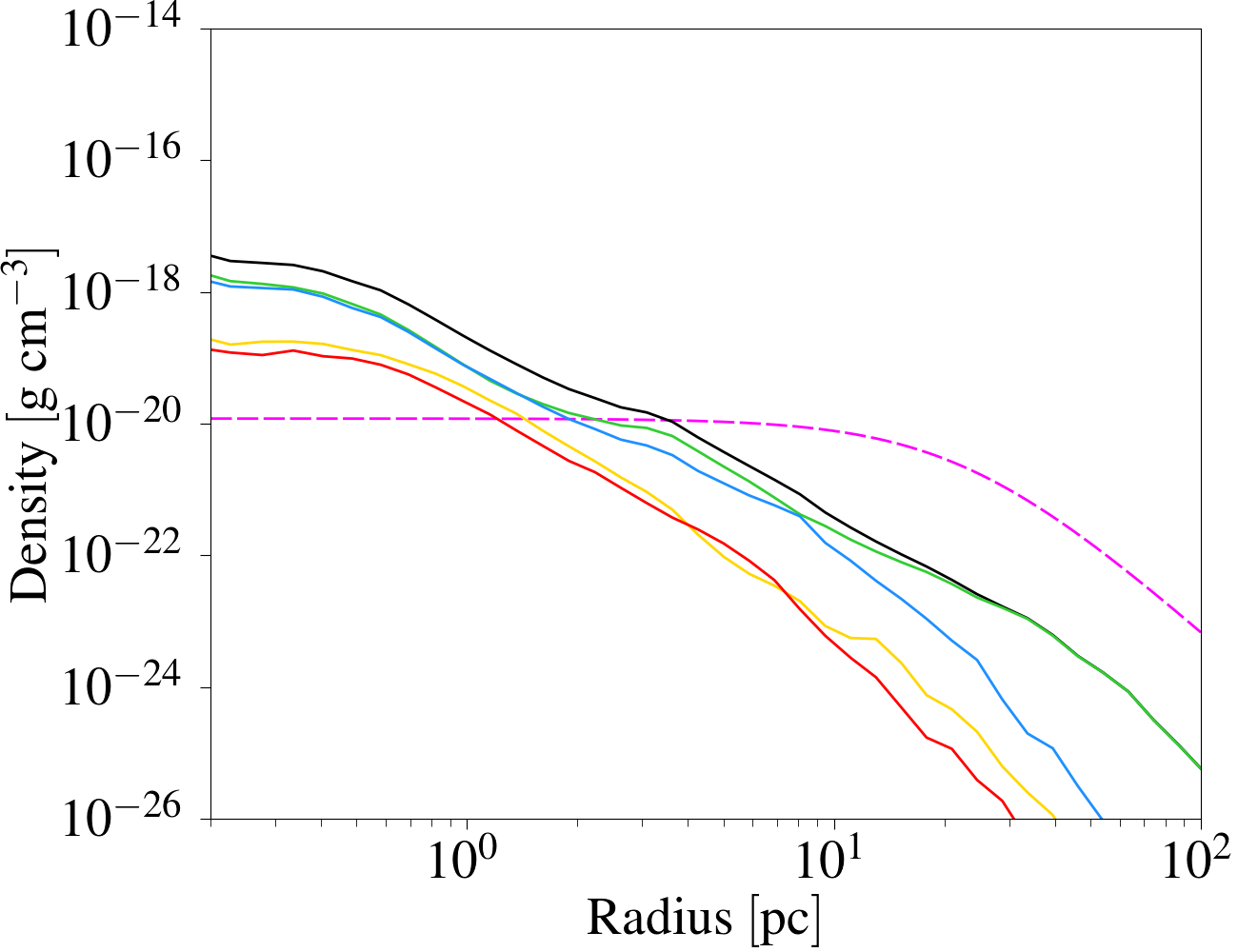}
        \includegraphics[width=0.192\textwidth]{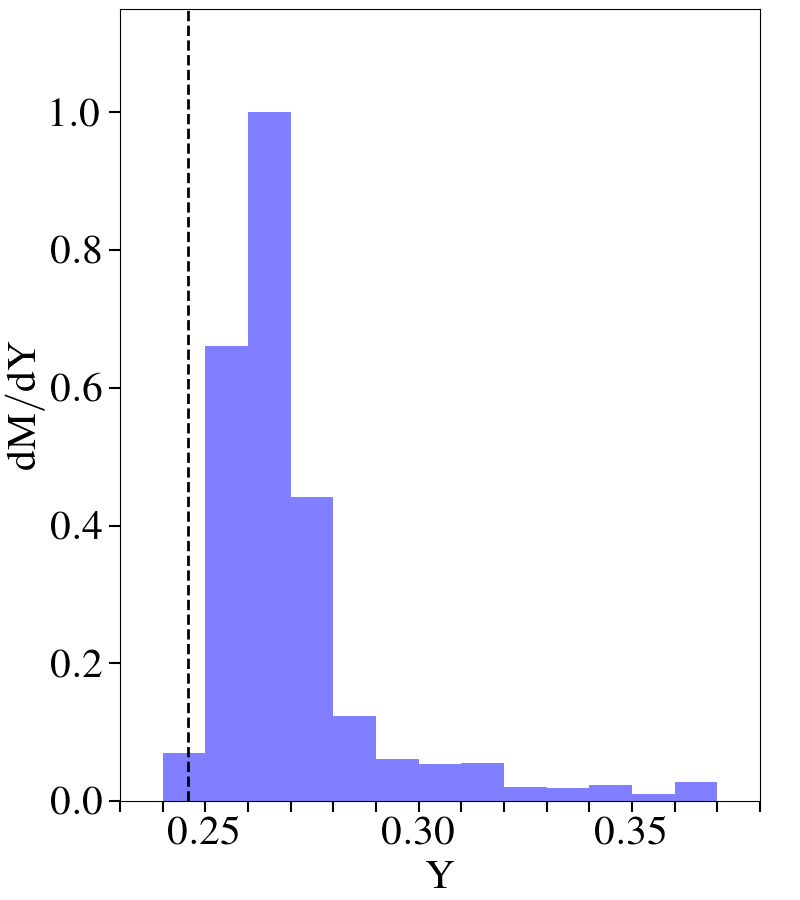}
        \hspace{0.08cm}
        \includegraphics[width=0.287\textwidth]{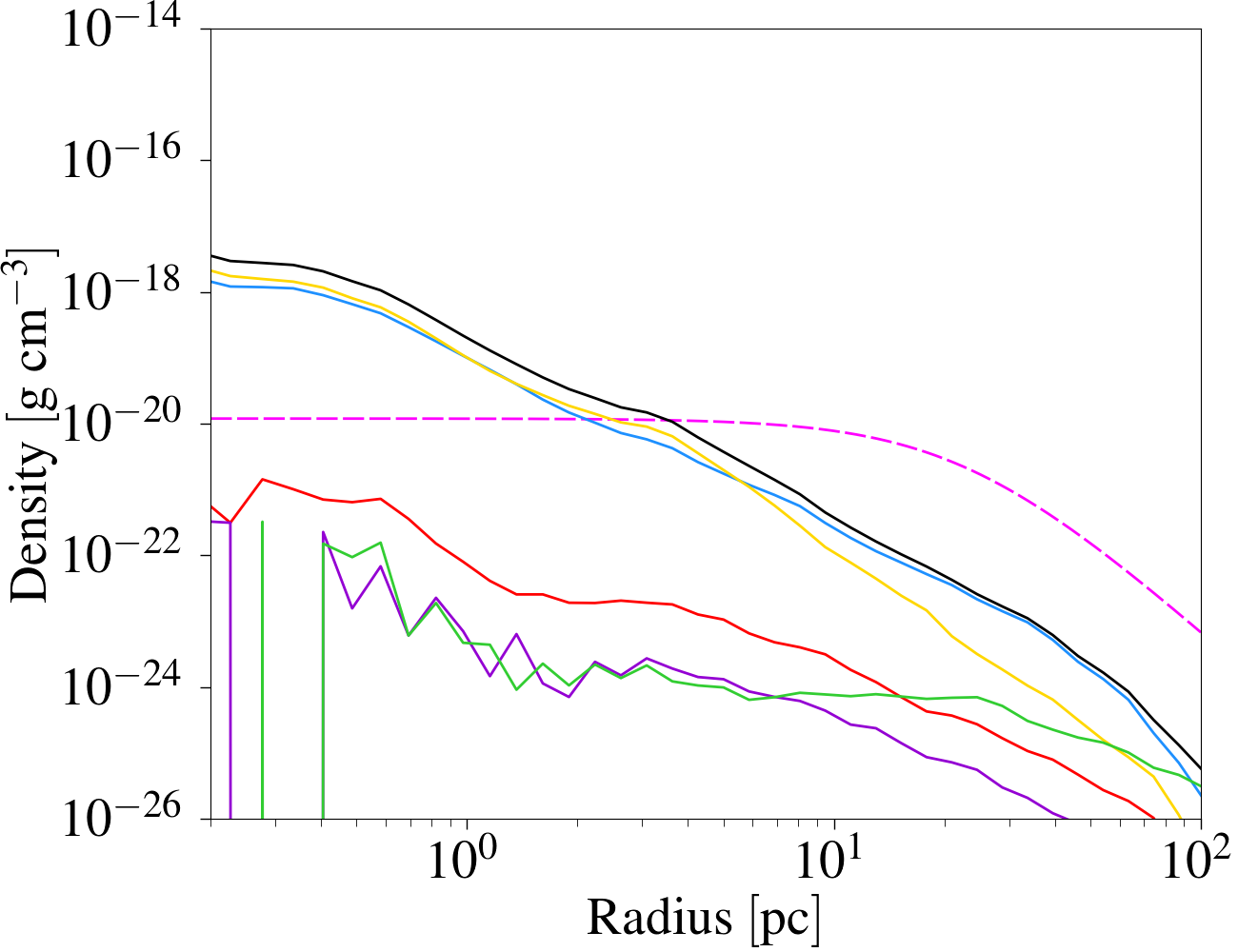}
        \includegraphics[width=0.216\textwidth]{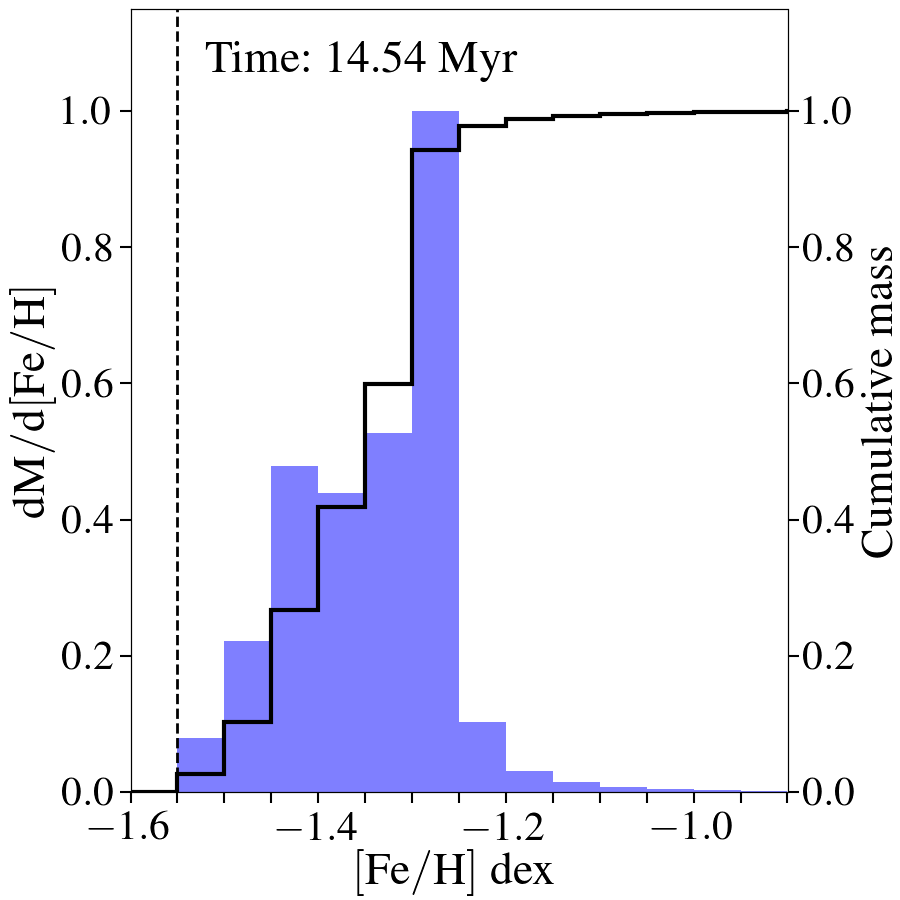}
        \\
        \includegraphics[width=0.287\textwidth]{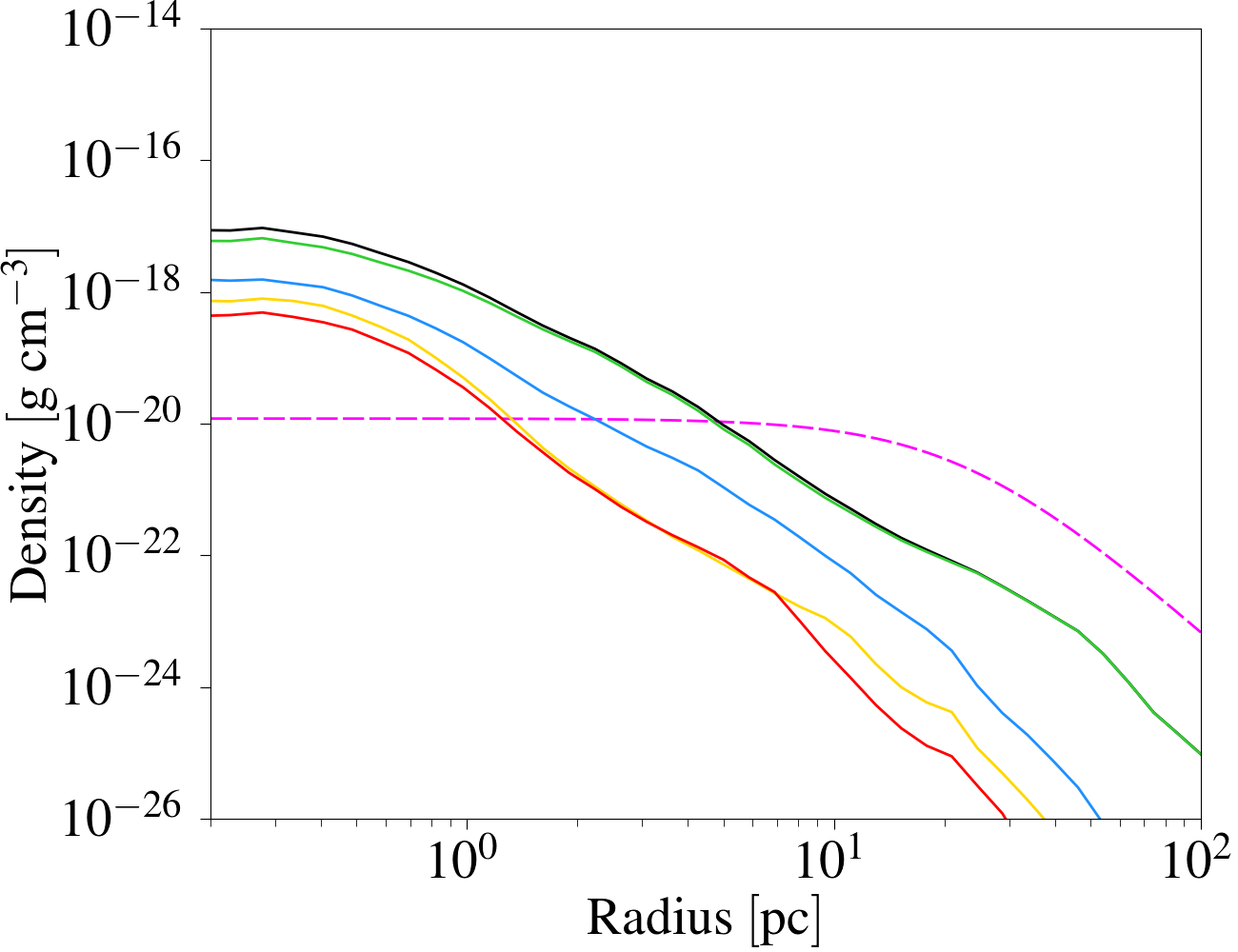}
        \includegraphics[width=0.192\textwidth]{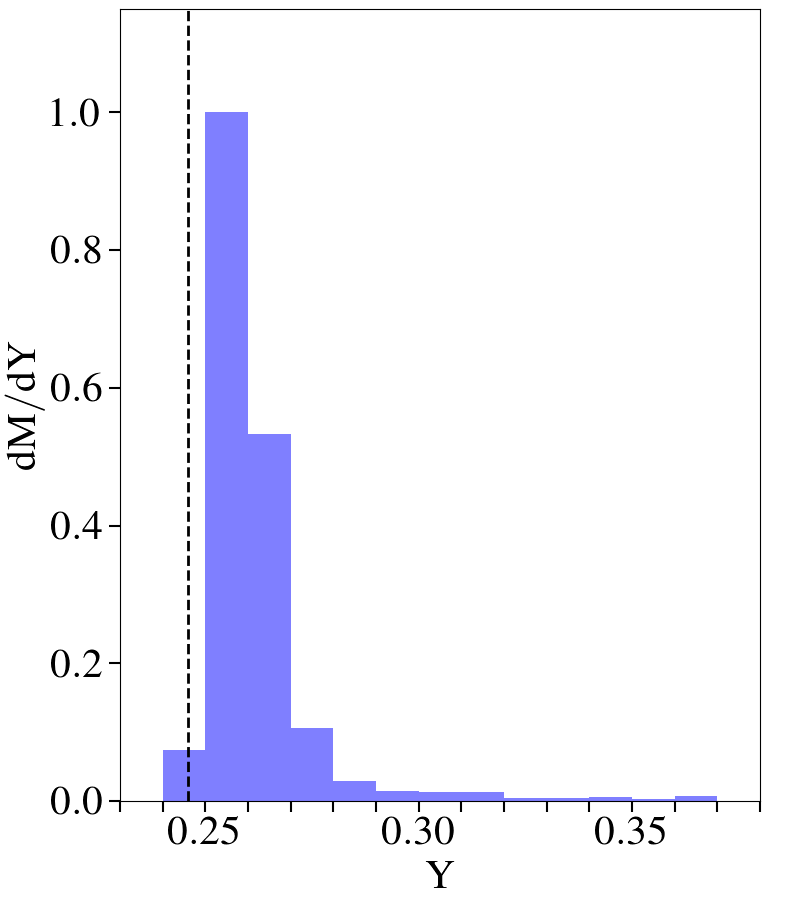}
        \hspace{0.08cm}
        \includegraphics[width=0.287\textwidth]{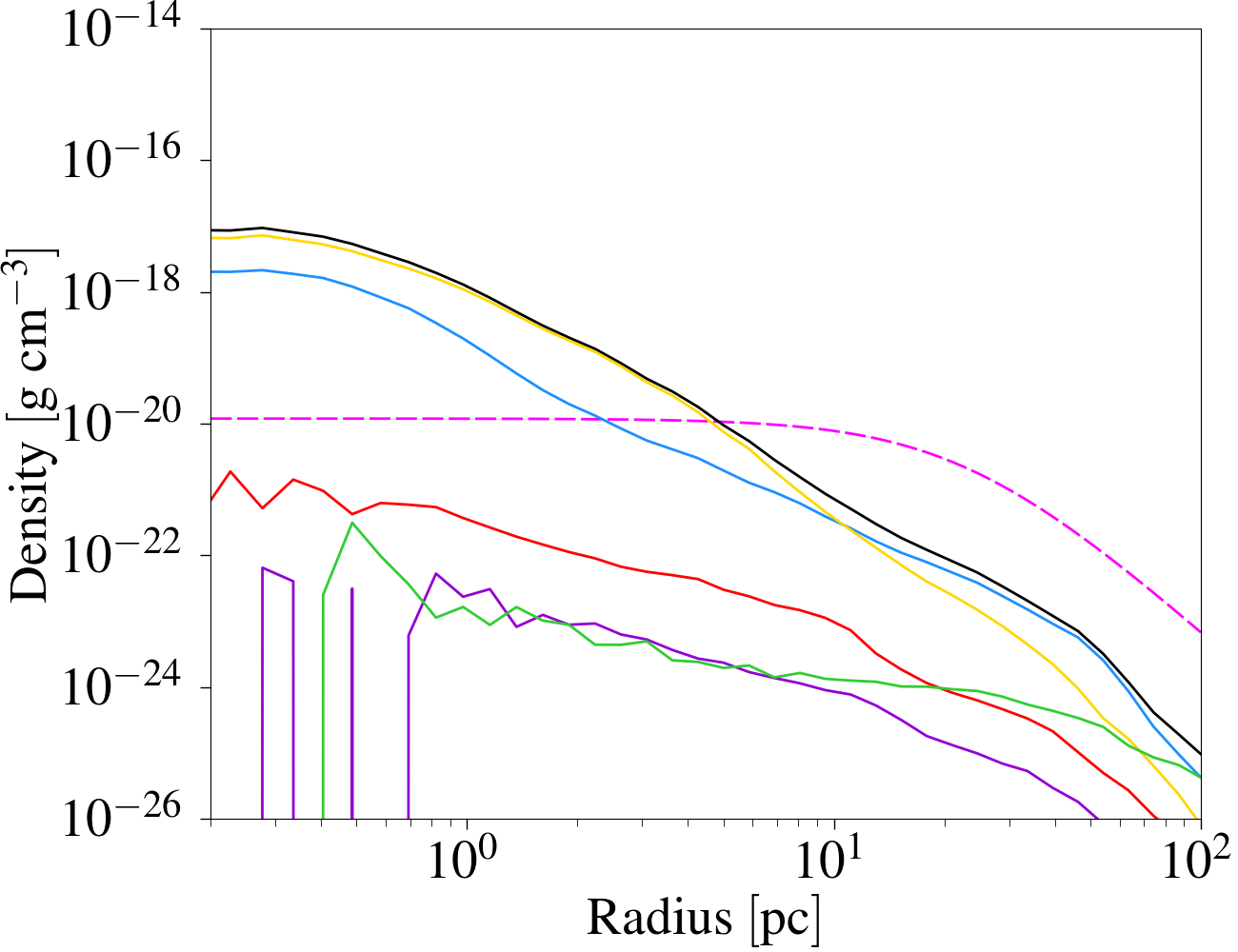}
        \includegraphics[width=0.216\textwidth]{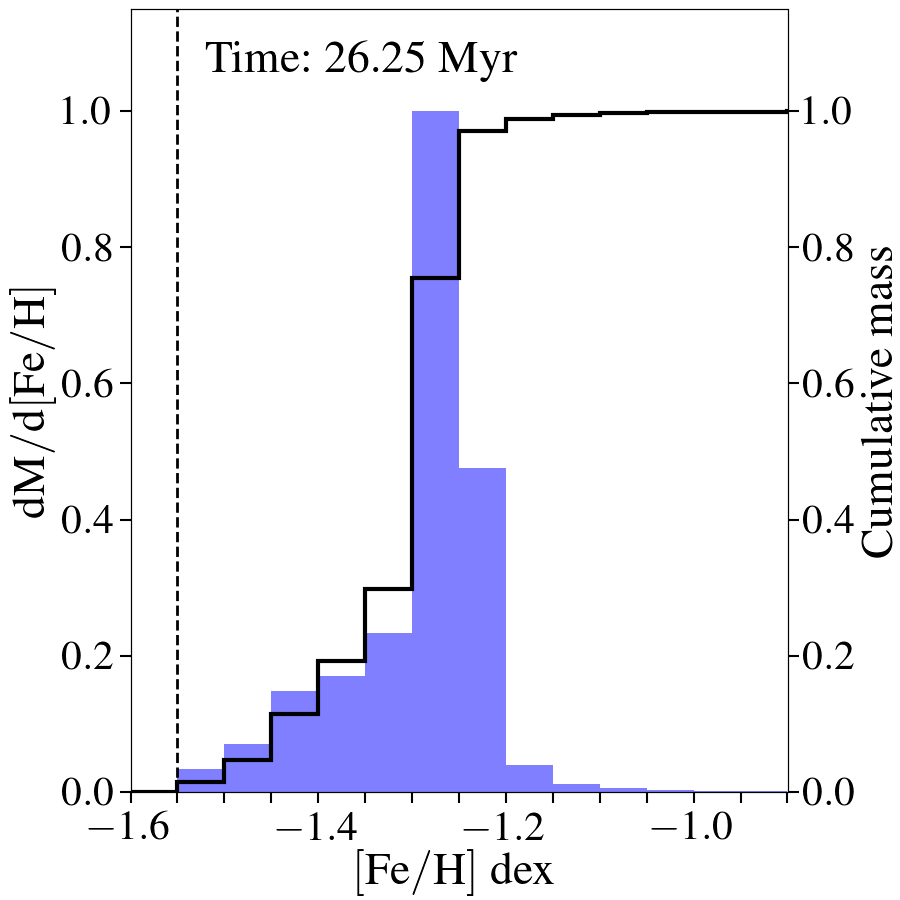}
        \\
        \includegraphics[width=0.287\textwidth]{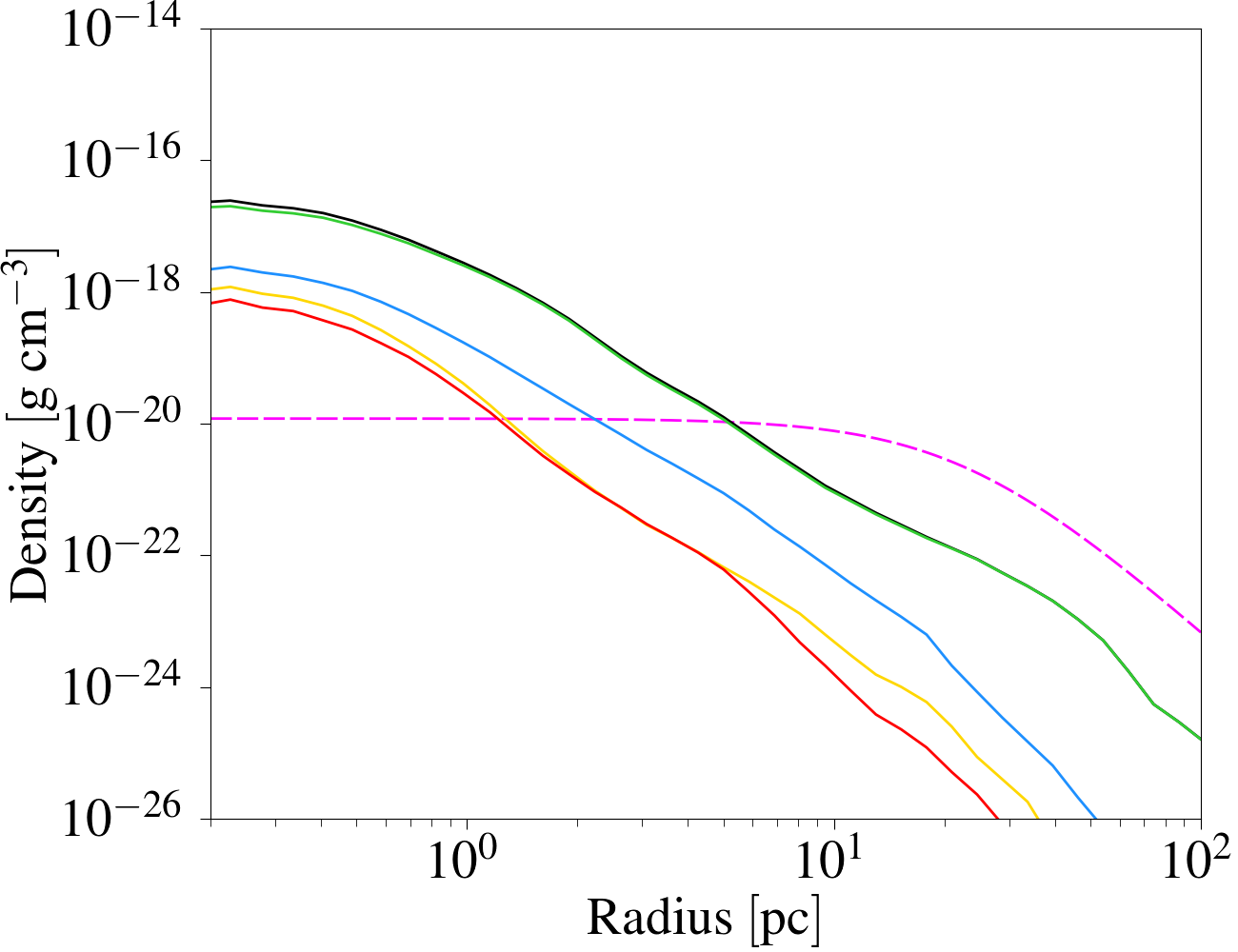}
        \includegraphics[width=0.192\textwidth]{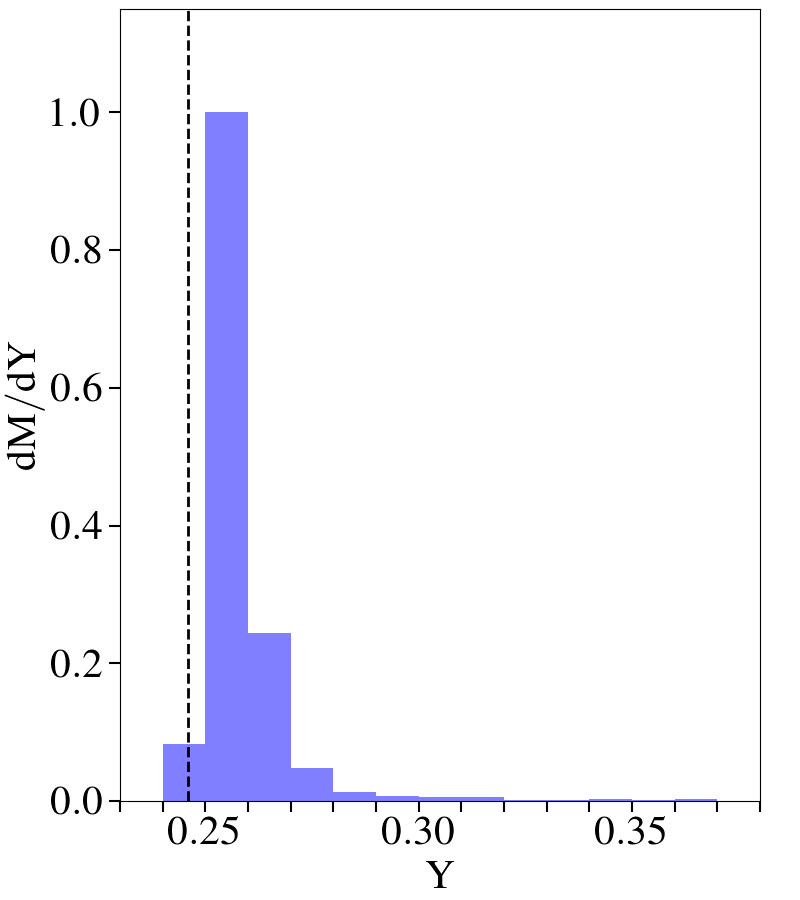}
        \hspace{0.08cm}
        \includegraphics[width=0.287\textwidth]{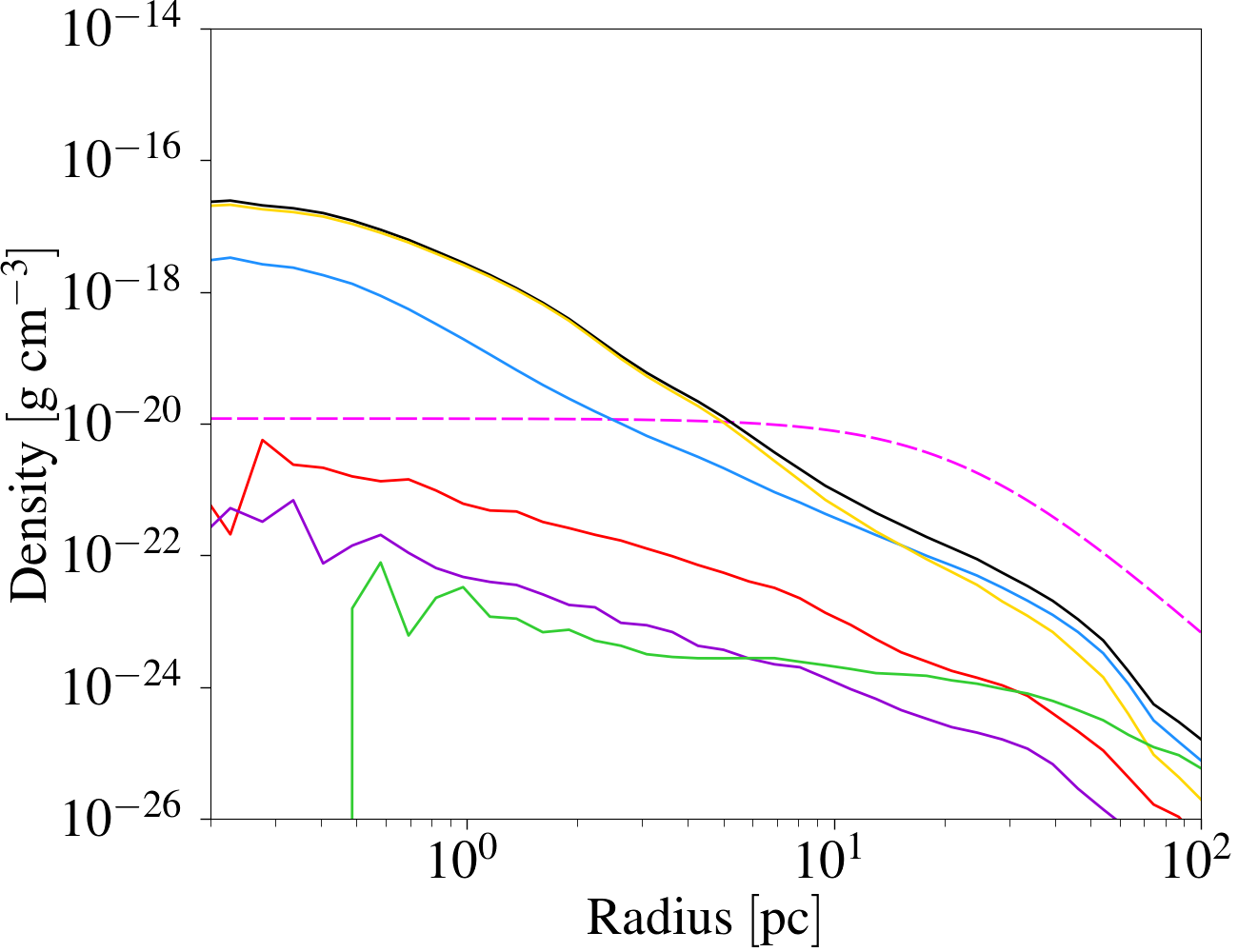}
        \includegraphics[width=0.216\textwidth]{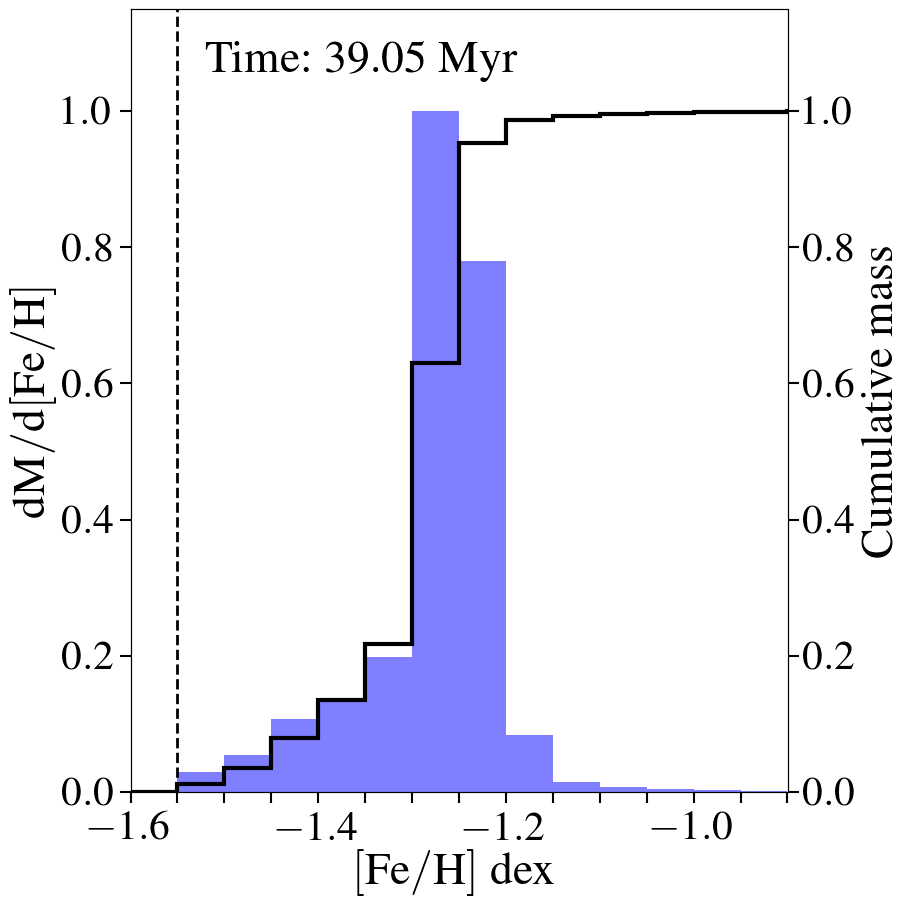}

 \caption{First and third columns: total density profile of SG stars at $t=6, 14, 26, 39$ Myr and density profiles of SG stars for several ranges of the helium mass fraction Y and the [Fe/H] ratio, respectively, for the high-density model (HD). The FG density profile is also plotted (see the legend for the details). Second and forth columns: the mass distribution of Y and [Fe/H] ratio, respectively, in the SG stars at the aforementioned evolutionary times (reported in each panel). Other symbols and lines as in Figure \ref{fig:den&mdf_lowND}.}   
   \label{fig:den&mdf_highND}
\end{figure*}

\subsection{High-density model}
\subsubsection{Dynamical evolution of the gas}
In our high density simulations we have assumed that the density of the pristine gas is $\rho_{\rm pg}=10^{-23} \mathrm{g\ cm^{-3}}$, a value 10 times greater than that adopted in the lower density model. Through Equation \ref{eq:infall}, we have derived that, for this value of the pristine gas density, the infall starts around $40$ Myr after the FG formation which corresponds to $\sim 1$ Myr after the starting point of our simulations. 
In Figure \ref{fig:maps_high_ND}, we show the density and temperature maps on the x-y plane computed at the same evolutionary times as \citetalias{calura2019}, with the only exception of the last map. The times associated to the four snapshots are $t=6$ Myr, $t=14$ Myr, $t=26$ Myr and $t=39$ Myr.

A comparison of these maps with the ones obtained for the LD case, shown in Figure \ref{fig:maps_lowND}, shows that, in the HD model, the infalling pristine gas is able to penetrate deeper into the system, limiting the expansion of the SN bubbles. The pristine gas is surrounding the system, at variance with the low density case in which it remains confined at the border of the computational box. In addition, a significantly larger number of stars are formed, not only in the very central part of the cluster, but also along the accretion column, absent in the LD model, downstream of the system.

At $6$ Myr, the pristine gas has already crossed the system, however, the cold, dense and narrow tail formed in \citetalias{calura2019} simulation is not clearly seen here because it is constantly disrupted by the continuous Type Ia SN explosions. Dense regions can be found also around the shell of the SN bubbles, where the swept up gas collides with the pristine gas. Here, stars can be formed far from the cluster centre, however, most of the stars are formed in the central region and along the perturbed tail.

At later times, at $14$ Myr, the overlap of various bubbles has created an extended region of hot and rarefied gas surrounding the dense, cold stream visible along the x-axis. A large number of stars are formed in the central part of the system, but also in the dense region corresponding to the tail location, where a flow of dense material is pointing towards the centre of the cluster. The infalling pristine gas is acting against the bubbles expansion, in particular pushing them rightwards. The stellar component has expanded with respect to the previous snapshot, especially downstream of the system.

At $26$ Myr, hot bubbles are present as a result of recent SN explosions near the centre of the system. Nevertheless the bubbles are not expanding through the entire computational box as it happens in the low density case (Figure \ref{fig:maps_lowND}, top right panel). The bubbles are more confined and their shape is far from being spherical. This is the result of the ram-pressure of the pristine gas which is highly effective in the HD model. The velocity field reveals also in this case the presence of a flow of gas extending in a tail. A large number of stars are formed in this region of cold and dense gas, even though the bulk of stars are formed in the proximity of the centre of the system.

At $39$ Myr, the system has not changed much; cold and dense gas is forming new stars both in the central regions and along the accretion column which is still present at this time. The gas heated up by SNe explosions is still confined by the pristine gas.

\subsubsection{Evolution of the stellar component}

In Figure \ref{fig:den&mdf_highND} we plot the density profiles of the SG for various ranges of Y and [Fe/H] ratio at different evolutionary times for the high-density model, together with the helium and [Fe/H] ratio mass distribution functions. 

At $6$ Myr, the shock caused by the infall of pristine gas has already crossed the system and stars are formed from a mixed gas. In the external region stars with no helium enrichment, and therefore formed mainly out of pristine gas, are dominant, while in the central region the AGB ejecta have been less diluted and stars display extreme helium abundances. However, the component not enriched in helium is dominant in mass as it can be seen from the Y mass distribution where three peaks at low, intermediate and extreme helium enrichment can be clearly seen. On the other hand, the iron mass distribution is peaked around [Fe/H] $\sim -1.45$ with a long tail towards higher values, which however contributes almost negligibly to the total stellar mass. 
The density of the SG is higher than the FG one in the central region but it is lower than the case without Type Ia SNe studied by \citetalias{calura2019} (${\rm HD\_C19}$).

At $14$ Myr, stars with almost no helium enhancement (with respect to FG stars) have become dominant at all radii, while those with high helium abundances give the lowest contribution, as illustrated by both the density profiles and the Y distribution. Therefore, the pristine gas comprises a large fraction of the gas out of which SG form, at variance with the LD case, but in agreement with ${\rm HD\_C19}$. 
Conversely, the [Fe/H] ratio of the bulk of the stars is increased: this means that, even though the pristine gas is significantly diluting the AGB ejecta, iron from Type Ia SNe is retained by the system and recycled to form new stars.
Most of the stars show a significant iron enrichment, in particular in the central region, while poorly iron enriched stars represent a very small fraction of the total stellar mass.

Later, at $26$ Myr, the SG is dominated at all radii by stars with modest helium enrichment. In this scenario, Type Ia SNe are not able to confine the infalling gas like in the low-density models, as it is visible from the two-dimensional maps.
The [Fe/H] mass distribution becomes narrower and peaks around $-1.25$ dex, with stars falling in it being the dominant component in the central part of the system. The density in such region is similar to what has been obtained by \citetalias{calura2019}. With the exception of a large number of Fe-enriched stars formed in our simulation, our results are similar to ${\rm HD\_C19}$, meaning that SN effects in a dense medium are significantly reduced with respect to the low-density case.

At $39$ Myr, the density in the centre of the cluster is slightly increased and the peaks both in the Y and [Fe/H] distributions are sharper than at 26 Myr. The bulk of the stars are mildly enriched in helium but remarkably enriched in iron, in particular in the central region.

The average stellar [Fe/H] ratio of SG stars at the end of the simulation is $-1.28$ dex, slightly larger than the LD value, with a dispersion of ${\rm \sigma^{\rm SG}_{[Fe/H]}=0.08 }$ dex. As for the previous two models, we have estimated the internal iron dispersion of the whole cluster assuming that $70\%$ of stars belong to the SG. We derive ${\rm \sigma_{[Fe/H]}= 0.14 }$ dex, which agrees with the typical spread found in Type II clusters \citep{milone2017,johnson2015}. As for the helium abundance, its spectroscopic determination  is much more challenging than the iron one, therefore, most of the studies rely on photometric determination of the relative helium abundance between multiple populations (\citealt{milone2015,milone2018,martins2021} and references therein). We derive the average mass fraction $\overline{Y}_{\rm SG}=0.258$ for the SG stars, which leads to a spread between second and first generation of ${ \Delta Y_{ \rm SG-FG}}=0.012$ in very good agreement with observations. Some of the Type II GCs analyzed by \citet{milone2018}, in fact, display small helium spreads between the two generations, in some cases even smaller than what we have derived for the HD model.




\begin{figure*}
        \centering

        \includegraphics[width=0.287\textwidth]{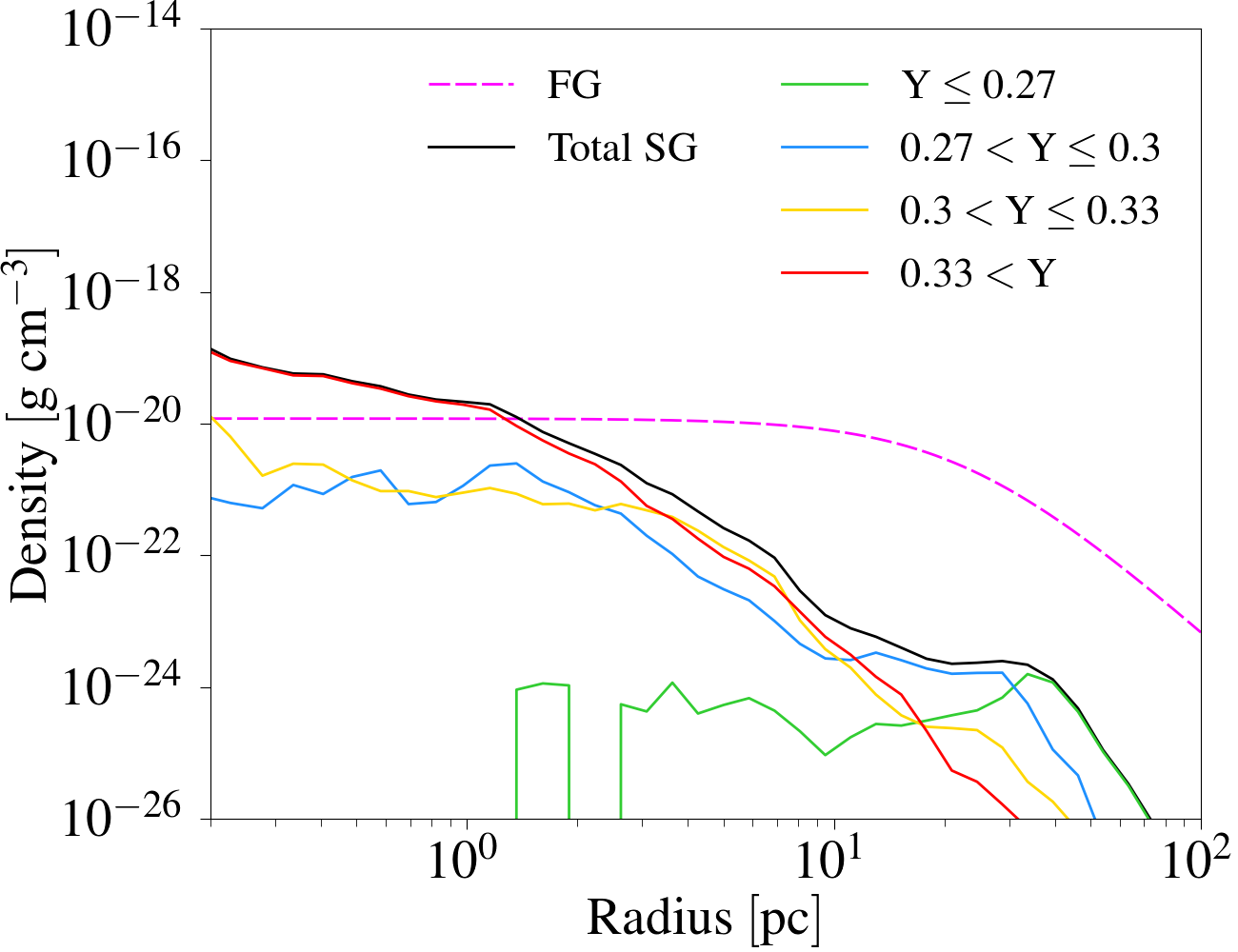}
        \includegraphics[width=0.192\textwidth]{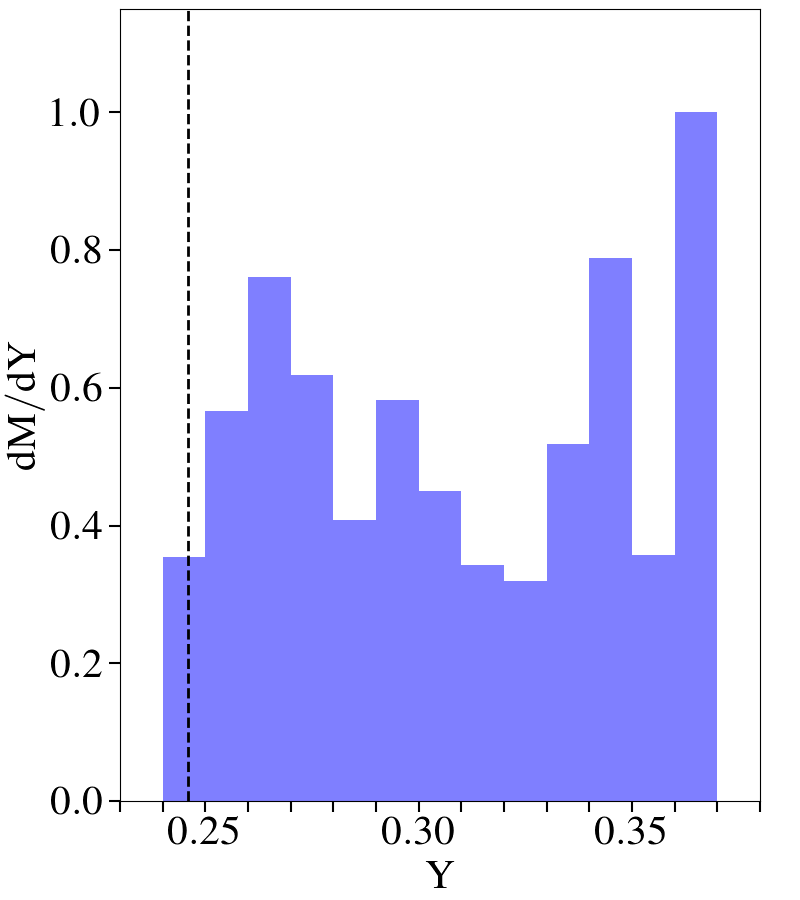}
        \hspace{0.08cm}
        \includegraphics[width=0.287\textwidth]{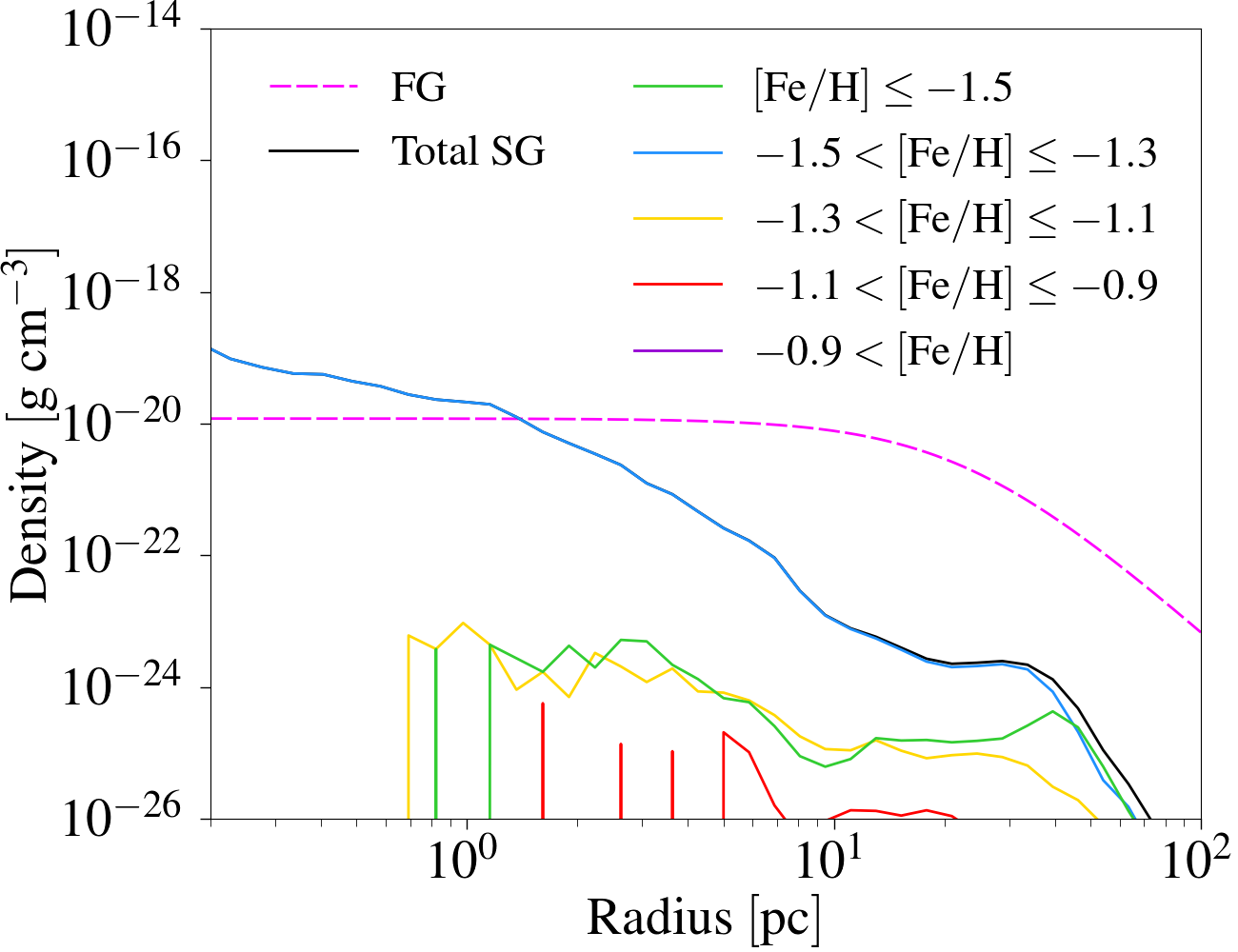}   \includegraphics[width=0.216\textwidth]{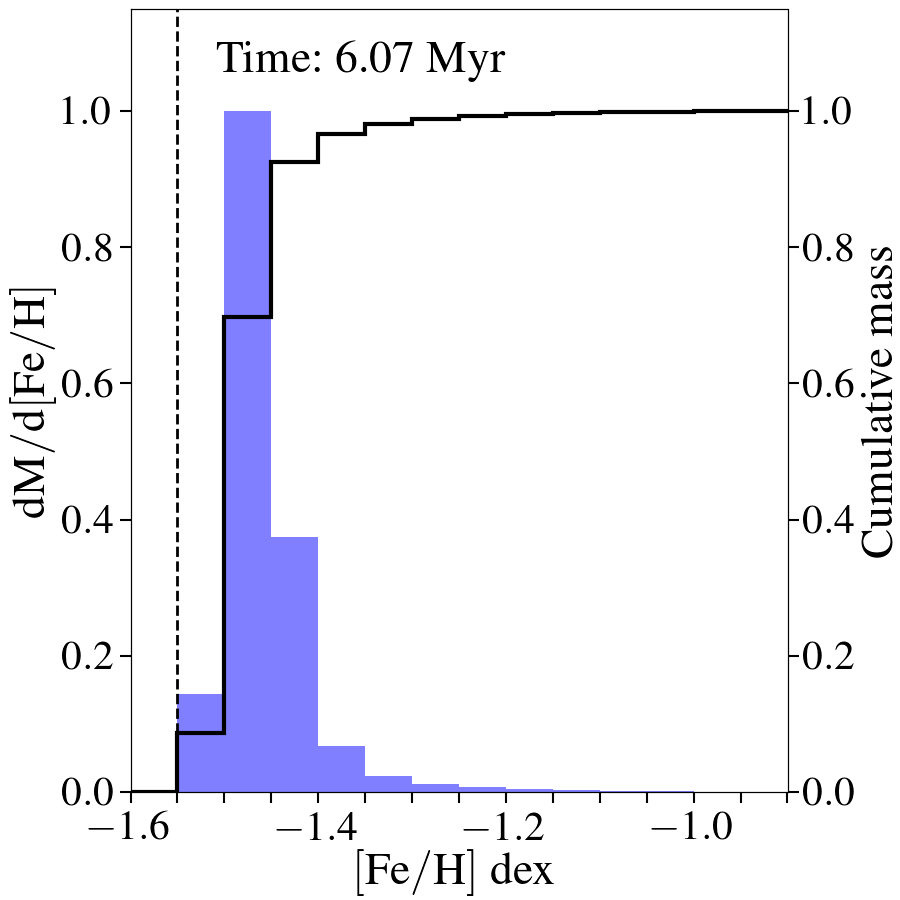}
        \\
        \includegraphics[width=0.287\textwidth]{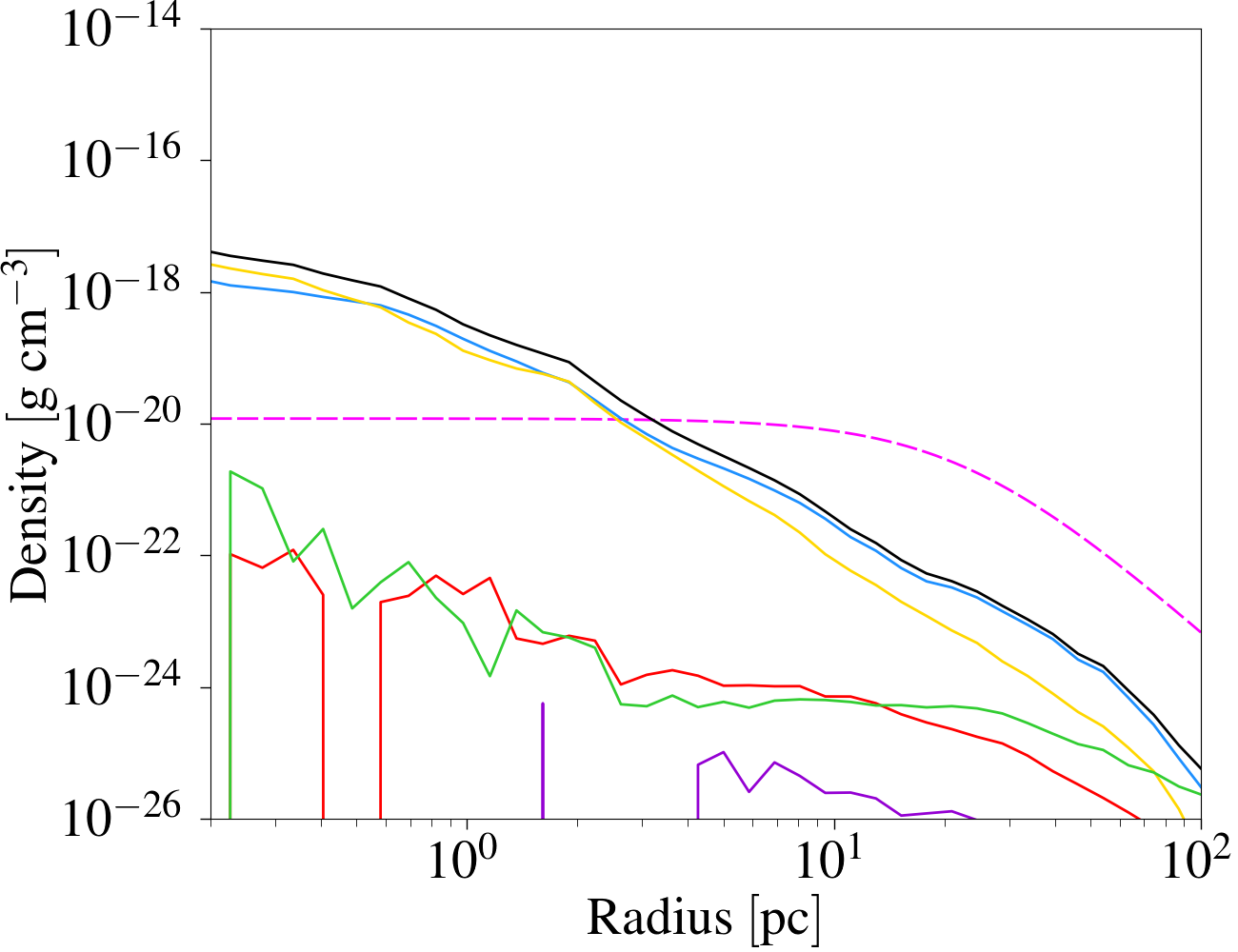}
        \includegraphics[width=0.192\textwidth]{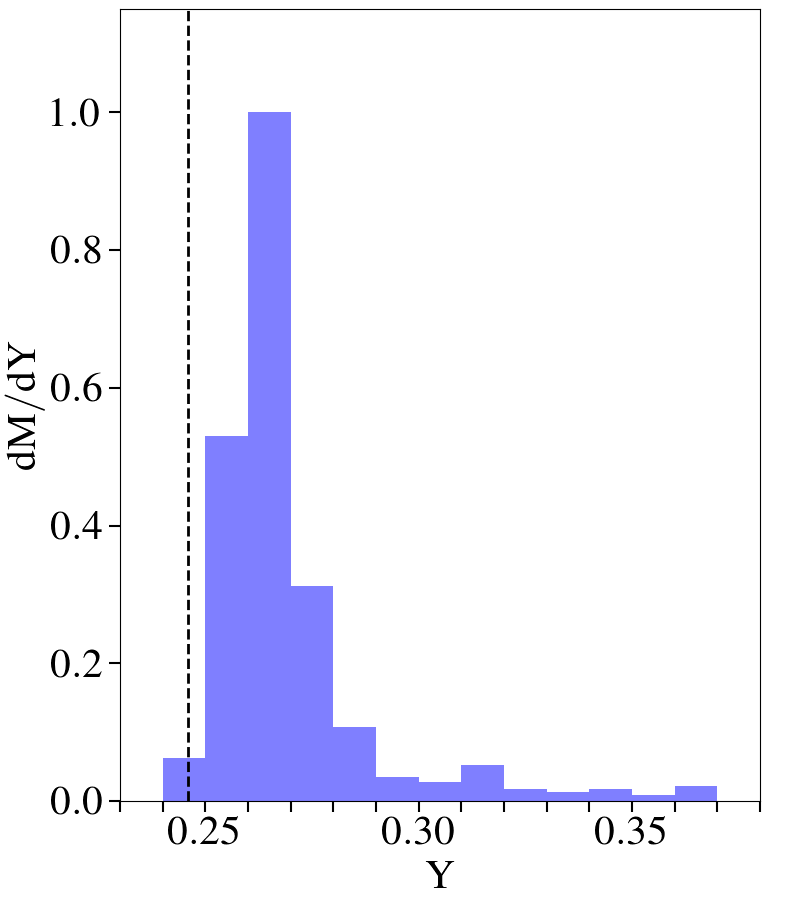}
        \hspace{0.08cm}
        \includegraphics[width=0.287\textwidth]{den_prof_fe00031.png}
        \includegraphics[width=0.216\textwidth]{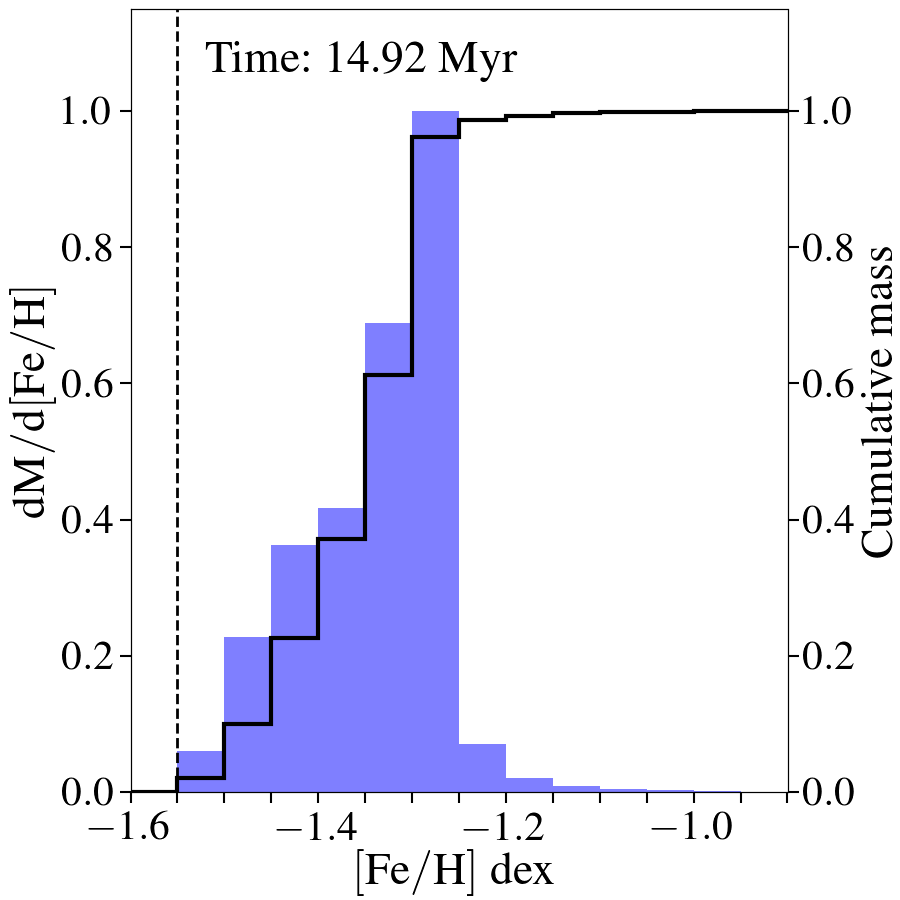}
        \\
        \includegraphics[width=0.287\textwidth]{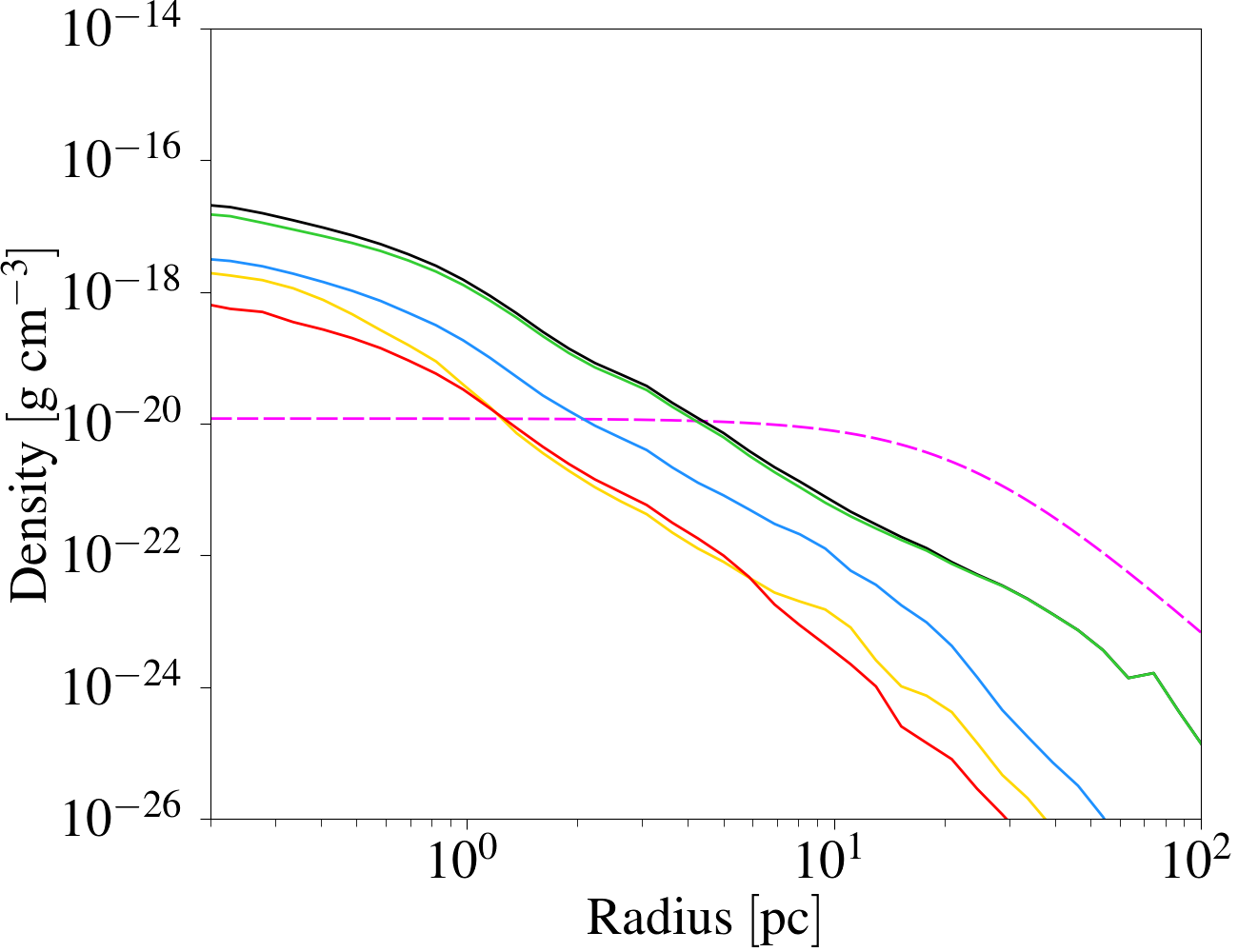}
        \includegraphics[width=0.192\textwidth]{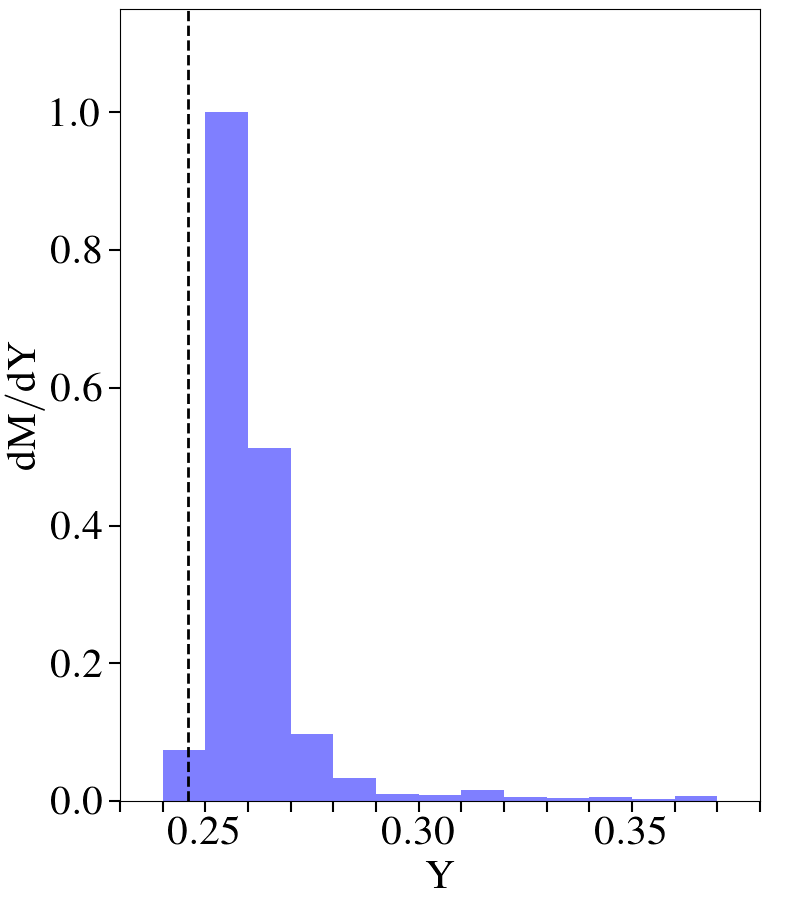}
        \hspace{0.08cm}
        \includegraphics[width=0.287\textwidth]{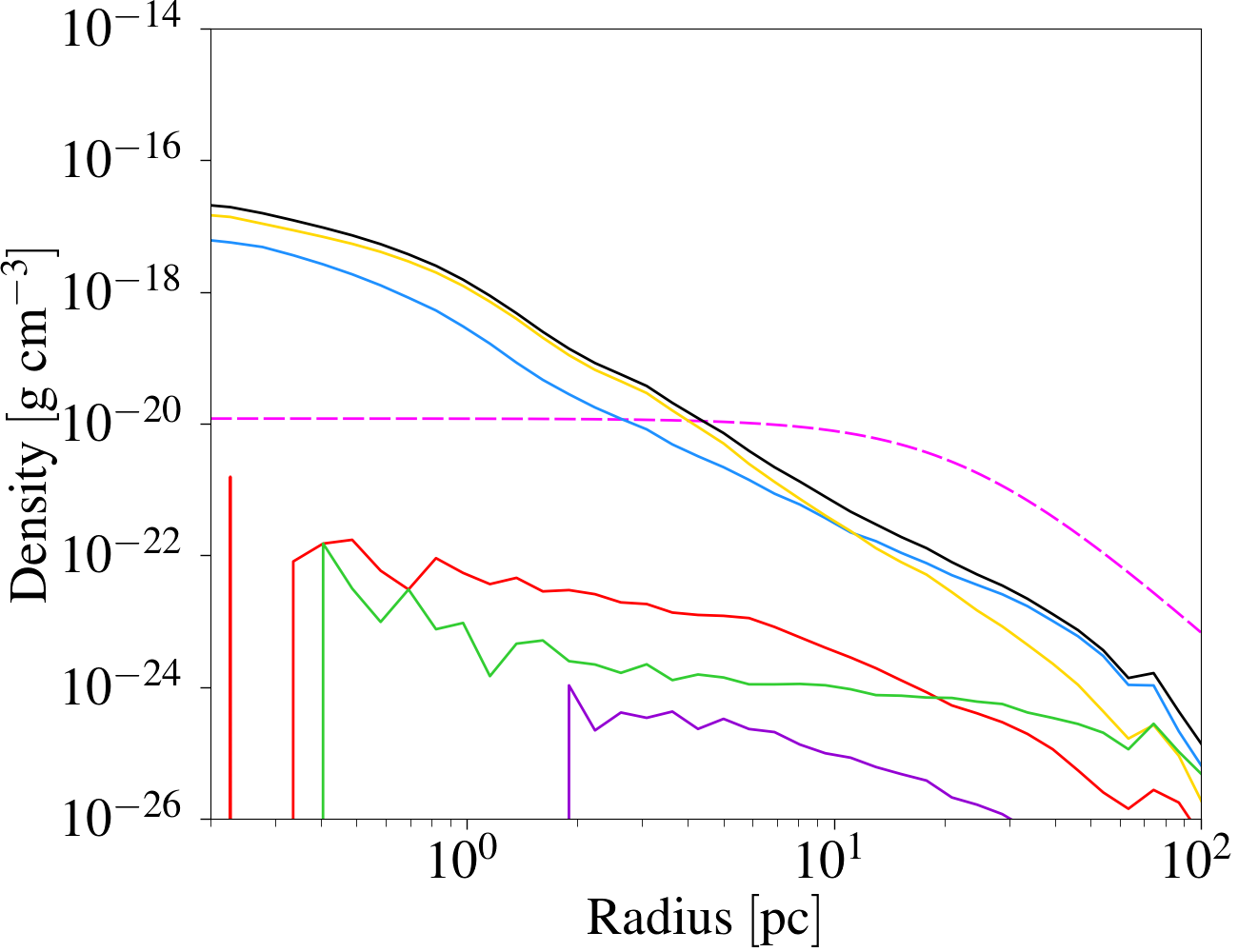}
        \includegraphics[width=0.216\textwidth]{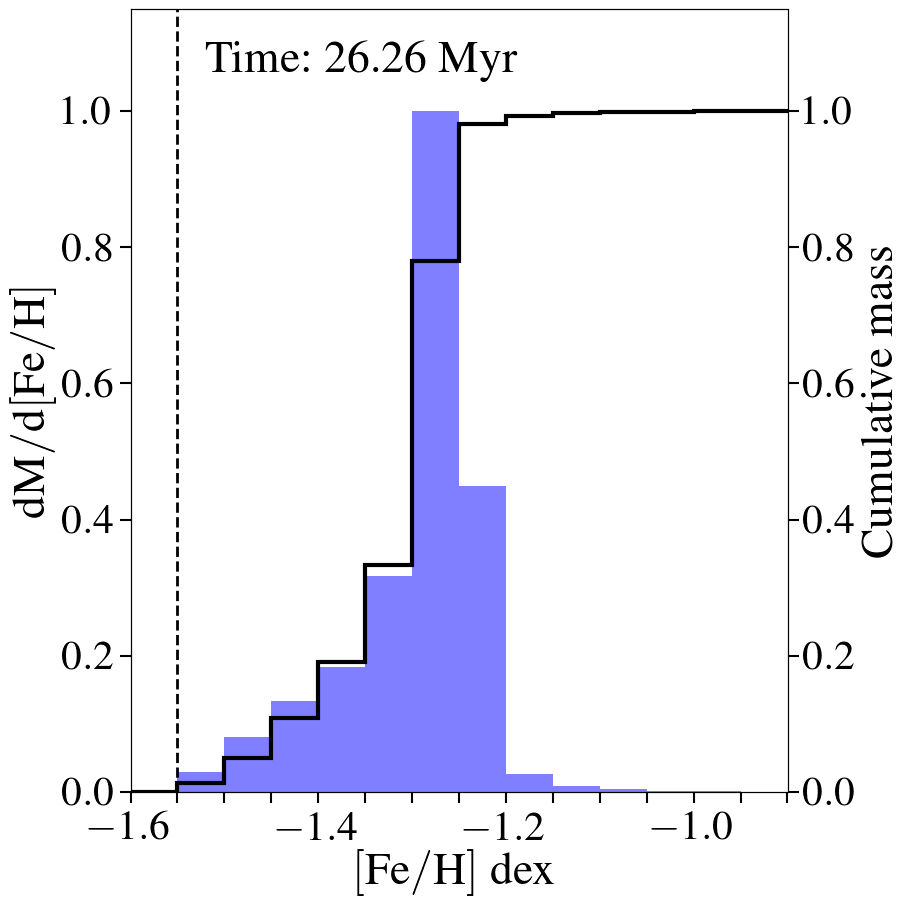}
        \\
        \includegraphics[width=0.287\textwidth]{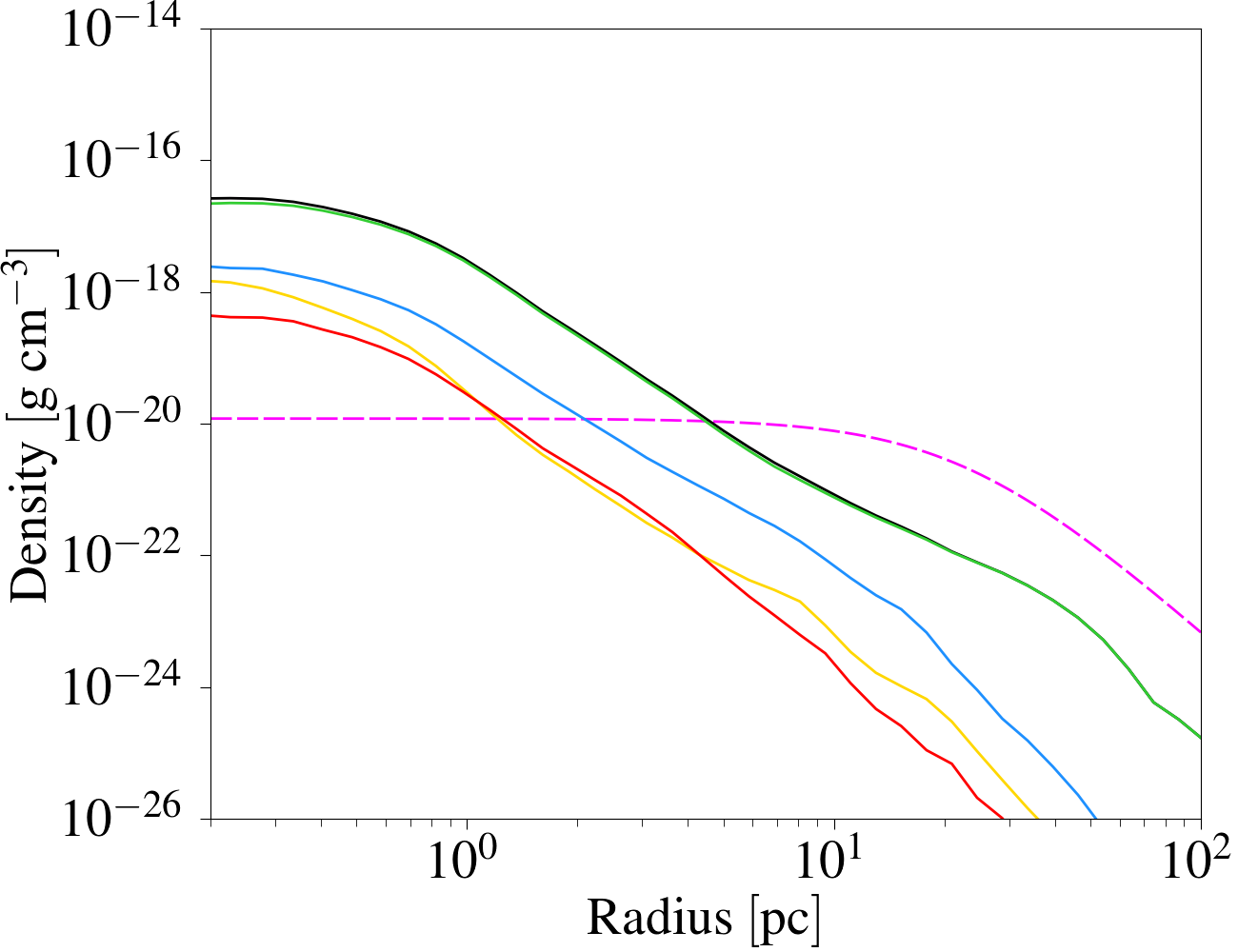}
        \includegraphics[width=0.192\textwidth]{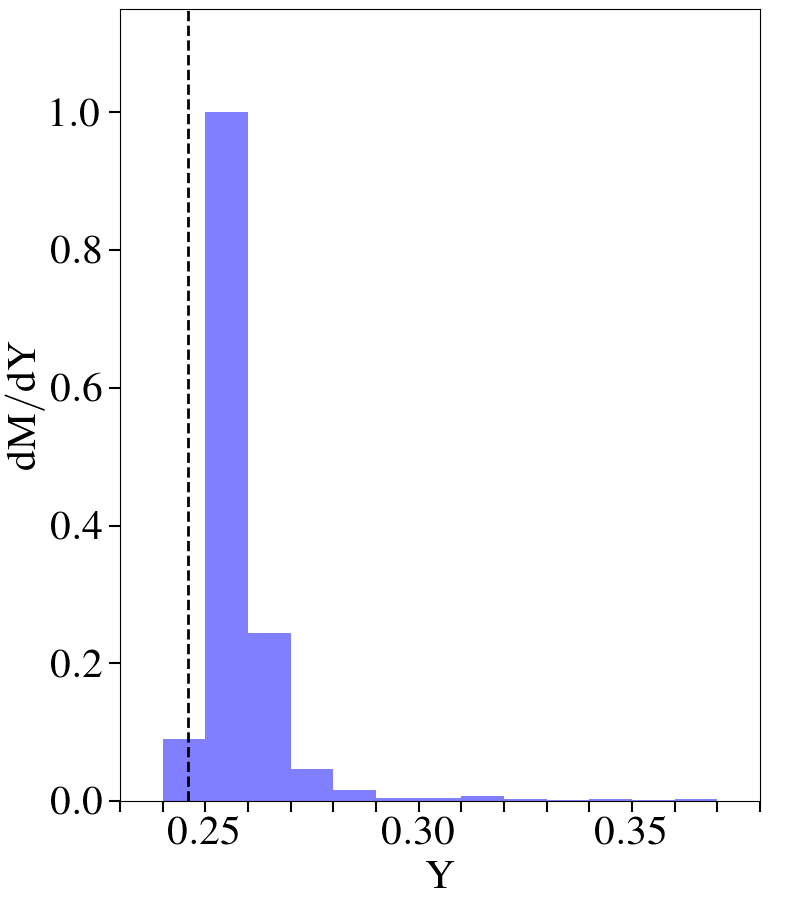}
        \hspace{0.08cm}
        \includegraphics[width=0.287\textwidth]{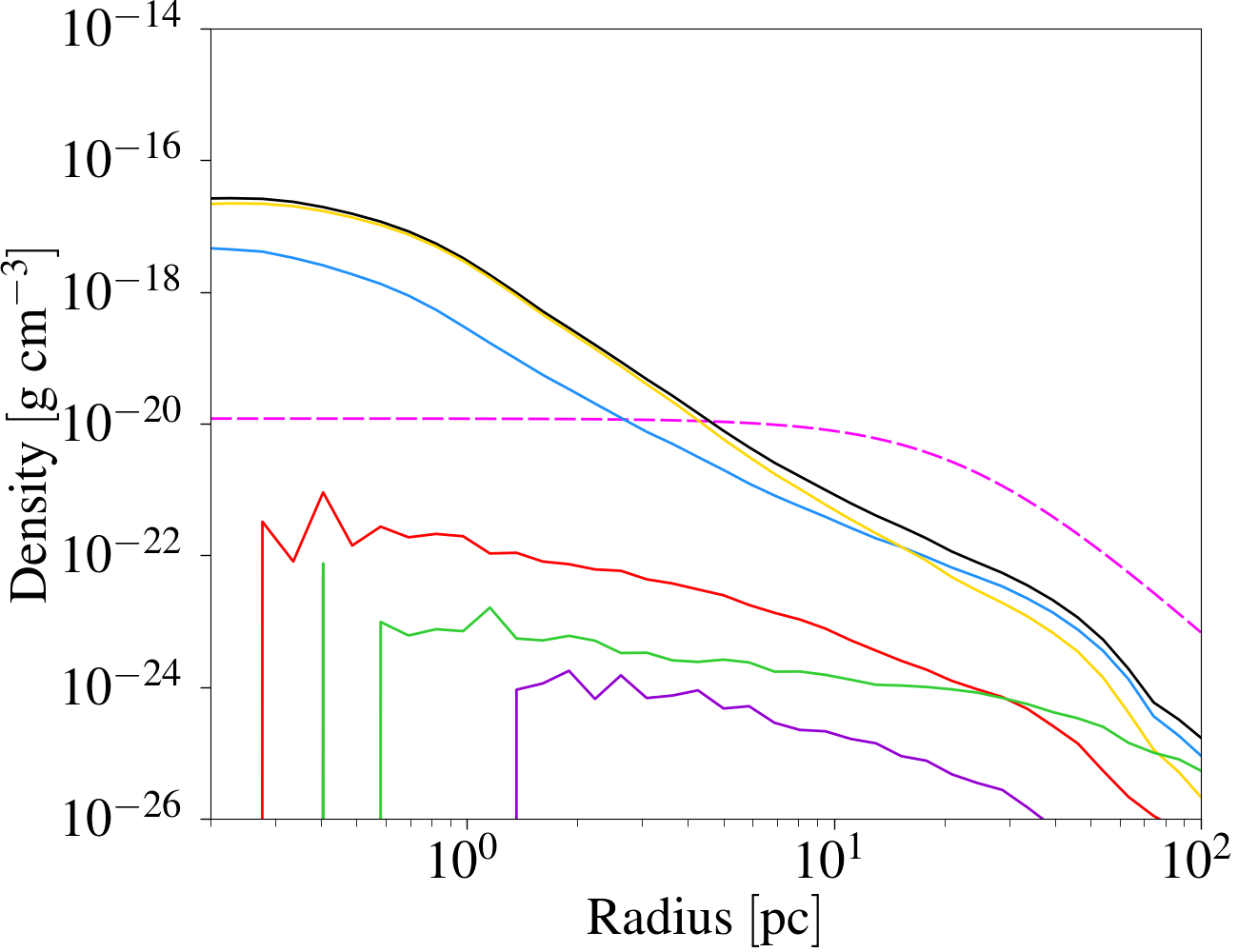}
        \includegraphics[width=0.216\textwidth]{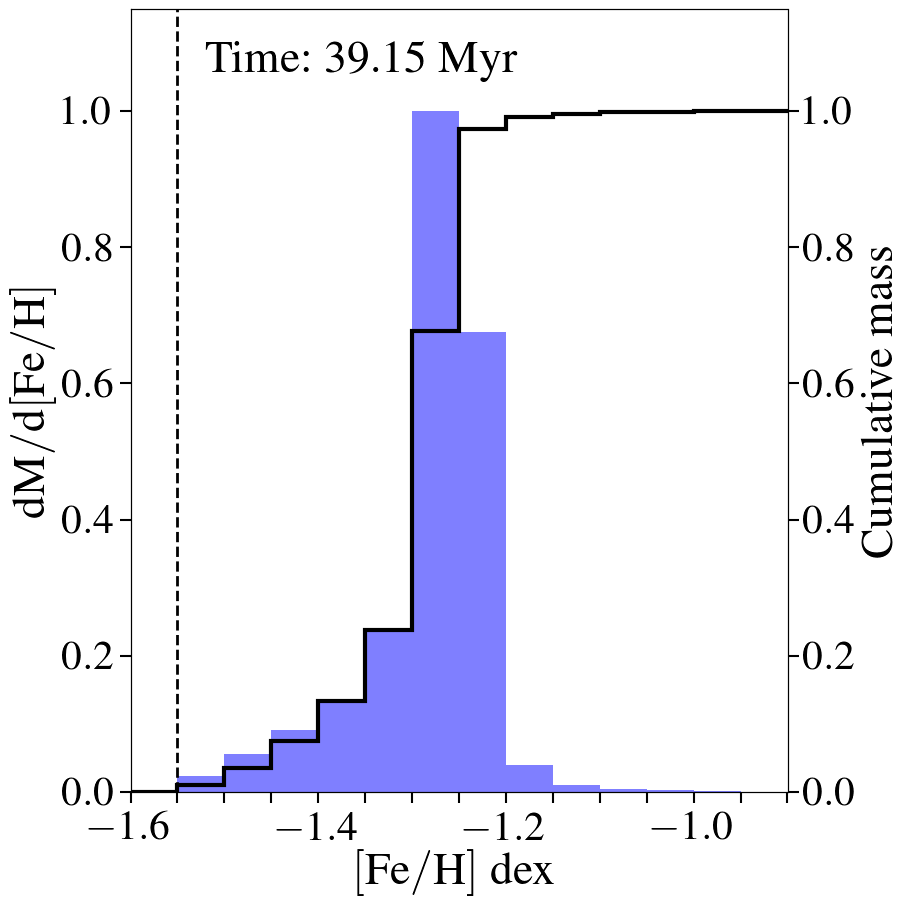}

 \caption{First and third columns: total density profile of SG stars at $t=6, 14, 26, 39$ Myr and density profiles of SG stars for several ranges of the helium mass fraction Y and the [Fe/H] ratio, respectively, for the high-density model with delayed cooling (${\rm HD\_DC}$). The FG density profile is also plotted (see the legend for the details). Second and forth columns: the mass distribution of Y and [Fe/H] ratio, respectively, in the SG stars at the aforementioned evolutionary times (reported in each panel). Other symbols and lines as in Figure \ref{fig:den&mdf_lowND}.}   
   \label{fig:den&mdf_highDC}
\end{figure*}

\subsection{High-density model with delayed cooling}

As shown in Figure \ref{fig:k&o15_high}, in the HD model, $3.5\%$ of all the SNe do not meet the \citet{kim&ostriker2015} criterion, hence they are basically not resolved. One potential risk of this result is that, in the HD simulation, the global effects of SN feedback might be underestimated.
In order to better assess the impact of this issue on our simulation and to be sure that this does not lead to a severe underestimation of SN feedback, we performed another simulation where radiative cooling has been delayed, i.e. temporarily switched off, in all the locations where SNe explode.
We have followed the prescriptions described in \citet{teyssier2013} to artificially turn off cooling. Here we focus on the density profiles of the SG and on its abundance patternsa shown in Figure \ref{fig:den&mdf_highDC}, which are quantities not particularly sensitive to the stochasticity of processes such as star formation or to the local effects of a small number of SN explosions.

As for such quantities, no substantial differences can be seen with respect to the HD model, which denotes that even though some SNe were not satisfying the condition of \citet{kim&ostriker2015}, the cluster capability to accreate gas and form new stars is not affected. We have also verified that the SNe not meeting the criterion are located far from the centre of the system, and therefore their effect, in the ${\rm HD\_DC}$ scenario, is not significantly perturbing the gas in the inner regions.


\section{Discussion}
\label{sec:discuss}
\begin{figure*}
        \centering

        \includegraphics[width=\textwidth]{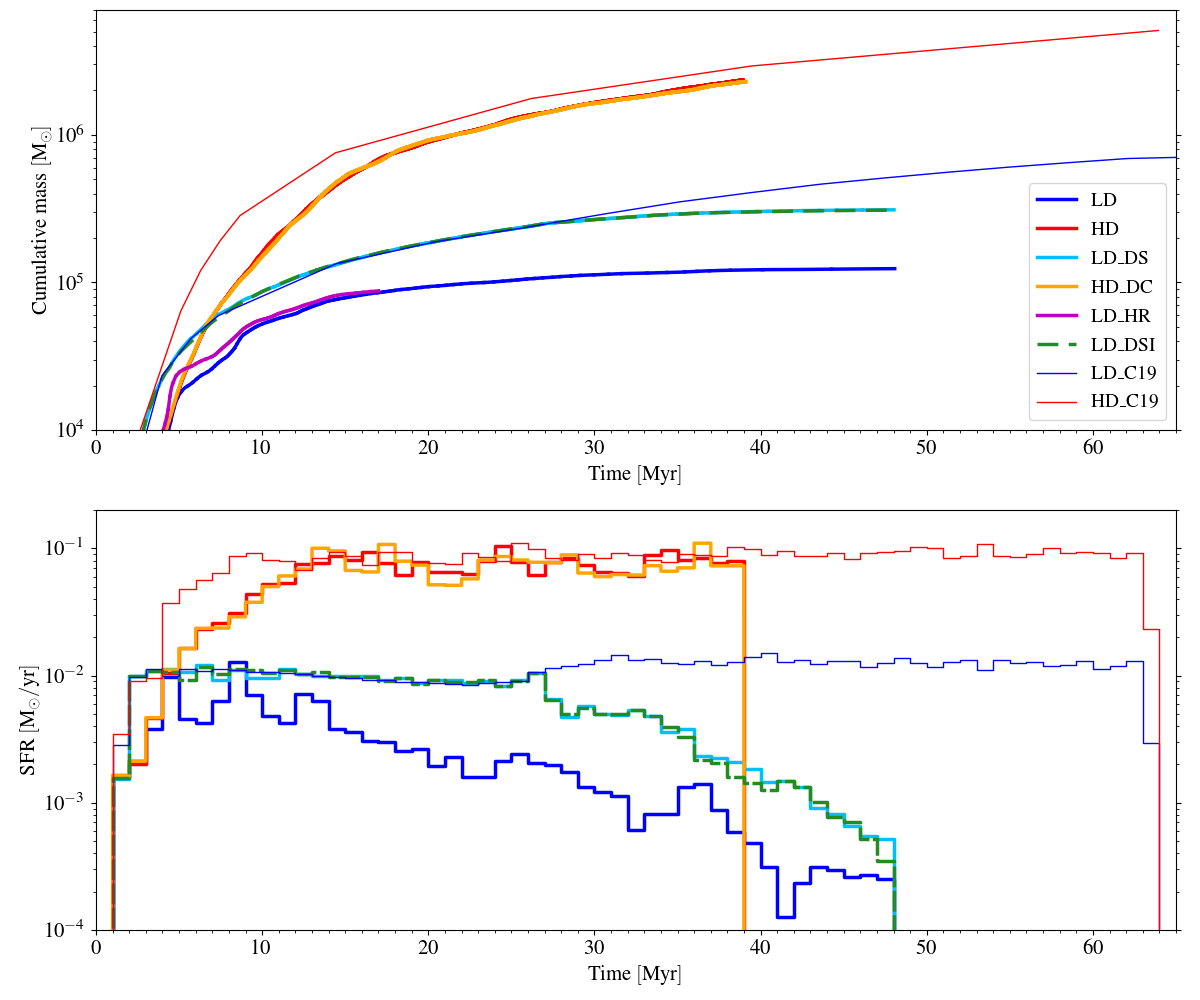}

\caption{Upper panel: evolution of the SG stellar mass for all our simulations. Lower panel: evolution of the SFR for all our models. The blue lines represent the standard low-density models while the red lines the high-density ones. The thin lines show the results obtained by \citetalias{calura2019} for their high resolution simulations. The purple line in the upper panel represents the result of the low-density model at high resolution (${\rm LD\_HR}$). The orange lines represent the high-density model with delayed cooling while the light-blue lines represent the low-density model with delayed Type Ia SNe. The green dashed lines represent the low density model where the infall is switched off once the first SN bubble have reached the negative $x$ boundary.} 
  \label{fig:sfr}
\end{figure*}

With the exception of \citet{dercole2008} who, by means of 1D simulations, found that a few SN Ia are able to halt the SF immediately after their explosion, our study is the first that addresses the effects of Type Ia SNe on the star formation in GCs through hydrodynamic simulations. 
It is however important to stress that Type Ia SNe do not play any role in regulating the star formation history of FG stars, which is expected to have lasted only a few Myr.

In Figure \ref{fig:sfr} we show the evolution of the cumulative mass in the upper panel and of the SFR in the lower one and compare our results with the ones of \citetalias{calura2019}. 

 As discussed in Section \ref{sec:results_LD}, with the short delay time adopted in this work the LD models are strongly affected by SN Ia feedback, which prevents the reaccretion of pristine gas. In the LD model, SN Ia feedback halts the SG star formation early limiting the total mass of the SG population. The infall of pristine gas, which starts at 21 Myr, does not have a significant effect on the evolution of the system, as shown in Figure \ref{fig:maps_lowND} and \ref{fig:den&mdf_lowND}.

Focusing on the SFR, it can be seen that its evolution is characterized by three major peaks in the first 15 Myr which are responsible for the peaks in the [Fe/H] mass distribution. Only during these peaks the SFR is comparable with the one of the model  SNe (${\rm LD\_C19}$), while between the peaks it is almost halved. After 15 Myr, the SFR decreases with several peaks at later times which, however, do not affect significantly the total stellar mass.

We have also plotted the time evolution of the cumulative mass for the ${\rm LD\_HR}$ model, namely the low density LD model at higher resolution (0.3 pc). At the beginning, it deviates from the result obtained at lower resolution, but this discrepancy has to be ascribed to the stochasticity of the SF process. At later times, in fact, the gap between the two models decreases, meaning that in our simulations the numerical convergence is satisfactory.

 Moving to the ${\rm LD\_DS}$ model, at the beginning both the cumulative mass and SFR evolution have the same trends of ${\rm LD\_C19}$, with small differences due to the stochastic method used to associate a mass to each newborn star, as we have described in Section \ref{sec:phy_ing}. At around 27 Myr, when the first supernovae start exploding, the SFR starts decreasing significantly, whereas it remains almost constant in the case without SNe. 

Once Type Ia start exploding, a comparison of the SFR evolution in the two low-density models shows that the one with delayed SNe is characterized by an overall smoother SFR decline. This difference is due to the lower gas density achieved in the LD case, which leads to a more inhomogeneous distribution of gas.

In Figure \ref{fig:sfr} we report also the results obtained for ${\rm LD\_DSI}$, a model similar to ${\rm LD\_DS}$, but where we stopped the infall at around $\sim 28$ Myr. The two models show a similar evolution of the SFR and cumulative mass. We can therefore conclude that our simplistic assumption regarding the infall implementation does not affect the evolution of the system. This is true even as far as the particle density profiles and the mass distribution of Y and [Fe/H] are concerned. 

On the other hand, our HD run predicts a lower SFR in the first 10 Myr than ${\rm HD\_C19}$, namely the case without SNe, which translates in a lower cumulative mass. However, after this first phase, both the SFR and the cumulative mass are similar to the values obtained without SNe. The total stellar mass at the end of our simulation is, in fact, ${\rm  2.3 \times 10^6 M_{\odot}}$ slightly smaller than the one obtained by \citetalias{calura2019} at the same evolutionary time. It results that the fraction of the initial FG over the total SG mass is equal to 4, which is in agreement with previous studies within the AGB scenario \citep{dantona2013,ventura2014} . However, it has to be noted that in this model the cumulative mass is still increasing at 39 Myr, therefore this ratio has to be taken as an upper limit.

Finally, the ${\rm HD\_DC}$ model shows a similar evolution of the cumulative stellar mass with the HD model, reaching the same final SG stellar mass. Small variations can be seen instead in the SFR evolution. The reason of such small differences lies in the stochastic method used in the SF implementation whose effects are more evident in the SFR than in the cumulative mass evolution.

To compare our results with observations we should refer to clusters of the same age, namely young massive clusters (YMCs), which are sometimes regarded as proto-GCs. \citet{bastian2013} did not find any evidence of ongoing star-formation in almost 130 YMCs with an age range between 10 Myr and 1Gyr. Similar results have been found by \citet{cabreraziri2014,cabreraziri2016}, however, YMC younger than 2 Gyr do not show any sign of multiple populations \citep{martocchia2018}. A comparison with YMCs of the same age is therefore not feasible.
As for the recently discovered high-redshift progenitors of GC, precise measures of their star formation rates and ages are still challenging to derive. To shed important light on the SF history of young GCs, a still star-forming system which already contains multiple populations should be found at high redshift. It is however not achievable with present instruments but perhaps possible in the future, in particular with the ELT with the aid of specific techniques, such as adaptive optics \citep{fiorentino2017,vanzella2019}.

\section{Conclusions}
\label{sec:conclu}
In this paper we have modelled, by means of grid-based, three-dimensional hydrodynamic simulations, the formation of SG stars in a massive cluster moving through a uniform gas distribution, including the feedback from FG Type Ia SNe, which represents the novelty of our work. Our aim is, specifically, to explore the role of Type Ia SN feedback
in determining the duration of SG formation epoch and study the effect of SN Ia ejecta on the chemical properties of the SG population.
Our simulations start at the end of the FG Type II SN epoch, when the ejecta of the most massive AGB stars are released in the cluster and the SG formation starts. During this stage, Type Ia SNe belonging to the FG start exploding and external gas with pristine (i.e. same as FG stars) composition is accreted from the surroundings. SG stars are therefore formed out a mixed gas whose composition depends on which of these three sources is dominant.
In all our simulations, the cluster is already in place composed only by FG low- and intermediate-mass stars and surrounded by an extended cavity created by the explosions of FG Type II SNe which is later replenished by the collapse of interstellar gas. 
We have studied models with two different values for the external pristine gas density: a low-density one of ${\rm \rho_{pg}=10^{-24} g\ cm^{-3}} $ and a high-density one of ${\rm \rho_{pg}=10^{-23} g\ cm^{-3}} $. Different values of the pristine gas density lead to different infall times, which, in turn, have a significant impact on the chemical composition and the stellar mass of the SG stars. 
Moreover, in this paper we have focused our attention on models with short SN Ia delay times. Given the large uncertainties still present on the timing of Type Ia SN explosions, we have considered two delay time distributions: firstly, we have
assumed that SNe Ia start exploding and releasing gas at the beginning of the SG formation epoch together with the most massive AGBs; then, we have performed a simulation, in the low-density scenario, where we have delayed the beginning of the SN Ia by 25 Myr.

Finally, we have run a simulation in which cooling is switched off locally, to assess the possible effects of the cooling overestimation in high density regions, a typical artificial phenomenon in numerical simulations. 

We summarize here our main results:
\begin{enumerate}[i)]
    \item In the low-density model with SN Ia explosions starting at the
beginning of the SG star formation epoch, the continuous explosions of Type Ia SNe do not halt the SG formation abruptly, as in 1D simulations by \citet{dercole2008}, but they significantly lower the SFR. At the beginning cooling flows are still formed but Type Ia SNe increasingly diminish their formation. Most of the SG stellar mass is formed, in fact, in the first 20 Myr, before the beginning of pristine gas accretion. SNe explosions confine the infall gas far from the system and therefore no accretion column is formed at variance with the case without Type Ia SNe. This implies that the AGB ejecta are poorly diluted with pristine gas, a necessary requirement for the AGB scenario in order to reproduce the observed anticorrelations. Moreover, the negligible dilution leads to the formation of a SG dominated by stars with extreme helium enrichment, at variance with observation, making this model not viable.  
    \item In the low-density model with Type Ia SNe starting 25 Myr after
the beginning of the SG formation, the SFR decreases after the first Type Ia SN explosions but it is always greater than the case without delay. The infall of pristine gas starts before the SN explosions, so some of it is accreted by the system and dilutes the AGB ejecta. At variance with the previous model, not all the stars show an extreme helium enrichment; a small fraction of stars display an intermediate helium enrichment as a result of a dilution and are spatially less concentrated than the extreme
population. The accretion of pristine gas continues for about 5 Myr but, eventually, Type Ia SNe confine the gas far from the cluster centre as it happened in the model without delay.  However, dilution is still too low and the spread in helium between the two generations is still much larger than the observed values. Therefore, longer delay times for the onset of Type Ia SNe are needed by the low density models in order to increase the contribution of the pristine gas and therefore the extent of dilution.    
    
    \item In the high-density model, Type Ia SNe have only mild effects both on the SFR evolution and on the helium enrichment. The final SG mass is only 1.2 times smaller than without Type Ia SNe. The higher density assumed for the pristine gas prevents Type Ia from confining it in the cluster outer regions, as found in the low-density models. In this model, the pristine gas slows the SN bubble expansion decreasing the growth of their filling factor. In this case an accretion column is present, even though it is continuously perturbed by SN explosions. The overall continuous accretion of gas leads to a SG dominated by stars with very modest helium enrichment  ($\overline{Y}_{\rm SG}=0.258$ ), similarly to what is obtained without Type Ia SNe. On the contrary, the SG iron distribution is significantly affected by the presence of Type Ia SNe. Since their bubbles are confined by the infalling gas, most of the iron released during their explosions is retained by the system and then recycled to form new stars which show a significant iron enrichment. The mean [Fe/H] ratio of SG stars is, in fact, equal to [Fe/H]$=-1.28$ dex. Assuming that, after the subsequent evolution, SG stars account for $70\%$ of the whole cluster, we derive an internal iron dispersion of ${\rm \sigma_{[Fe/H]}=0.14\ dex}$, in good agreement with the typical spread observed in Type II GCs.
    
    \item No substantial difference has been found between the standard high-density model and the one with a delayed cooling meaning that, overall, the impact of Type Ia SNe is not artificially suppressed by cooling. 
    
\end{enumerate}

It has to be noted that in all our simulations we do not include ionizing feedback from SG stars which could lead to a steeper decrease of the SFR \citep{chantereau2020}. Massive stars belonging to the SG, if any, should contribute to heat up the ISM further, reducing the amount of gas eligible for star formation. However, the SG formation is supposed to take place in an environment already crowded by the FG stars, a scenario in which the SF process is still unexplored \citep{renzini2015}. The shape of the IMF characterizing the SG is not known so far, even though some studies have suggested that it could be truncated at around ${m \sim \rm 8-10 M_{\odot}}$ \citep{dercole2010,bekki2019}.



Our simulations have shown that the SG formation can continue during the Type Ia SNe epoch and the ejecta of these supernovae can contribute to the mix of gas out of which SG stars form. The formation of SG stars from mixed gas of SN Ia ejecta and AGB gas revealed by our study indicates a possible avenue for the formation of Galactic GC with multiple population characterized by a spread in Fe; these clusters represent about 20 per cent of the Galactic globular cluster system (see \citealt{marino2015,johnson2015,milone2017,marino2019}) and have been shown to host stellar populations enriched also in s-process elements which could be produced by low-mass AGBs (see e.g. \citealt{dantona2016}). Moreover, as shown by \citet{gratton2019}, Type II GCs are relatively massive, which might suggests a correlation between the iron enrichment and the cluster mass. We intend to further extend the initial set of simulations presented in this paper to study a broader range of initial conditions  and explore the possible dynamics and formation history of these complex clusters.

\section*{Acknowledgements}
The authors thank the anonymous referee for having carefully read the manuscript and provided suggestions that helped to improve it. 
We are grateful to L. Greggio, A. Milone and E. Dalessandro for useful discussions and suggestions.  FC acknowledges support from the INAF mainstream
(1.05.01.86.31) and from PRIN INAF 1.05.01.85.01.  EV acknowledges support from NSF grant AST-2009193. We acknowledge the computing centre  of Cineca and INAF, under the coordination of the "Accordo Quadro MoU per lo svolgimento di attività congiunta di ricerca Nuove frontiere in Astrofisica: HPC e Data Exploration di nuova generazione", for the availability of computing resources and support. We acknowledge the use of computational resources from the parallel computing cluster of the Open Physics Hub (https://site.unibo.it/openphysicshub/en) at the Physics and Astronomy Department in Bologna. This research was supported in part by Lilly Endowment, Inc., through its support for the Indiana University Pervasive Technology Institute, and in part by the Indiana METACyt Initiative. The Indiana METACyt Initiative at IU is also supported in part by Lilly Endowment, Inc.


\section*{Data Availability}

 The data underlying this article will be shared on reasonable request to the corresponding author.



\bibliographystyle{mnras}
\bibliography{sne1a_gc} 








\bsp	
\label{lastpage}
\end{document}